\documentclass[]{aa}

\usepackage[utf8]{inputenc}
\usepackage{graphicx}
\graphicspath{{../figures/}}
\usepackage{natbib}
\usepackage{hyperref}
\hypersetup{pdfpagemode = {UseNone},
            pdftitle = {title},
            pdfauthor = {authors},
            pdfsubject = {},
            pdfview = {FitH},
            pdfstartview = {FitH},
            colorlinks = {true},
            linkcolor = [rgb]{0,0.35,0.7},
            citecolor = [rgb]{0,0.35,0.7},
            filecolor = [rgb]{0.61,0,0},
            urlcolor = [rgb]{0.61,0,0},
           }
\usepackage{txfonts}
\usepackage{enumitem}
\usepackage{placeins}
\usepackage{listings}

\newcommand{\codestyle}[1]{\texttt{#1}}
\newcommand{\fargo}{\textsc{FARGO} }
\newcommand{\fargocpt}{\textsc{FargoCPT} }
\newcommand{\fargoadsg}{\textsc{FARGOADSG} }
\newcommand{\fargothreed}{\textsc{FARGO3D} }
\newcommand{\fargotwodoned}{\textsc{FARGO\-2D1D} }
\newcommand{\fargoca}{\textsc{FARGOCA} }
\newcommand{\fargothorin}{\textsc{FARGO\_THORIN} }
\newcommand{\gfargo}{\textsc{GFARGO} }
\newcommand{\rebound}{\textsc{Rebound} }

\newcommand{\pluto}{\textsc{PLUTO} }

\begin{document}

\title{\fargocpt: A 2D Multi-Physics Code for Simulating the Interaction of Disks with Stars, Planets and Particles}
\titlerunning{\fargocpt: A 2D Multi-Physics Code for Planet-Disk Interaction Simulations}

\author{Thomas~Rometsch\inst{1}
	\and
	Lucas~M.~Jordan\inst{2}
	\and
	Tobias~W.~Moldenhauer\inst{2}
	\and
	Dennis~Wehner\inst{3}
	\and
	Steven~Rendon~Restrepo\inst{4}
	\and
	Tobias~W.\,A.~M\"uller
	\and
	Giovanni~Picogna\inst{5}
	\and
	Wilhelm~Kley
	\thanks{deceased}\fnmsep\inst{2}
	\and
	Cornelis~P.~Dullemond\inst{1}
}

\institute{Institut für Theoretische Astrophysik, Zentrum für Astronomie (ZAH), Universität Heidelberg,
	Albert-Ueberle-Str. 2, 69120 Heidelberg, Germany\and
	Institut für Astronomie und Astrophysik, Universit\"at T\"ubingen,
  	Auf der Morgenstelle 10, 72076 T\"ubingen, Germany
  	\and
	Fakultät für Physik, Universität Duisburg-Essen, Lotharstraße 1, 47057 Duisburg, Germany 
	\and
	Leibniz-Institut für Astrophysik Potsdam (AIP), An der Sternwarte 16, 14482 Potsdam, Germany
	\and
	Universitäts-Sternwarte, Fakultät für Physik, Ludwig-Maximilians-Universität München, Scheinerstr. 1, 81679 München, Germany\\
	\email{rometsch@uni-heidelberg.de} 
}

\date{\today}
%BEGIN_FOLD
\abstract
{% context heading (optional)
Planet-disk interactions play a crucial role in the understanding of planet formation and disk evolution. 
There are multiple numerical tools available to simulate these interactions, including the often-used \fargo code and its variants.
Many of the codes were extended over time to include additional physical processes with a focus on their accurate modeling.
}
{% aims heading (mandatory)
We introduce \textsc{FargoCPT}, an updated version of \fargo incorporating other previous enhancements to the code, to provide a simulation environment tailored to study interactions between stars, planets, and disks, ensuring accurate representation of planet systems, hydrodynamics, and dust dynamics with a focus on usability.
}
{% methods heading (mandatory)
The radiation-hydrodynamics part of \fargocpt uses a second-order upwind scheme in 2D polar coordinates supporting multiple equations of state, radiation transport, heating and cooling, and self-gravity. Shocks are considered using artificial viscosity. 
Integration of the N-body system is achieved by leveraging the REBOUND code. 
The dust module utilizes massless tracer particles, adapting to drag laws for the Stokes and Epstein regimes. 
Moreover, \fargocpt provides mechanisms to simulate accretion onto the stars and planets.
}
{% results heading (mandatory)
The code has been tested in practice by its use in various publications.
Additionally, it comes with an automated test suite to test the physics modules.
}
{% conclusions heading (optional), leave it empty if necessary
\fargocpt offers a unique set of simulation capabilities within the current landscape of publicly available planet-disk interaction simulation tools.
Its structured interface and underlying technical updates are intended to assist researchers in the ongoing exploration of planet formation.
}
%END_FOLD


\maketitle
%
%________________________________________________________________

%%%%%%%%%%%%%%%%%%%%%%%%%%%%%%%%%%%%%%%%
%%%%%%%%%%%%%%%%%%%%%%%%%%%%%%%%%%%%%%%%
%%%%%%%%%%%%%%%%%%%%%%%%%%%%%%%%%%%%%%%%
\section{Introduction}
%%%%%%%%%%%%%%%%%%%%%%%%%%%%%%%%%%%%%%%%
%%%%%%%%%%%%%%%%%%%%%%%%%%%%%%%%%%%%%%%%
%%%%%%%%%%%%%%%%%%%%%%%%%%%%%%%%%%%%%%%%

Planet migration is a crucial part of our understanding of planet formation.
Besides analytical or semi-analytical calculations, one way to study it is by
 hydrodynamics calculations coupled with gravitational N-body simulations.
Specifically, in the context of planet formation, the shape of the hydrodynamic
object resembles a thin, flared disk. \fargo \citep{masset_fargo_2000} is a
computer program to simulate these protoplanetary disks by numerically solving
the hydrodynamics equations on a staggered grid using a second-order upwind scheme.
It includes a special algorithm, with the same name, which can relax the time step
constraint by making use of the axial symmetry of the disk flow, thereby reducing
the computational cost of simulations. The specific form of the hydrodynamics equations
solved by the program is stated later in Sect.~\ref{sec:physics}.

Because \fargo is tailored to the study of protoplanetary disks, the code uses a cylindrical grid, reflecting the major symmetry of the system.
One main assumption used in the code is that the simulated disks are thin in the sense that their vertical extent is small compared to the radial distance from the central star.
This is the justification for approximating the three-dimensional disk with a two-dimensional representation.
Using this approximation significantly reduces the required time for simulations of protoplanetary disks and enables the long-term study of these systems.
In its original form, the code employed the \mbox{(locally-)}isothermal assumption, in which the temperature is assumed to be fixed in time and only dependent on the distance from the central start.
This allows solving the energy equation analytically and additionally reduces the computational cost of the simulation.

In the past two decades, studies have shown that additional effects play an important role in the evolution of protoplanetary disks and their interaction with embedded planets \citep[for overviews see][]{kley_planet-disk_2012,baruteau_planet-disk_2014,paardekooper_planet-disk_2023}.
Relevant physical processes include the heating and cooling of the gas \citep[e.g.][]{baruteauCorotationTorqueRadiatively2008a, kley_migration_2008, paardekooper_disc_2008,lega_migration_2014,masset_coorbital_2017}, self-gravity \citep[e.g.][]{pierens_how_2005,baruteauTypePlanetaryMigration2008}, and radiation transport \citep[e.g.][]{morohoshi_gravitational_2003,paardekooper_halting_2006}.
We have added all of these effects to our version of the \fargo code and made them fully controllable from the input file and verified their correctness through extensive testing. We also extended the number of physical quantities that are evaluated during the simulation and written to output files.

While there are multiple options to run planet-disk interaction simulations including the aforementioned effects, the authors recognized a need for a simulation code that is not only able to simulate the relevant physical processes accurately but is also easy to use.
This is especially important for students and researchers who are not experts in modifying and compiling C/C++ or FORTRAN codes in a typical Linux environment.
These skills can pose a significant hurdle to run simulations.

The primary aim behind publishing this code is to provide a comprehensive simulation tool for planet-disk interactions that remains accessible and useful to both students and more senior researchers.
To this end, the testing suite for the physical modules is fully automated with pass-fail tests so it can be run after every future modification of the code to ensure that it is still working as intended.

On the physics side, a significant enhancement is the incorporation of radiation physics.
Furthermore, the introduction of a particle module enables in-depth studies on the impact of embedded planet systems on the structure and dust distribution of planet-forming disks, a task that is currently often performed to model disk observations.
We also added cooling and mass inflow functions and an updated equation of state specifically tailored to study cataclysmic variable systems.
Additionally, the code can simulate the disk centered around any combination of N-body objects.
This enables, for instance, simulations of circumbinary disks in the center of mass-frame or simulations of a circumsecondary disk in a binary star system.

Simulation programs for planet-disk interaction can, broadly speaking be organized into two main categories: Lagrangian and Eulerian methods.
Lagrangian methods trace the dynamics of the disk by following the motion of a large number of particles.
The most common example of this is the Smoothed Particle Hydrodynamics (SPH) method with a prominent example being the PHANTOM code \citep{price_phantom_2018}.

The Eulerian methods solve the hydrodynamics equations on a grid, which can either be fixed or dynamic.
In this category, again, we can distinguish between two main approaches used in astrophysics:
methods that require artificial viscosity and Godunov methods \citep{woodward1984art_vis_and_godunov}.
Either of these methods can be implemented as a finite-volume or finite-difference method.
Godunov methods solve the Riemann problem at the cell interfaces to compute the fluxes through the interfaces which
allows for an accurate treatment of shocks and an approximate treatment of smooth flows.
Prominent examples include the \pluto code \cite{mignone2007pluto} and the Athena++ code \citep{stone_athena_2020}.

Schemes based on artificial viscosity can be derived as a
finite-difference scheme from Taylor series expansions of the Euler equations \citep{woodward1984art_vis_and_godunov}. 
This technique assumes that the solution is smooth, which is not the case at 
discontinuities such as shocks. It approximates shocks by smearing out discontinuities into smooth
regions of strong gradients via artificial viscosity.
The ZEUS code \citep{stone1992zeus1} and the \fargo code (which is based on the ZEUS code)
are prominent examples of this approach.
While these codes were classified as finite-difference schemes, they employ a staggered grid and implement their advection step by splitting the domain into cells and computing the fluxes through the cell interfaces.
The cell as a basic geometrical unit, and the sharing of fluxes among adjacent cells which ensures conservation of the advected quantities to machine-precision by construction, warrant the classification of the advection scheme as finite-volume \citep{anderson_computational_2020}.
The authors lean towards this classification because it rests on an intrinsic property of the scheme.
However, the crucial distinction seems to be the one between Godunov methods and methods that require artificial viscosity and the \fargo code family falls into the latter category.

While formally the Godunov methods are more accurate in handling discontinuities like shocks,
the artificial viscosity methods are better at modeling smooth flows and, in practice,
are often more robust and easier to use.
Both approaches have their advantages and disadvantages and the choice of which
method to use depends on the specific problem at hand. 
For a specific example, see \citet{ziampras_buoyancy_2023} who used, both, \pluto and \fargothreed to investigate buoyancy waves in the coorbital region of small-mass planets.

The original \fargo code was introduced by \citet{masset_fargo_2000}.
Based on this code, other groups developed their version of the \fargo code.
Examples include \fargothorin \citep{chrenkoEccentricityExcitationMerging2017}, \fargoca \citep{lega_migration_2014}, \fargotwodoned \citep{cridaSimulatingPlanetMigration2007b}, and the \gfargo code \citep{regalyPossiblePlanetformingRegions2012a,massetGfargoFargoGpu2015}, the predecessor of our code \fargoadsg \citep{baruteauCorotationTorqueRadiatively2008a,baruteauTypePlanetaryMigration2008,baruteauGasDustHydrodynamical2016a}, as well as the official successor \fargothreed \citep{benitez-llambayFARGO3DNewGPUoriented2016}.

In this work, we present \textsc{FargoCPT}, the \fargo version developed by members of the Computational Physics group at the University of Tübingen since 2012.
Publications from this group that used the code include \citet{mullerCircumstellarDisksBinary2012,mullerModellingAccretionTransitional2013},
\citet{picogna_how_2015}, \citet{rometsch_migration_2020,rometschSurvivalPlanetinducedVortices2021b}, and \citet{jordanDisksCloseBinary2021}.

Furthermore, we present novel approaches for handling challenging aspects of planet-disk interaction simulations. We introduce a new method for handling the indirect term suitable for simulations centered on individual stars in binary systems (Sect.~\ref{sec:shift-based-indirect-term}), a local viscosity stabilizer for use in the context of cataclysmic variables (Sect.~\ref{sec:viscosity_stabilizer}) and a correction for dust diffusion treated by stochastic kicks (Sect.~\ref{sec:dust_diffusion}).

This document is structured as follows.
First, the physical problem at hand is sketched in Sect.~\ref{sec:physics}.
Then, the newly introduced physics modules are described in Sect.~\ref{sec:improvements}.
Software engineering and usability aspects of the code are discussed in Sect.~\ref{sec:software}.
Finally, the paper is concluded in Sect.~\ref{sec:discussion} with a discussion of the code.
The Appendix includes a presentation of various test cases included in the automatic test suite.

%%%%%%%%%%%%%%%%%%%%%%%%%%%%%%%%%%%%%%%%
%%%%%%%%%%%%%%%%%%%%%%%%%%%%%%%%%%%%%%%%
%%%%%%%%%%%%%%%%%%%%%%%%%%%%%%%%%%%%%%%%
\section{Physical system}
%%%%%%%%%%%%%%%%%%%%%%%%%%%%%%%%%%%%%%%%
%%%%%%%%%%%%%%%%%%%%%%%%%%%%%%%%%%%%%%%%
%%%%%%%%%%%%%%%%%%%%%%%%%%%%%%%%%%%%%%%%
\label{sec:physics}

The \fargocpt code is a computer program that numerically solves the vertically-integrated radiation-hydrodynamics equations in the one-temperature approximation coupled with an N-body system.
They read
\begin{align}
	\label{eq:hydroeq_mass}
	\frac{\partial \Sigma}{\partial t} + \nabla \cdot (\Sigma \vec{u}) &= 0 \\
	\label{eq:hydroeq_momentum}
	\frac{\partial \Sigma \vec{u}}{\partial t} + \nabla \cdot (\Sigma \vec{u} \otimes \vec{u}) &= -\nabla P + \Sigma \vec{k} + \nabla \tau \\
	\label{eq:hydroeq_energy}
	\frac{\partial e}{\partial t} + \nabla \cdot (e \vec{u}) &= -P \nabla \vec{u} + \mathcal{S} + \mathcal{RT}
\end{align}
with the surface density $\Sigma = \int_{-\infty}^{\infty} \rho \,\mathrm{d}z$ as the vertically integrated gas volume density $\rho$, the gas velocity $\vec{u}$, the vertically integrated internal energy density $e = \Sigma \epsilon$ with the specific internal energy $\epsilon$,
the vertically integrated pressure $P = \int_{-\infty}^{\infty} p \,\mathrm{d}z$, accelerations $\vec{k}$ due to external forces (e.g. due to gravity), the viscous stress tensor $\tau$, and heat sinks and sources $\mathcal{S}$ (see Sect.~\ref{sec:energy_equation}).
The last term represents radiation transport
\begin{align}
	\mathcal{RT} = - \int_{-\infty}^{\infty} \nabla \vec{F} \mathrm{d}z
\end{align}
with the radiation flux $\vec{F}$ in three dimensions.
In the code, the vertical component of the radiation flux is split off and treated with an effective model, see Sect.~\ref{sec:energy_equation}.
Please note, that the code handles radiation hydrodynamics using the one-temperature approach which is discussed in more detail in Sect.~\ref{sec:fld}.

In polar coordinates $(r, \phi)$, to which the code is tailored, the equations read \citep[e.g.][]{massetCoorbitalCorotationTorque2002}
\begin{align}
% density
\frac{\partial \Sigma}{\partial t}+\frac{1}{r} \frac{\partial\left(r u_r \Sigma\right)}{\partial r}+\frac{1}{r} \frac{\partial\left(u_\phi \Sigma\right)}{\partial \phi} &=0\,,\label{eq:density_eq_polar}  \\
% velocity r
\frac{\partial u_r}{\partial t}+u_r \frac{\partial u_r}{\partial r}+\frac{u_\phi}{r} \frac{\partial u_r}{\partial \phi}-\frac{u_\phi^2}{r} 
&=-\frac{1}{\Sigma} \frac{\partial P}{\partial r}+ k_r+\frac{f_r}{\Sigma}\,, \label{eq:momentum_r_eq_polar} \\
% velocity phi
\frac{\partial u_\phi}{\partial t}+u_r \frac{\partial u_\phi}{\partial r}+\frac{u_\phi}{r} \frac{\partial u_\phi}{\partial \phi}+\frac{u_r u_\phi}{r} 
&=-\frac{1}{\Sigma r} \frac{\partial P}{\partial \phi} + k_\phi+\frac{f_\phi}{\Sigma}\,, \label{eq:momentum_phi_eq_polar} \\
% energy
\frac{\partial e}{\partial t} + \frac{1}{r} \frac{\partial\left(r u_r e\right)}{\partial r}+\frac{1}{r} \frac{\partial\left(u_\phi e\right)}{\partial \phi}
&= -\frac{P}{r} \frac{\partial \left(r u_r\right)}{\partial r} - \frac{P}{r}\frac{\partial u_\phi}{\partial \phi} + \mathcal{S} + \mathcal{RT} \label{eq:energy_eq_polar}
\end{align}
where $f_r$ and $f_\phi$ are the radial and azimuthal forces per unit area due to viscosity (see Eqs.~\eqref{eq:visocisty_momentum_update_r} and \eqref{eq:visocisty_momentum_update_phi}).
For a rotating coordinate system, we follow the conservative formulation of \citet{kley1998coriolis} and add the respective terms to the two momentum equations.

The l.h.s. of these equations are the transport step and the r.h.s. are the source terms.
Following the scheme of the ZEUS code \citep{stone1992zeus1}, the transport step is solved by a finite-volume method based on an upwind scheme with a second-order slope limiter \citep{van1977towards} and the code can make use of the \fargo algorithm \citep{masset_fargo_2000} to speed up the simulation.
The source terms are updated as described in Sect.~\ref{sec:order_of_operations} using first-order Euler steps or implicit updates.
The definitions for the heating and cooling term $\mathcal{S}$ and the radiation transport term $\mathcal{RT}$ are given in Sect.~\ref{sec:energy_equation}.

The external accelerations $\vec{k}$ are due to the gravitational forces from the star(s) and planets, correction terms in case of a non-inertial frame, and the self-gravity of the disk
\begin{align}\label{eq:external_forces}
\vec{k} = - \nabla (\Phi_\mathrm{Nb} + \Phi_\mathrm{Ind}) - a_\mathrm{SG}\,.
\end{align}
Interaction of the disk with the N-body objects is considered via the gravitational potential, see Eq.~\eqref{eq:gravitational_potential_nbody}.
Self-gravity of the gas is considered as an acceleration (see Sect.~\ref{sec:self-gravity}).

The N-body objects feel the gravitational acceleration $\vec{a}$ exerted by the gas.
This is computed by summation of the smoothed gravitational acceleration over all grid cells.
Refer to Sect.~\ref{sec:gravitational_smoothing} for formulas and details about the smoothing.
Because the simulation can be run in a non-inertial frame of reference, correction terms are applied as detailed in Sect.~\ref{sec:shift-based-indirect-term}.

Finally, \fargocpt features a particle module based on Lagrangian super-particles, where a single particle might represent any number of physical dust particles or solid bodies.
The particles feel the gravity from the N-bodies and interact with the gas through a gas-drag law.
Additionally, dust diffusion is modeled to consider the effects of gas turbulence.
The particle module is described in detail in section \ref{sec:dust}.

The next section describes the various physics modules that were added to the code.

%%%%%%%%%%%%%%%%%%%%%%%%%%%%%%%%%%%%%%%%
%%%%%%%%%%%%%%%%%%%%%%%%%%%%%%%%%%%%%%%%
%%%%%%%%%%%%%%%%%%%%%%%%%%%%%%%%%%%%%%%%
\section{Improvements of the physics modules}
%%%%%%%%%%%%%%%%%%%%%%%%%%%%%%%%%%%%%%%%
%%%%%%%%%%%%%%%%%%%%%%%%%%%%%%%%%%%%%%%%
%%%%%%%%%%%%%%%%%%%%%%%%%%%%%%%%%%%%%%%%
\label{sec:improvements}

We built upon the original \fargo code by \citet{masset_fargo_2000} and its improved version \fargoadsg.
It already included treatment of the energy equation \citep{baruteauCorotationTorqueRadiatively2008a}, self-gravity \citep{baruteauTypePlanetaryMigration2008}, and a Lagrangian particle module \citep{baruteauGasDustHydrodynamical2016a}.
We advanced these modules, added new physics modules, and added features to improve the usability of the code.
For a detailed description of the \fargoadsg code, and with it of the underlying hydrodynamics part of the code presented here, we recommend Chapter 3 of \citet{baruteauPredictiveScenariosPlanetary2008a}.

This section starts with outlining the order of operations in the operator splitting approach and then describes the various new features and changes.

%%%%%%%%%%%%%%%%%%%%%%%%%%%%%%%%%%%%%%%%
\subsection{Order of operations and interaction of subsystems}
%%%%%%%%%%%%%%%%%%%%%%%%%%%%%%%%%%%%%%%%
\label{sec:order_of_operations}

This section details the order in which the physical processes are considered during one iteration in the code and how they interact.

For the update step, we used the sequential operator splitting (also known as Lie-Trotter splitting)
where possible, meaning that we always use the most up-to-date quantities from the previous operators when applying the current operator.
This is the simplest and oldest splitting scheme and has better accuracy than applying all operators using the quantities at the beginning of the step.
This scheme is called additive splitting \citep[e.g.][]{geiser2017new}.
Each time step starts with accretion onto the planets. Conceptually, this is the same as performing planet accretion at the end of the time step, except for the first and last iteration of the simulation.

Then, we compute the gravitational forces between the N-body objects and the gas, between the N-body objects and the dust particles, and the self-gravity of the disk.
At this stage, the indirect term, i.e. the corrections for the non-inertial frame, is computed and added to the gravitational interaction.
These are then applied to the subsystems by updating the velocity of the N-body objects, by updating the acceleration of the dust particles, and by updating the potential of the gas.
Experience shows, that for the interaction between N-body and gas the positions of the N-body objects have to be at the same time as the gas.
From this point on, the N-body system and the particle system evolve independently in time until the end of the time step.

The gas velocities are first updated by the self-gravity acceleration and then by the N-body gravity potential, the pressure gradient, and, in the case of the radial update, also the centrifugal acceleration.
At this step, we update the internal energy of the gas with compression heating.
It is important to perform the energy update at this step before the viscosity is applied to avoid instabilities.
Then, we sequentially update velocities and the internal energy further by artificial viscosity, then viscosity (both of which depend on the gas velocities), and then apply heating and cooling terms as well as radiation transport.
The heating and cooling steps are applied simultaneously for numerical stability.
Finally, the internal energy is updated by radiation transport.

Once all the forces and source terms are applied, we conduct the transport step. 
Boundary conditions and the wave-damping zone are applied at the appropriate sub-steps throughout the hydrodynamics step.

\begin{figure*}
	\begin{center}
	\includegraphics[width=\linewidth]{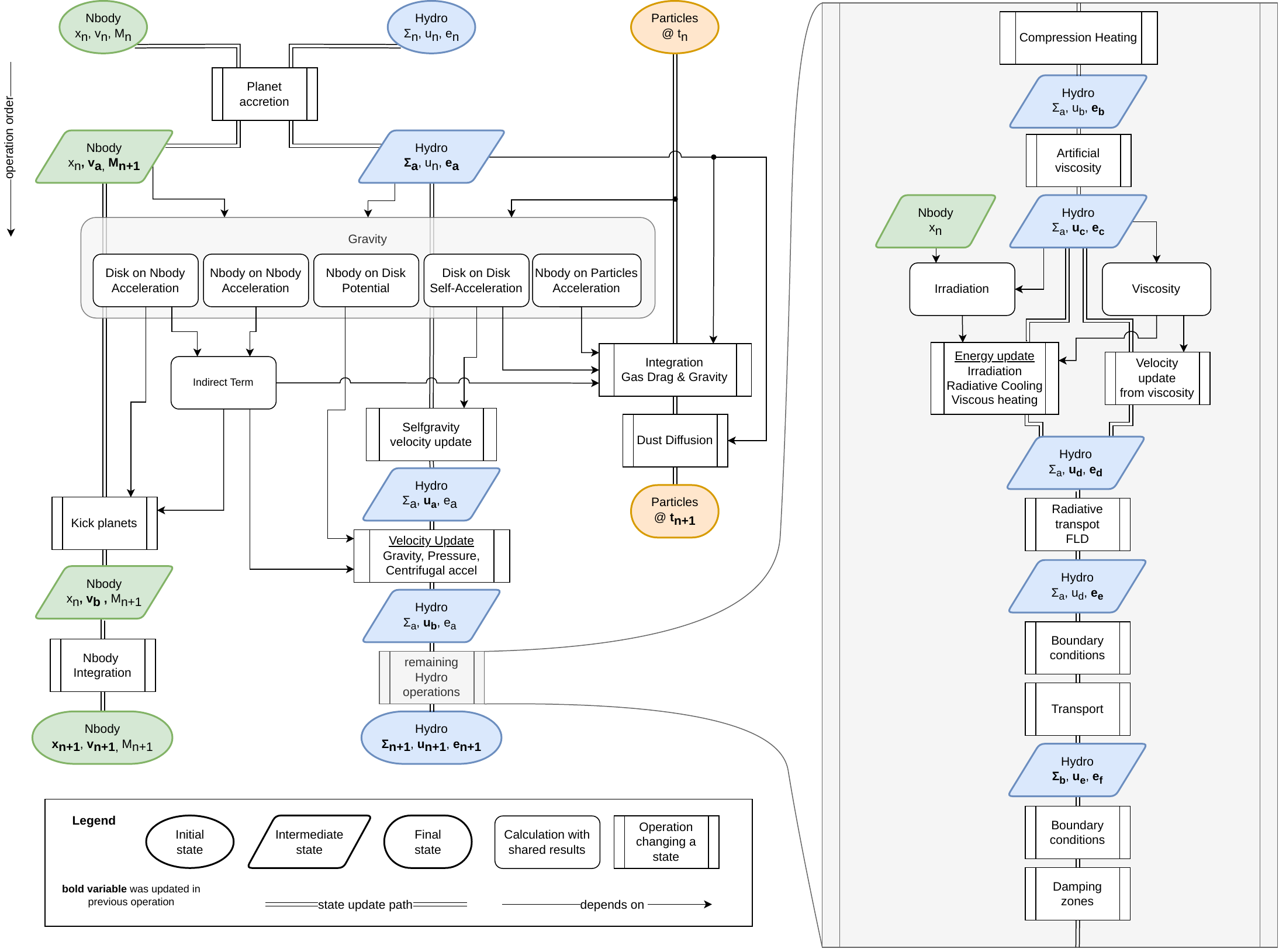}
	\caption{Order of operations in the operator splitting scheme showing the evolution of the N-body, hydro, and particle subsystems throughout one iteration step. 
	Single-line arrows mean that the originating object is used in the calculation of the destination.
	Double lines indicated the evolution of one of the subsystems from the initial state (ellipse) over intermediate states (parallelogram) to the final state (round shape).
	Each rectangle with rounded corners is a computation of an intermediate quantity using the state of the subsystem to which it is connected.
	The rectangles with bars on the sides indicate an operation that changes the state of a system.
	The variables that were changed by an operation are indicated in boldface in the intermediate states.
	\label{fig:flow}}
	\end{center}
\end{figure*}

A detailed diagram of the order of the operations during each time step and the interactions between the subsystems are shown in Fig.~\ref{fig:flow}.
The diagram illustrates how each of the subsystems is advanced in time and between which sub-steps information is exchanged between the subsystems.
The next paragraphs guide through the diagram starting from the top, explaining the meaning of the differently shaped patches, arrows and lines.

Each of the three subsystems is colored differently: the N-body system is shown in green, the hydrodynamics in blue and the dust particles in orange.
A colored patch represents the \emph{state} of a subsystem.
Elliptical patches indicate the initial state at the beginning of a time step, parallelograms indicate intermediate states and colored rectangles with rounded corners indicate the final state at the end of the time step.
Each state is labeled with variable names.
The respective subscript indicates the the sub-step during the operator splitting.
For example, the energy is updated from $e_n$, first to $e_a$, then to $e_b$ and after additional steps finally to $e_{n+1}$.
These variables will be used in Sect.~\ref{sec:energy_equation} to refer to the sub-steps.
A variable printed in bold text indicates that the variable was changed by the last operation.

A white rectangle with rounded corners indicates an \emph{intermediate calculation}, the result of which is used in multiple operations.
For example, the calculation of the indirect term is used in all three subsystems.
Finally, the rectangles with bars on the sides indicate an \emph{operation} that changes the state of a system.

Double lines trace the change of a subsystem throughout the time step.
Lines with arrows attached at the end emerge from a state patch or an intermediate calculation and end in an operation or intermediate calculation patch.

The large rectangle with bars shaded in grey at the right side of the diagram illustrates the sub-steps involved in the hydrodynamics part of the simulation.
Except for the irradiation operation which requires the position of the N-bodies, this part of the simulation is independent of the N-body system.

The shaded rectangle with rounded corners in the upper third of the diagram illustrates the various gravitational interactions between the subsystems.
The information indicated by arrows that end at the borders of this patch can be applied in any of the contained intermediate calculations.

The operations are ordered from the top to bottom in the order they are applied during the time step.
This means, that an operation or a state is generally only influenced by states above it or at the same level.
Please follow the arrows to get a sense of the order of operations.

\subsection{N-body module}

The original \fargo code includes a Runge-Kutta fifth-order N-body integrator, which was used with the same time step as the hydro simulations.
This could lead to some situations in which the time step was too large for the N-body simulation, for instance, during close encounters in simulations of multiple migrating planets.
Instead of implementing our N-body code with adaptive time-stepping, we incorporated the well-established and well-tested \rebound code \citep{rein_rebound_2012} with its 15th order IAS15 integrator \citep{rein_ias15_2015}
into \fargocpt.
\rebound is called during every integration step of the hydro system to advance the N-body system for the length of the CFL time step.
During this time, multiple N-body steps might be performed as needed.
The interaction with the gas is incorporated by adding $\Delta t\,\vec{a}$ to the velocities of the bodies before the N-body integration.

We changed the central object from an implicit object, which was assumed to be of mass 1 in code units placed at the origin, to a moving point mass.
It is now treated exactly like any other N-body object, notably including gravitational smoothing, which was not applied to the star before
This solves an inconsistency between the interaction of disk and planets, and disk and star.
This is necessary as the smoothing simulates the effect of the disk being stratified in the vertical direction and makes the 
potential behave more like it would in 3D \citep{mullerTreatingGravityThindisk2012}.

Changing the star from being an implicit object to an explicit one, we changed all the relevant equations, which now include the mass of the central object and the distance to it.

%%%%%%%%%%%%%%%%%%%%%%%%%%%%%%%%%%%%%%%%
\subsection{Gravitational interactions}
%%%%%%%%%%%%%%%%%%%%%%%%%%%%%%%%%%%%%%%%
\label{sec:gravitational_interactions}

The dominating physical interaction between planets and a disk is gravity.
In principle, any object that has mass causes a gravitational acceleration on any other object.

In our simulation, N-body objects are considered to have a mass and particles are considered to be massless.
The gas disk can be configured in one of three states.
The gas disk can
\begin{enumerate}
\item be massless and not accelerate the N-bodies.
\item have mass, accelerate the N-bodies, the particles and itself (self-gravity included).
\item have mass, accelerate the N-bodies, but not itself or the particles (self-gravity ignored).
\end{enumerate}

The first two cases are consistent whereas the third case is inconsistent.
However, it is a commonly used approximation.
The reason for this is that the computation of the (self-)gravity of the gas disk is computationally expensive.

In the third case, when the forces on the N-bodies due to the gas are computed using the full surface density, the N-bodies feel the full mass of the disk interior to their orbit while gas on the same orbit does not feel this mass.
Consequently, their respective equilibrium state angular velocities differ which causes a nonphysical shift in resonance locations and thus in the torque and migration rate experienced by the planets \citep{baruteauTypePlanetaryMigration2008}.
This mismatch can be alleviated by simply subtracting the azimuthal average of the surface density in the calculation of the force.
This is done in the code in the case that self-gravity is disabled.

Our code does not support a configuration in which only one of the N-bodies feels the disk, such as a situation in which the planet feels the disk but the star does not.
This is done intentionally to avoid nonphysical systems.
Likewise, the indirect term caused by the disk is always included if the disk has mass (cases 2 and 3).

Whether the disk is considered to have mass is configured by the \emph{DiskFeedback} option.
If this is set to \emph{yes}, the disk is considered to have mass and vice-versa.
The three cases and the value of the code parameters are summarized in Table~\ref{tab:disk_gravity_cases}.

\begin{table*}
\begin{center}
\caption{\label{tab:disk_gravity_cases}Possible cases for the disk gravity.}
\begin{tabular}{c|ccc|cc}
	Case & N-body & Particles & Disk & \emph{DiskFeedback} & \emph{SelfGravity} \\
	\hline
	massless & no & no & no & no & no \\
	massive w/o sg & yes & no & no & yes & no \\
	massive & yes & yes & yes & yes & yes \\
\end{tabular}
\tablefoot{The first group of columns indicates whether or not gravity from the disk is applied to the respective system (N-body, particles, disk) in the simulation. The second group shows the configuration of the code parameters.}
\end{center}
\end{table*}

%%%%%%%%%%%%%%%%%%%%%%%%%%%%%%%%%%%%%%%%
\subsection{Coordinate center}
%%%%%%%%%%%%%%%%%%%%%%%%%%%%%%%%%%%%%%%%
\label{sec:coordinate_center}

Because the star is treated like any other N-body object and can move, we need to define the coordinate center of the hydro domain.
In principle, we can choose an arbitrary reference point.
However, because the N-body system dominates the gravity, we choose the center of mass of any combination of N-bodies as the coordinate center.

With the N-body objects located at $\vec{r}_i$, with $i=1,2,\dots,N_p$, the coordinate center $\vec{R}$ can now be set as
\begin{align}\label{eqn:coordinate_center}
	\vec{R} = \frac{1}{\sum_{i=1}^{N_c} M_i}\sum_{i=1}^{N_c} M_i\,\vec{r_i}\,,
\end{align}
where $N_c$ is the number of point masses whose center of mass is chosen to be the coordinate center.
For example, $N_c = N_p$ selects the full N-body system as a reference, $N_c = 1$ selects the primary as a center, and $N_c = 2$ selects a stellar binary or the center of mass of a star-planet system.

The center of mass of the full N-body system is likely the most desirable of such choices because it can be an inertial frame that does not require considering fictitious forces.
This eliminates the need for an extra acceleration term to correct for the non-inertial frame, the so-called indirect term, which can be the source of numerical instabilities.
At least it eliminates the need for the indirect term resulting from the N-body system which generally dominates the indirect term, though contributions from the disk gravity might still need to be considered.

An application is the simulation of circumbinary disks, in which the natural coordinate center is the center of mass of the binary stars.

%%%%%%%%%%%%%%%%%%%%%%%%%%%%%%%%%%%%%%%%
\subsection{Shift-based indirect term}
%%%%%%%%%%%%%%%%%%%%%%%%%%%%%%%%%%%%%%%%
\label{sec:shift-based-indirect-term}

The correction forces needed because of a non-inertial frame are typically called the indirect term in the context of planet-disk interaction simulations.
Based on the definition of the coordinate center in Eq.~\eqref{eqn:coordinate_center}, the corresponding acceleration can formally be written as 
\begin{align}\label{eqn:kick-indirect-term}
    \vec{a}_\text{Ind} = - \frac{\text{d}^2}{\text{d}t^2} \vec{R} = - \frac{1}{M_c} \sum_{i=1}^{N_c} M_i\, \vec{a_i}\,,
\end{align}
where we used $\vec{a}_i = \ddot{\vec{r}_i}$, defined $M_c = \sum_{i=1}^{N_c} M_i$, and assumed that the derivatives of the masses are negligible.
The individual $\vec{a}_i$ should include all accelerations from forces acting on a specific N-body $i$, including gravity from all sources, other N-bodies or the disk.

Usually, this correction term is evaluated at the start of a time step, yielding a vector $\vec{a}_\text{Ind}^t$.
As outlined in Sect.~\ref{sec:physics}, this is then added to the velocities of the N-bodies as a sort of Euler step $\vec{v}_i^{t\prime} = \vec{v}_i + \Delta t \vec{a}_\text{Ind}^t$ before integration of the N-body system.
For the disk, the indirect term is applied through the potential.
This update is first-order in time.
As long as the magnitude of the indirect term is small compared to direct gravity, this choice appears to be good enough.

Using an Euler step as above is not sufficient in simulations in which the indirect term becomes stronger than the direct gravity.
This can be the case in simulations of circumbinary disks centered on one of the stars which are useful for the study of circumstellar disks in binary star systems.
In such simulations, the disk can become unstable with the Euler method.

As a solution, we compute the indirect term as the average acceleration felt by the coordinate center during the whole time step.
The average acceleration is computed from the actual shift needed to keep the coordinate center put.
Therefore we call this method the \emph{shift-based indirect term}.
In our code, this is achieved by making a copy of the N-body system after the N-body velocities have been updated by disk gravity, integrating the copy in time and computing the net acceleration from the velocities as
\begin{equation}\label{eq:shift-based-indirect-term}
    \vec{a}_\mathrm{Ind} = - \frac{\vec{v}_c^{t+\Delta t} - \vec{v}_c^t}{\Delta t}.
\end{equation}
where $\vec{v}_c = \frac{\text{d}}{\text{d}t} \vec{R}$ is the velocity of the coordinate center.
We then discard the N-body system copy, apply this indirect term to the N-body, hydro and particle systems and integrate them.
This way, the acceleration of the center and the indirect term cancel out and the system stays put much better compared to the old method.
Note, that the computation of the acceleration is handled this way because the IAS15 is a predictor-corrector scheme that does not produce an effective acceleration that would be accessible from the outside. Using the acceleration from an explicit Runge-Kutta scheme would produce the same results.

We tested, both, the old and the shift-based indirect term implementations and could not find any discernable difference in the dynamics of systems with embedded planets as massive as 10 Jupiter masses around a solar mass star.
However, the shift-based indirect term enables simulations of circumbinary disks centered on one of the stars which was previously not possible.
Even resolved simulations of circumsecondary disks (centered on the secondary star) when the latter is less massive than the primary star are possible.
For the impact of this new scheme on the simulation of circumbinary disks, see Jordan et al. (2024, in prep).

%%%%%%%%%%%%%%%%%%%%%%%%%%%%%%%%%%%%%%%%
\subsection{Gravity and Smoothing}
%%%%%%%%%%%%%%%%%%%%%%%%%%%%%%%%%%%%%%%%
\label{sec:gravitational_smoothing}
For modeling gravity, we use the common scale-height-dependent smoothing approach.
This approach accounts for the fact that a gas cell in 2D, that represents a vertically extended disk, feels a weaker gravitational pull from an N-body object compared to the same cell for a truly razor thin 2D disk \citep{mullerTreatingGravityThindisk2012}.
The potential due to a point mass $k$ with mass $M_k$ at a distance $d_{ik}$ from the center of a cell $i$ is then given by
\begin{equation}
\label{eq:potential_2}
 \Phi_k = -\frac{\mathrm{G} M_k}{\sqrt{d_{ik}^2 + \epsilon_i^2}}
\end{equation}
where $\mathrm{G}$ is the gravitational constant, $d_{ij}$ is the distance between the point mass and the cell, and $\epsilon_i = \alpha_\text{sm} H_i$ is the smoothing length with the smoothing parameter $\alpha_\text{sm}$ and the cell scale height $H_i$.
According to \citet{mullerTreatingGravityThindisk2012},
$\alpha_\text{sm}$ should be between 0.6 -- 0.7 to accurately describe gravitational forces and torques, which is why we apply it to all N-bodies, even the central star.
\citet{massetCoorbitalCorotationTorque2002} found that a factor of 0.76 most closely reproduces type I migration rates of 3D simulations.
Please note that $\epsilon_i$ has to be evaluated with the scale height at the location of the cell, not the location of the N-body object \citep{mullerTreatingGravityThindisk2012}.

When the gravity of planets inside the disk is also considered for computing the scale height (Sect.\,\ref{sec:aspect_ratio}),
the scale height and the smoothing length in Eq.\,\eqref{eq:potential_2} become smaller in the vicinity of the planet.
There, the smoothing can then become too small such that numerical instabilities occur.

To remedy this issue, we added an additional smoothing adapted from \citet{klahr2006}.
We found this to be a necessity when simulating binaries where one component enters the simulation domain due to an eccentric orbit or small inner domain radius.
It is applied as a factor to the potential in Eq.\,\ref{eq:potential_2} and is given by
\begin{equation}
\label{eq:potential_smoothed}
 \Phi_k^\mathrm{sm} = \Phi_k
 \begin{cases}
	\left[ \left(\frac{s_{ik}}{\epsilon_i}\right)^4 - 2 \left(\frac{s_{ik}}{\epsilon_i}\right)^3 + 2 \left(\frac{s_{ik}}{\epsilon_i}\right)\right] & \text{if } s_{ik} < d_{\text{sm},k}\\
	1 & \text{otherwise}
  \end{cases} 
\end{equation}
where $s_{ik} = \sqrt{d_{ik}^2 + \epsilon_i^2}$ is the $\epsilon$-smoothed distance and $d_{\text{sm},k}$ is another smoothing length, separate from $\epsilon_i$.
This smoothing is purely numerically motivated and guarantees numerical stability close to the planet. 
It has no effect outside $d_{\text{sm},k}$.
The smoothing length $d_{\text{sm},k}$ is a fraction of the Roche radius $R_\text{Roche}$, which is the distance from the point mass to its L1 point with respect to point mass $k=1$.
For the case that the mass of the N-body particles changes, for example, with planet accretion, we compute $R_\text{Roche}$ using one Newton-Raphson iteration to calculate the dimensionless Roche radius, starting from its value during the last iteration.
For low planet masses, $R_\text{Roche}$ reduces to the Hill radius.

In many Fargo code variants, the gravity of the central star is not smoothed.
However, the smoothing is required because the disk is vertically extended.
The gravitational potential of the central star should therefore also be smoothed, otherwise, it is overestimated.
This results in a deviation of order $10^{-4}$ for the azimuthal velocity compared to a non-smoothed stellar potential.
Although this deviation is small, we believe that the stellar potential should be smoothed for consistency.
A downside of this choice is that there exists no known analytical solution for the equilibrium state of a disk around a central star with a smoothed potential.

The full N-body system with $N_\mathrm{NB}$ members then has the total potential
\begin{align}\label{eq:gravitational_potential_nbody}
\Phi_\text{NB} = \sum_{k=1}^{N_\text{NB}} \Phi^\text{sm}_k\,.
\end{align}
Having covered how the point masses affect the gas, the following paragraphs describe how the planets are affected by the gas.
For a given point mass $k$ at position $\vec{r}_k$, the gravitational acceleration exerted onto the point mass by the disk is
\begin{align}
\vec{a}_{k}^\mathrm{gas} = - \mathrm{G} \sum_{i=1}^{N_\mathrm{cell}} f_\text{sm} \frac{m_i}{s_{ik}^3} \vec{d}_{ik}\,, \label{eq:gravitational_acceleration_gas}
\end{align}
where $\vec{d}_{ik} = \vec{r}_k - \vec{r}_i$ is the distance vector between the planet and the cell, $m_i$ is the mass of grid-cell $i$, and $s_{ik}$ is the smoothed distance between the cell and the point mass.
It is given by $s_{ik} = \sqrt{d_{ik}^2 + \epsilon_i^2}$ with $d_{ik} = |\vec{d}_{ik}|$.

Again, the acceleration includes $\epsilon_i$ to account for the finite vertical extent of the disk.
The same coefficient is used for the potential.
The additional factor is the analog to Eq.~\eqref{eq:potential_smoothed} and is given by
\begin{align}
	f_\text{sm} = \begin{cases}
		4 \left(\frac{s_{ik}}{l_1}\right)^3 - 3 \left(\frac{s_{ik}}{l_1}\right)^4 \,, & s_{ik} \leq l_1 \\
		1	& s_{ik} > l_1 \\
	\end{cases}\,. \label{eq:gravitational_acceleration_gas_smoothing_factor}
\end{align}

Please note, that the interaction is not symmetric.
The acceleration of N-body objects due to the gas is computed using direct summation while the gravitational potential in Eq.~\eqref{eq:gravitational_potential_nbody} enters into the momentum equation Eq.~\eqref{eq:momentum_r_eq_polar} and \eqref{eq:momentum_phi_eq_polar} via a numerical differentiation.
Because we compute the smoothing length using the scale height at the location of the cell, as we argue one should do on physical grounds \citep[e.g.][]{mullerTreatingGravityThindisk2012}, 
the differentiation causes extra terms that depend on the smoothing length.

To alleviate this issue, we implemented computing the acceleration of the gas due to the N-body objects by direct summation, which has negligible computational overhead.
In this way, the interaction is fully symmetric and no additional terms are introduced.

As of now, we are unaware of how much this asymmetry affects the results of simulations of planet-disk interaction and the issue is left for future work.

%%%%%%%%%%%%%%%%%%%%%%%%%%%%%%%%%%%%%%%%
\subsection{Self-gravity}
%%%%%%%%%%%%%%%%%%%%%%%%%%%%%%%%%%%%%%%%
\label{sec:self-gravity}

Calculating the gravitational potential of a thin (2D) disk is a complex task.
Indeed, it requires the vertical averaging of Poisson's equation, which in general is not feasible.
Due to this limitation, in thin disk simulations, we often resort to a Plummer potential approximation for the gas, taking the following form:
\begin{align}
\Psi_\text{sg} (\vec{r}) = - \mathrm{G} \iint \frac{\Sigma(\vec{r}') }{s^2 + \epsilon_\text{sg}^2} \, \mathrm{d} \vec{r}' \label{eq:sg_potential}
\end{align}
with $s = ||\vec{r} - \vec{r}'||$, the gravitational constant $\mathrm{G}$ and a smoothing length $\epsilon_\text{sg}$.
Contrary, to a common belief the role of this smoothing length is not to avoid numerical divergences at the singularity, $s=0$.
While it indeed fulfills this function, its main purpose is to account for the vertical stratification of the disk.
In other terms, it permits gathering the combined effects of all disk vertical layers in the midplane.
Without such a smoothing length, the magnitude of the self-gravity (SG) acceleration would be overestimated.
In this context, many smoothing-lengths have been proposed but the most widely used is the one proposed by \citet{mullerTreatingGravityThindisk2012}.
Based on an analytic approach, they suggested that the softening should be proportional to the gas scale height, $\epsilon_{sg}=1.2 H_g$, to correctly capture SG at large distances.

Direct computation of the potential according to Eq. \eqref{eq:sg_potential} is prohibitive since it requires $N^2$ operations.
Fortunately, assuming a logarithmic spacing in the radial direction and a constant disk aspect ratio, $h=H/r=\text{const.}$, the potential can be recast as a convolution product, which can be efficiently computed in order $N\,\log(N)$ operations \citep{binneyGalacticDynamics1987} thanks to Fast-Fourier methods \citep{FFTW05}.
Such a method was implemented for the SG accelerations by \citet{baruteauPredictiveScenariosPlanetary2008a}.
We use the same method and the module used in \fargocpt is based on the implementation in FargoADSG \citep{baruteauTypePlanetaryMigration2008}.

The traditional choice for the smoothing length is $\epsilon^2_\text{sg} = B\,r^2$, where $B$ is a constant.
This choice, however, has two drawbacks.
First, it breaks the $r-r'$ symmetry of the gravitational interaction, which violates Newton's third law of motion.
As a consequence, a nonphysical acceleration in the radial direction manifests \citep{baruteauPredictiveScenariosPlanetary2008a}. 
Second, even if the choice of $B=1.2\,h$ minimizes the errors at large distances \citep{mullerTreatingGravityThindisk2012} this, nonetheless, results in SG underestimation at small distances, independent of the value of $B$ \citep{rendonrestrepoSelfgravityThindiscSimulations2023}.
This underestimation can quench gravitational collapse.

\fargocpt includes two improvements to alleviate those problems.
The asymmetry problem can be alleviated by choosing a smoothing length that fulfills the symmetry requirement $\epsilon^2_\mathrm{sg} (\vec{r}, \vec{r}^\prime) = \epsilon^2_\mathrm{sg} (\vec{r}^\prime, \vec{r})$.
Additionally, it can be shown that the Fourier scheme is still applicable for a more general form of the smoothing length $\epsilon^2_\mathrm{sg} (\vec{r}, \vec{r}^\prime) = rr^\prime \, f(r/r^\prime, \phi - \phi^\prime)$.
This is fulfilled if the smoothing length is a Laurent series in the ratio $\frac{r}{r^\prime}$ and a Fourier series in $\phi - \phi^\prime$ (which additionally captures the $2\pi$ periodicity in $\phi$).
Testing has shown, that the azimuthal dependence is negligible, so we only consider the constant term from the Fourier series.
Furthermore, the radial dependence is only weak, so we only consider the first two terms of the Laurent series.
This leads to the following form of the smoothing length
\begin{align}\label{eq:sg_smoothing_length_result}
	\epsilon^2_\mathrm{sg} (\vec{r}, \vec{r}^\prime) = \chi^2 r r^\prime + \lambda^2 (r - r^\prime)^2\,,
\end{align}
with two positively defined coefficients $\chi$ and $\lambda$.
These two parameters depend on the aspect ratio $h$ and can be precomputed for a given grid size by numerically minimizing the error between the 2D approximation and the full 3D summation of the gravitational acceleration.
This requires specifying the vertical stratification of the disk.
Assuming that the gravity from the central object is negligible compared to the disk SG, the vertical stratification is a Spitzer profile
\begin{align}
	\rho(z) = \frac{\Sigma}{2 H_\mathrm{sp}} \frac{1}{ \cosh^2(z/H_\mathrm{sp})}\,,& &H_\mathrm{sp} = \frac{c_\mathrm{iso}^2}{\pi \mathrm{G} \Sigma} \approx Q H = Q h r\,,
\end{align}
with the Toomre parameter 
\begin{align}\label{eq:toomre_parameter}
	Q = \frac{c_s \kappa_e}{\pi \mathrm{G} \Sigma} \approx \frac{c_s \Omega_\text{K}} {\pi \mathrm{G} \Sigma}\,,
\end{align}
with the epicycle frequency $\kappa_e$.
Furthermore assuming a grid with $r_\mathrm{max}/r_\mathrm{min} = 250/20$, the fit formula for the coefficients are $\chi(h) = -0.7543 h^2 + 0.6472 h$ and $\lambda(h) = 0.4571 h + 0.6737 \sqrt{h}$.
This is the formula used in \fargocpt.
For grids with substantially different ratios of outer to inner boundary radius, the minimization procedure has to be repeated and the constants changed.
This smoothing length leads to a symmetric self-gravity force and has reduced errors for both small and large distances.

In the case of a non-constant aspect ratio, we use the mass-averaged aspect ratio which is recomputed after several time-steps.
An additional benefit of the symmetric formulation is improved conservation of angular momentum by way of removing the self-acceleration present when using the old smoothing length.

In the limit of weak self-gravity, $Q \geq 20$, \citet{rendonrestrepoSelfgravityThindiscSimulations2023} corrected the underestimation of SG at small distances introducing a space-dependent smoothing length, which matches the exact 3D SG force with an accuracy of 0.5\%.
However, they did not correct the symmetry issue.
Despite, this oversight they showed that their correction can lead to the gravitational collapse of a dust clump trapped inside a gaseous vortex \citep{2023_rendon_gressel_PPVII_poster} or maintain a fragment bound by gravity \citep{2023_rendon_restrepoBarge_rossbi3d}.
In their latest work, (Rendon Restrepo et al. 2024, in prep.) found the exact kernel for all SG regimes which makes the use of a smoothing length obsolete:
\begin{equation}\label{Eq: SGFC exact}
\begin{array}{lll}
L^{a b}_{sg} (d)
      & = & \displaystyle \sqrt{\pi} \frac{d^2}{8} 
            \displaystyle \exp\left(\frac{d^2}{8} \right) \left[ K_1\left(\frac{d^2}{8} \right) - K_0\left(\frac{d^2}{8} \right) \right]
\end{array}
\end{equation}
where $K_0$ and $K_1$ are modified Bessel functions of the second kind, $d=||\vec{r}-\vec{r}'||/ \langle H\rangle(r,r')$ and $\langle H\rangle(\vec{r}, \vec{r}') = \sqrt{\frac{H_\text{sg}^2(\vec{r}') + H_\text{sg}^2(\vec{r})}{2}}$ with $H_\text{sg}$ defined in Eq.~\eqref{eq:scale_height_sg}.
This Kernel remains compatible with the aforementioned convolution product and Fast Fourier methods.
Although it is computationally expensive to compute the kernel using Bessel functions, it can be precomputed for locally isothermal simulations, thus it has to be computed only once.
For radiative simulations in which the aspect ratio changes, it can be updated only every so often making the method computationally feasible.
Finally, this solution shares the properties of the solution presented above making the SG acceleration symmetric and removing the self-acceleration.

When considering SG, the balance between vertical gravity and pressure that sets the scale height needs to include the SG component as well.
This effect can be included in the standard vertical density stratification $\rho(z) = \Sigma/(2\pi\,H) \,\exp\left(-1/2 (z/H)^2 \right)$ to good approximation by adjusting the definition of the scale height \citep[][see their Appendix A]{bertin_class_1999}.
In the case that the SG option is turned on, the standard scale height is replaced by
\begin{align}\label{eq:scale_height_sg}
	H_\text{sg} & =     \sqrt{\frac{2}{\pi}} \, H \, f(Q)\,, \\
	f(Q)       & =     \frac{\pi}{4 Q} \left[ \sqrt{1+\frac{8 Q^2}{\pi}} -1 \right]\,,
\end{align}
with the Toomre $Q$ parameter from Eq. \eqref{eq:toomre_parameter}.
In the code, we multiply the result of the standard scale height computation by the factor $\sqrt{2/\pi} f(Q)$.
The epicycle frequency $\kappa_e$ in the Toomre parameter in Eq.~\eqref{eq:toomre_parameter} is calculated as $\kappa_e^2 = 1/r^3 \mathrm{d}((r^2 \Omega)^2)/\mathrm{d}r$ \citep[e.g.][]{binneyGalacticDynamics1987} with the angular velocity of the gas $\Omega$.

For simulations of collapse in self-gravitating disks, we added the OpenSimplex algorithm to our code to initialize noise in the density distribution. 
The noise is intended to break the axial symmetry of the disk and thereby help gravitational instabilities to develop.

%%%%%%%%%%%%%%%%%%%%%%%%%%%%%%%%%%%%%%%%
\subsection{Energy equation and radiative processes}
%%%%%%%%%%%%%%%%%%%%%%%%%%%%%%%%%%%%%%%%
\label{sec:energy_equation}

The original \fargo \citep{masset_fargo_2000} code did not include treatment of the energy equation.
This was added in various later incarnations of the \fargo code, including FargoADSG \citep{baruteauCorotationTorqueRadiatively2008a} on which this code is based.
In this section, we outline the procedure of how the energy update is performed in \fargocpt.
Refer to the grey box on the right in Fig.~\ref{fig:flow} for a visual representation.
The subscripts of the variables in this section refer to the ones in Fig.~\ref{fig:flow}, intended to help the reader in locating the sub-steps.

The energy update step due to compression or expansion, shock heating, viscosity, irradiation, and radiative cooling or $\beta$-cooling is implemented using operator splitting.
Our scheme consists of a mix of implicit update steps.

In principle, we perform the energy update on the sum of internal energy density $e$ and the radiation energy density $e_\text{rad} = a T^4$, thus assuming a perfect coupling between the ideal and the photon gas, the so-called one-temperature approximation (see Sect.~\ref{sec:fld} for more detail).
Separately, these quantities change as
\begin{align}
\frac{\partial e}{\partial t} &= -p \nabla \vec{u} + Q_\text{shock} + Q_\text{visc}\\
\frac{\partial e_\text{rad}}{\partial t} &= Q_\text{irr} - \int_{-\infty}^\infty \nabla \cdot \vec{F} \mathrm{dz} \\
&= Q_\text{irr} - Q_\text{cool} - \int_{-\infty}^\infty \nabla \cdot \left(\vec{F} - \frac{\partial F_z}{\partial z} \right) \mathrm{dz}
\end{align}
with the three-dimensional radiation energy flux $\vec{F}$, viscous heating $Q_\text{visc}$, shock-heating as captured by artificial viscosity $Q_\text{shock}$, irradiation $Q_\text{irr}$, and radiative losses at the disk surfaces $Q_\text{cool} = \int_{-\infty}^{\infty} \frac{\partial F_z}{\partial z} \,\mathrm{d}z$.
Note, that the vertical cooling part $Q_\text{cool}$ is split off from the integral over $\nabla \cdot \vec{F}$.

In principle, the total energy update is then
\begin{equation}\label{eq:total_energy_update}
	\begin{gathered}
		\begin{aligned}
\frac{\partial (e + e_\text{rad})}{\partial t} = &-p \nabla \vec{u} + Q_\text{shock} \\ &+ Q_\text{visc} + Q_\text{irr} - Q_\text{cool} \\ &- \int_{-\infty}^\infty \nabla \cdot \vec{F} - \frac{\partial F_z}{\partial z} \mathrm{dz}
		\end{aligned}
	\end{gathered}
\end{equation}
We split this update into three main parts applied in the following order.
The first line represents shock heating and compression heating (see Sect.~\ref{sec:compression_shock_heating}), the second line represents heating and cooling (see Sect.~\ref{sec:heating_cooling}), and the third line represents the radiation transport (see Sect.~\ref{sec:fld}).

The following subsections go over the details of this process.

%%%%%%%%%%%%%%%%%%%%%%%%%%%%%%%%%%%%%%%%
\subsubsection{Compression heating and shock heating}
%%%%%%%%%%%%%%%%%%%%%%%%%%%%%%%%%%%%%%%%
\label{sec:compression_shock_heating}

The pressure term is updated first following \citet{dangelo_thermohydrodynamics_2003} with an implicit step using an exponential decay ensuring stability (see their Eq.~24).
The update reads
\begin{equation}
	\label{eq:pressure_energy_update}
	e_b = e_a\, \mathrm{exp}\left[-(\gamma - 1) \Delta t \nabla \vec{u}_n \right]\,.
\end{equation}
As a next step, shock heating is treated in the form of heating from artificial viscosity
\begin{align}\label{eq:artificial_viscosity_energy_update}
	e_c = e_b + Q_\text{shock} \, \Delta t\,,
\end{align}
with $Q_\text{shock}$ being the right-hand-side of Eq.~\eqref{eq:artificial_viscosity_heating} where $\Sigma = \Sigma_a$ and $\vec{u} = \vec{u}_b$.

%%%%%%%%%%%%%%%%%%%%%%%%%%%%%%%%%%%%%%%%
\subsubsection{Heating and cooling}
%%%%%%%%%%%%%%%%%%%%%%%%%%%%%%%%%%%%%%%%
\label{sec:heating_cooling}

This step considers the update due to viscosity, irradiation, and cooling.
The relevant part from Eq.~\eqref{eq:total_energy_update} is
\begin{align}
	\frac{\partial (e + e_\text{rad})}{\partial t} = Q_\text{visc} + Q_\text{irr} - Q_\text{cool}\,.
\end{align}
To arrive at the expression for this update step, we first insert explicit expressions for the two energies.
The internal energy density is $e = \rho c_V T$ with the specific heat capacity $c_V = \frac{\mathrm{k}_\mathrm{B}}{\mathrm{m}_\mathrm{H} \mu (\gamma - 1)}$, where $\mathrm{k}_\mathrm{B}$ is the Boltzmann constant, $\mathrm{m}_\mathrm{H}$ is the mass of a hydrogen atom, $\mu$ is the mean molecular weight, and $\gamma$ the adiabatic index.
The radiation energy density is given by $e_\mathrm{rad} = \frac{4\sigma_\mathrm{SB}}{c} T^4$ with the Stefan-Boltzmann constant $\sigma_\mathrm{SB}$ and the speed of light $c$.
Then, we use a first-order discretization for the time derivative and rearrange it according to the time index of $T$.
Finally, we convert to an expression for the internal energy $e$ arriving at
\begin{align}
	\label{eqn:energy_update_implicit}
	e_d = e_c +  \Delta t \frac{Q_\text{visc} + Q_\text{irr} - Q_\text{cool}}{\alpha}
\end{align}
with 
\begin{align}
	\alpha = 1 + 8H\big|_{e=e_b} \frac{\sigma_\mathrm{SB}}{c} \left(\frac{m_\mu(\gamma-1)}{R\Sigma_a} \right)^4 e_b^3\,.
\end{align}
This implicit energy update is stable in our testing.
The heating and cooling rates are calculated as described in the following sections.
Each of the terms is optional and can be configured by the user.
Cooling rates are discussed in Sect.~\ref{sec:cooling} and \ref{sec:s-curve-cooling}, irradiation is discussed in Sect.~\ref{sec:irradiation}. and viscous heating is discussed in Sect.~\ref{sec:viscosity}.

%%%%%%%%%%%%%%%%%%%%%%%%%%%%%%%%%%%%%%%%
\subsubsection{Radiative Cooling}
%%%%%%%%%%%%%%%%%%%%%%%%%%%%%%%%%%%%%%%%
\label{sec:cooling}

For radiative cooling, we take the same approach as \citep{mullerCircumstellarDisksBinary2012} where energy can escape through the disk surfaces and the energy transport from the disk midplane to the surfaces is modeled using an effective opacity.
The associated cooling term is
\begin{align}\label{eq:cooling_energy_update}
Q_\text{cool} = 2 \sigma_\text{SB} \frac{T^4 - T_\text{min}^4}{\tau_\text{eff}}\,.
\end{align}
The minimum temperature $T_\text{min}$, defaulting to 4\,K, is set to take into account that the disk does not cool towards a zero Kelvin region but rather towards a cold environment slightly warmer than the cosmic microwave background.
This can be important in the outer regions of the disk and effectively sets a temperature floor.
For the effective opacity we use \citep[][]{hubenyVerticalStructureAccretion1990,mullerCircumstellarDisksBinary2012,dangelo2012outward_migration}
\begin{align}\label{eq:effective_opacity}
\tau_\text{eff} = \frac{3}{8}\tau + \frac{k}{4} + \frac{1}{4\tau + \tau_\text{min}}
\end{align}
with where $k=2$ for an irradiated disk and $k=\sqrt{3}$ for a non-irradiated disk \citep{dangelo2012outward_migration} 
, $\tau = \kappa \Sigma / \sqrt{8\pi}$ is the optical depth calculated from the Rosseland mean opacity $\kappa$, and $\tau_\text{min} = 0.01$ is a floor value to capture optically very thin cases in which line opacities become dominant \citep{hubenyVerticalStructureAccretion1990}.
There are three options to compute the opacity. For the first two, see \citet{mullerCircumstellarDisksBinary2012} for more detail.
\begin{itemize}
\item \texttt{lin}: \citet{linDynamicalOriginSolar1985}
\item \texttt{bell}: \citet{bellUsingFUOrionis1994}
\item \texttt{constant}: A constant opacity.
\end{itemize}
Additional opacity laws can be easily implemented by expanding the code by a function that returns the opacity $\kappa(\rho, T)$ as a function of temperature and volume density.

In case of $\beta$-cooling \citep{gammieNonlinearOutcomeGravitational2001},
\begin{align}\label{eq:beta-cooling}
	Q_\beta = (e-e_\text{ref}) \frac{\Omega_K}{\beta}
\end{align}
is added to $Q_\text{cool}$ where $e_\text{ref}$ is the reference energy to which the energy is relaxed.
This reference can either be the initial value or prescribed by a locally isothermal temperature profile.
Please note, that $\beta$-cooling is a misnomer as it is not exclusively a cooling process but rather a relaxation process towards a reference state.
If the temperature is lower than in the reference state, the energy is increased, i.e. heating occurs.

%%%%%%%%%%%%%%%%%%%%%%%%%%%%%%%%%%%%%%%%
\subsubsection{S-curve cooling}
%%%%%%%%%%%%%%%%%%%%%%%%%%%%%%%%%%%%%%%%
\label{sec:s-curve-cooling}

For the specific case of simulating cataclysmic variables, we also included the option to compute $Q_\mathrm{cool}$ according to 
\cite{ichikawa1992dwarfnova} and \citet{kimura2020tilt}. 
Their model splits the cooling function into a cold,
radiative branch and a hot, convective branch.
They used opacities based on \citet{cox1969opacities} and a vertical radiative flux, i.e. the radiative loss from one surface of the disk, of
\begin{equation}
	F = \tau \sigma T^4
\end{equation}
for the optically thin regime to derive the radiative flux of the cold branch:
\begin{equation}
	\begin{aligned}
	\mathrm{log} F_\mathrm{cool,cgs} &= 9.49\, \mathrm{log}\, T_\mathrm{cgs} + 0.62\, \mathrm{log}\, \Omega_\mathrm{cgs}\\
	&+ 1.62\, \mathrm{log}\, \Sigma_\mathrm{cgs} + 0.31\, \mathrm{log}\, \mu - 25.48
	\end{aligned}
\end{equation}
where the subscript cgs indicates the value of the respective quantity in cgs units.
The cold branch is valid for temperatures $T < T_A$, where $T_A$
is the temperature at which
\begin{equation}
	\tau \sigma T_A^4 = F_\mathrm{cool}(T_A) = F_A.
\end{equation}
The radiative flux of the hot branch is derived by assuming that disk quantities do not vary in the vertical direction (one-zone model) for an optically thick disk:
\begin{equation}
	F = \frac{16 \sigma T^4}{3 \kappa \rho H}.
\end{equation}
Using Kramer's law for the opacity of ionized gas, the radiative
flux can be approximated as
\begin{equation}
	\begin{aligned}
	\mathrm{log} F_\mathrm{hot,cgs} &= 8\, \mathrm{log}\, T_\mathrm{cgs} + \mathrm{log}\, \Omega_\mathrm{cgs}
	+ 2\, \mathrm{log}\, \Sigma_\mathrm{cgs}  \\
	&+ 0.5\, \mathrm{log}\, \mu - \text{c}_\mathrm{hot}
	\end{aligned}
\end{equation}
where the constant $\text{c}_\mathrm{hot} = 25.49$ was used in \citet{ichikawa1992dwarfnova}
and $\text{c}_\mathrm{hot} = 23.405$ in \citet{kimura2020tilt}.
The difference between the two constants is that the first leads to weaker cooling compared to the latter.
The hot branch is valid for temperatures $T > T_B$, where $T_B$ is the temperature
at which
\begin{equation}
	F_\mathrm{hot,cgs} (T_B) = F_B
\end{equation}
where $\log F_B = \mathrm{max} (K, \log F_A)$ with \citep[based on Fig.~(3) in][]{mineshigeDiskinstabilityModelOutbursts1983}
\begin{equation}
	K = 11 + 0.4\, \mathrm{log }\left[ \frac{2\cdot 10^{10}}{r_\mathrm{cm}} \right].
\end{equation}
The radiative flux in the intermediate branch is given by an interpolation
between the cold and hot branches:
\begin{equation}
	\mathrm{log}\, \mathrm{F}_\mathrm{int} = (\mathrm{log}\, F_A - \mathrm{log}\, F_B)
	\mathrm{log}\, \frac{T}{T_B} /\, \mathrm{log}\,\frac{T_A}{T_B} + \mathrm{log}\, F_B.
\end{equation}
The cooling term is then given by $Q^\mathrm{scurve}_\mathrm{cool} = 2 F$ due to the
radiation from both sides of the disk where $F$ is the radiative flux pieced together
from the cool, intermediate, and hot branches as described above.
These prescriptions were developed for the conditions inside an accretion disk,
and we found that it leads to numerical issues when applied to the low-density regions
outside a truncated disk. We therefore opted to modulate
the radiative flux with a square root function for densities below
$\Sigma_\mathrm{thresh} < 2 \, \mathrm{g \, cm^{-2}}$ and a square function
for temperatures below $T_\mathrm{thresh} < 1200 \, \mathrm{K}$:
\begin{align}\label{eqn:scurve_cooling_Q}
Q^\mathrm{scurve}_\mathrm{cool} &= 2 F(T_\mathrm{tmp}, \Sigma_\mathrm{tmp}) \cdot
\sqrt{\frac{\Sigma}{\Sigma_\mathrm{tmp}}} \cdot \left( \frac{T}{T_\mathrm{tmp}} \right)^2\\
\Sigma_\mathrm{tmp} &= \mathrm{max} \left(\Sigma, 2\,\mathrm{g \, cm^{-2}}\right)\\
T_\mathrm{tmp} &= \mathrm{max} \left(T, 1200\,\mathrm{K}\right)
\end{align}

%%%%%%%%%%%%%%%%%%%%%%%%%%%%%%%%%%%%%%%%
\subsubsection{In-plane radiation transport using FLD}
%%%%%%%%%%%%%%%%%%%%%%%%%%%%%%%%%%%%%%%%
\label{sec:fld}

The last term in Eq.~\eqref{eq:hydroeq_energy} or \eqref{eq:total_energy_update}, the in-plane radiation transport, is treated using the flux-limited diffusion (FLD) approach \citep{levermoreFluxlimitedDiffusionTheory1981,levermoreRelatingEddingtonFactors1984}.
The method allows to treat radiation transport as a diffusion process, both, in the optically thick and optically thin regime.
Our implementation builds upon \citet{kleyRadiationHydrodynamicsBoundary1989,kley_migration_2008,mullerPlanetsBinaryStar2013} and uses the successive over-relaxation (SOR) method to solve the linear equation system involved in the implicit energy update.

In principle, the process of radiation transport is described by the evolution of a two-component gas consisting of an ideal gas and a photon gas which have thermal energy densities $e = e_\text{gas}$ and $e_\text{rad}$, respectively \citep{mihalasFoundationsRadiationHydrodynamics1984}.
In the flux-limited diffusion approximation, their coupled evolution is given by \citep{kleyRadiationHydrodynamicsBoundary1989,commerconRadiationHydrodynamicsAdaptive2011,kolbRadiationHydrodynamicsIntegrated2013b}
\begin{align}
		\frac{\partial e_\text{rad}}{\partial t} -\nabla \cdot\left(\frac{c \lambda}{\kappa_{\mathrm{R}} \rho} \nabla e_\text{rad}\right) & = \ \ \kappa_{\mathrm{P}} \rho c\left(a_{\mathrm{R}} T^4-e_\text{rad}\right) \\
		\frac{\partial e}{\partial t}  & =-\kappa_{\mathrm{P}} \rho c\left(a_{\mathrm{R}} T^4-e_\text{rad}\right)\,,
\end{align}
where $\lambda$ is the flux limiter.
We use the flux limiter presented in \citet[][see this reference for alternatives]{kleyRadiationHydrodynamicsBoundary1989} which is given by
\begin{align}
	\lambda = \begin{cases}
		2 / (3 + \sqrt{9 + 10\,R^2}) &, 0 \leq R \leq 2 \\
		10 / (10\,R + 9 + \sqrt{180\,R + 81}) &, R > 2
	\end{cases}
\end{align}
with the dimensionless quantity
\begin{align}
R = \frac{1}{\rho\kappa}\frac{|\nabla e|}{e}\,.
\end{align}

In \textsc{FargoCPT}, we use the one-temperature approximation, meaning that we assume that the photon gas and the ideal gas equilibrate their temperatures instantaneously, allowing us to reduce both equations into a single equation for the total thermal energy $e + e_\text{rad}$.
We further assume that $e_\text{rad}$ is negligible against $e$ (and the same for their time derivatives) yielding
\begin{align}\label{eq:fld_equation}
\frac{\partial e}{\partial t} \approx \frac{\partial (e + e_\text{rad})}{\partial t} = \nabla \cdot \frac{\lambda c}{\rho \kappa} \, \nabla e_\text{rad} = - \nabla \cdot \frac{\lambda \,16\, \sigma_\text{SB}}{\rho \kappa} T_\text{rad}^3 \, \nabla T_\text{rad}\,.
\end{align}
To see that $e_\text{rad}$ is indeed negligible against $e$, consider their ratio
\begin{align}
	\frac{e_\text{rad}}{e} = \frac{a_\text{R} T^4}{\rho c_V T} = 2.15\times 10^{-23} \left( \frac{T}{1\,\mathrm{K}} \right)^3 \left( \frac{\rho}{\mathrm{g}/\mathrm{cm}^3} \right)^{-1}\,,
\end{align}
where we used $\mu = 2.35 \mathrm{m}_\mathrm{H}$ and $\gamma = 1.4$.
Assuming further a minimum mass solar nebula (MMSN) \citep{hayashiStructureSolarNebula1981} density with $\Sigma(r) = 1700\,\mathrm{g}/\mathrm{cm}^2 \left(r/\mathrm{au}\right)^{-3/2}$, the ratio becomes
\begin{align}\label{eq:s-curve-cooling-Q}
	\frac{e_\text{rad}}{e} = 4.46\times10^{-16} \left( \frac{T}{1\,\mathrm{K}} \right)^{3.5}\,.
\end{align}
The dependence on the radius cancels out in the calculation of $\rho$ from $\Sigma$ because of the specific exponent of $-3/2$ of the MMSN.
Now we can see that $e_\text{rad}/e$ ranges from $4.7\times10^{-9}$ at $T = 100\,\mathrm{K}$ to $1.6\times10^{-4}$ at $T = 2000\,\mathrm{K}$.
Thus, the approximation is justified in the context of planet-forming disks.

In principle, the ideal gas and the photon gas can have two separate temperatures $T_\text{gas} = T$ and $T_\text{rad}$.
They are connected to the energy densities via
\begin{align}
	e = c_V \rho T\,,\quad e_\text{rad} = a_R T_\text{rad}^4\,.
\end{align}
Now, we again make use of the assumption that the ideal gas and the photon gas equilibrate their temperatures instantaneously, i.e. $T = T_\text{rad}$ and use the fact that $\rho$ can be considered constant during the radiation transport part of the operator splitting scheme.
Furthermore, we assume the opacity $\kappa$ to be constant during the radiation transport step, although in principle it depends on $T$.

Equation~\eqref{eq:fld_equation} can then be recast into an equation for $T$ yielding
\begin{align}
\frac{\partial T}{\partial t} = - \frac{1}{\rho c_V} \nabla \cdot \frac{\lambda \,16\, \sigma_\text{SB}}{\rho \kappa} T^3 \, \nabla T = \frac{1}{\rho c_V} \nabla \cdot K \,\nabla T
\end{align}
with $K = \lambda \frac{16 \sigma_\text{SB}}{\rho \kappa}$.
This diffusion equation is then discretized and the resulting linear equation system is solved using the SOR method resulting in an updated temperature $T^\prime$.
Details of the discretization and implementation of the SOR solver can be found in Appendix 1 of \citet{mullerPlanetsBinaryStar2013}.
Finally, the internal energy of the gas is updated using the new temperature,
\begin{align}
e_e = \Sigma c_V T^\prime\,.
\end{align}
Here, $e_e$ is an energy surface density again (also see Fig.~\ref{fig:flow}), as opposed to the energy volume densities in the rest of the section above.

Note, that our implementation uses the 3D formulation.
Other implementations of midplane radiation transport \citep[also the one described in][]{mullerPlanetsBinaryStar2013} use the surface density instead and have to introduce a factor $\sqrt{2\pi}\,H$ (sometimes also chosen as $2H$) to link the surface density to the volume density which depends on the vertical stratification of the disk.
However, if we assume $\rho$ (or equivalently $H$) to be constant throughout the radiative transport step, this factor cancels out in the end and the two approaches are equivalent.
For the sake of simplicity, we use the 3D version and assume that all horizontal radiation transport is confined to the midplane.
Before the radiation transport step, we compute $\rho = \Sigma / (\sqrt{2\pi}\,H)$.

The FLD implementation is tested with two separate tests.
The first test, presented in Appendix~\ref{appendix:FLD_test}, shows a test for the physical part in which a disk equilibrates to two different temperatures enforced at the inner and outer boundaries.
The second test, presented in Appendix~\ref{appendix:sor_solver_test}, shows a test for the SOR diffusion solver which compares the numerical results of a 2D diffusion process against the available analytical solution.

%%%%%%%%%%%%%%%%%%%%%%%%%%%%%%%%%%%%%%%%
\subsubsection{Irradiation}
%%%%%%%%%%%%%%%%%%%%%%%%%%%%%%%%%%%%%%%%
\label{sec:irradiation}

The irradiation term is computed as the sum of the irradiation from all N-body objects.
This allows for simulations in which planets or a secondary star irradiate the disk.
An N-body object is considered to be irradiating, if it is assigned a temperature and radius in the config file.

For each single source with index $k$, the heating rate due to irradiation is computed following \citet{menou2004low_mass} and \citet{dangelo2012outward_migration} as
\begin{align}\label{eq:irradiation_energy_update}
	Q^k_\text{irr}
	= 2(1-\epsilon) \frac{L_k}{4 \pi d_k^2} W_G \frac{1}{\tau_\text{eff}}
\end{align}
with the disk albedo $\epsilon$ which is set to $1/2$, the luminosity of the source $L_k$, the distance to the source $d_k$, and the effective optical depth $\tau_\text{eff}$.
The luminosity is calculated as $L_k = 4\pi R_k \sigma_\mathrm{SB} T_k^4$ with the radius $R_k$, and the temperature $T_k$ of the source, and the effective opacity $\tau_\mathrm{eff}$ as given in Eq.~\eqref{eq:effective_opacity}.
The remaining factor $W_G$ is a geometrical factor that accounts for the disk geometry in the case of a central star \citep{chiang1997spectral_energy} and includes terms for close to the source (first term) and far from the source (second term).
It is given by
\begin{align}
	W_G = 0.4\left(\frac{R_k}{r}\right) + h \left( \frac{\mathrm{d\;log}\;H}{\mathrm{d\;log}\;r} - 1\right)\,.
\end{align}
We assume the flaring of the disk, $F = \frac{\mathrm{d\;log}\;H}{\mathrm{d\;log}\;r}- 1$, is constant in time and has the value of the free parameter specified for the initial conditions.
Properly accounting for the disk geometry would require ray-tracing from all sources to all grid cells which is computationally expensive.

Finally, the total irradiation heating rate is given by
\begin{align}\label{eq:total_irradiation_energy_update}
	Q_\text{irr} = \sum_{k} Q^k_\text{irr}
\end{align}
where the sum runs over all irradiating objects.

%%%%%%%%%%%%%%%%%%%%%%%%%%%%%%%%%%%%%%%%%%%%
\subsubsection{Non-constant Adiabatic Index}
\label{sec:nonconst-gamma}
In astrophysics, it is very common to treat matter as ideal gas, 
for which the following equation, also known as the ideal gas law, holds
\begin{equation}
    \label{eq:ideal_gas_law}
    p = \frac{k_\mathrm{B}}{\mu m_\mathrm{u}} \rho T
\end{equation}
with the pressure $p$, the Stefan Boltzmann constant $k_\mathrm{B}$, 
the mean molecular weight $\mu$, the density $\rho$, and the temperature $T$. 
In the case of an ideal gas, the assumption is that there are no 
interactions between the gas particles and the pressure is only exerted by
interactions with the boundary of a volume, containing the gas. This is
very often a good  approximation, especially in the case of accretion disks
where densities are low.  To relate some of the thermodynamic quantities,
it is useful to introduce the adiabatic index $\gamma$, given by 
\begin{equation}
    \gamma = \frac{c_p}{c_v}
\end{equation}
which is the ratio of the heat capacity at constant pressure with the one 
at constant volume. The assumption of a constant adiabatic index leads to the 
simple relation between pressure and internal energy
\begin{equation}
    \label{eq:pressure_energy_3D}
    p = (\gamma -1) \rho \epsilon
\end{equation}
and the relation for the sound speed
\begin{equation}
    \label{eq:soundspeed_3D}
    c_s = \sqrt{\gamma \frac{p}{\rho}}.
\end{equation}
Such an equation of state describes a perfect gas and has a very low computational
cost. However, contributions due to rotational degrees of freedom of the particles
or changes in the chemical composition, such as the transition from neutral to ionized
hydrogen, are not accounted for in this description. 

In the following, we outline how the adiabatic index can be replaced with other quantities that provide the relations between pressure, density, energy, and sound speed when considering the effects mentioned above.
Hence, in Eqs.~\eqref{eq:pressure_energy_3D} and \eqref{eq:soundspeed_3D} the constant $\gamma$
will be replaced by the effective adiabatic index $\gamma_\mathrm{eff}$ and the first 
adiabatic exponent $\Gamma_1$ respectively. \\
With these changes, the equation of state can account for the dissociation and 
ionization processes of 
hydrogen, as well as rotational and translational degrees of freedom at lower 
temperatures. Such an equation of state was already implemented in \pluto
by \citet{vaidya2015pvte} which serves as a basis for the changes in our code.
In the \pluto code, this equation of state is called the \textit{PVTE} equation of state which stands for pressure-volume-temperature-energy. We adopted the same name for the equation of state in \fargocpt. \\
We start by writing the total internal energy density $\rho \epsilon$ 
of an ideal gas as a summation of several contributions:
\begin{equation}
    \label{eq:total_internal_energy}
    \rho \epsilon = (\epsilon_{\text{H}_2} + \epsilon_{\text{HI}}+
     \epsilon_{\text{HII}} + \epsilon_{\text{H+H}} + \epsilon_{\text{He}}) R \rho T 
     = \sum_i \epsilon_i  R \rho T,
\end{equation}
with $R=k_B / m_H$. These contributions are given by
\citep[compare Table\,1 from][]{vaidya2015pvte}
\begin{equation*}
\begin{split}
\epsilon_{\text{HI}} &= \frac{3}{2} X (1 + x) y \ \text{(Translational energy for Hydrogen)} \\
\epsilon_{\text{He}} &= \frac{3}{8} Y \ \text{(Translational energy for Helium)} \\
\epsilon_{\text{H+H}} &= 4.48 \text{eV } X y / (2 k_{B} T)
 \ \text{(Dissociation energy for} \\
 & \;\;\;\;\;\;\;\;\;\;\;\;\;\;\;\;\;\;\;\;\;\;\;\;\;\;\;\;\;\;\;\;\;\text{molecular hydrogen)}\\
\epsilon_{\text{HII}} &= 13.6 \text{eV } X x y/(k_B T) \ 
\text{(Ionization energy for}\\
&  \;\;\;\;\;\;\;\;\;\;\;\;\;\;\;\;\;\;\;\;\;\;\;\;\;\;\;\;\;\;\;\;\;\text{atomic hydrogen)}\\
\epsilon_{\text{H}_2} &= \frac{X (1-y)}{2} \left[\frac{3}{2} + \frac{T}{\zeta_v}
\frac{\mathrm{d} \zeta_v}{\mathrm{d} T} + \frac{T}{\zeta_r}
\frac{\mathrm{d} \zeta_r}{\mathrm{d} T}\right] \ \text{(Internal} \\
& \text{energy for molecular hydrogen)}
\end{split}
\end{equation*}
where $X$ (defaulting to 0.75 in the code) and $Y=1-X$ are the Hydrogen and Helium mass fractions and $y$ and $x$ are the Hydrogen 
dissociation and ionization fractions, defined as
\begin{equation}
    y = \frac{\rho_\mathrm{HI}}{\rho_\mathrm{HI} + \rho_\mathrm{H_2}}
\end{equation}
and 
\begin{equation}
    x = \frac{\rho_\mathrm{HI}}{\rho_\mathrm{HI} + \rho_\mathrm{HII}},
\end{equation}
$\zeta_v$ and $\zeta_r$ are the partition functions of vibration and rotation
of the hydrogen molecule and are described in \citet{dangelo2013}.
If we assume local thermodynamic equilibrium, then $y$ and $x$ can be 
computed by using the following two
Saha equations:
\begin{equation}
\frac{x^2}{1-x} = \frac{m_H}{X \rho} \left( \frac{m_e k_B T}{2 \pi \hbar^2} 
\right)^{3/2} \exp\left(\frac{-13.60 \, \text{eV}}{k_B T}\right)
\end{equation}
\begin{equation}
\frac{y^2}{1-y} = \frac{m_H}{2 X \rho} \left(  \frac{m_H k_B T}{4 \pi \hbar^2}
\right)^{3/2} \exp\left(\frac{-4.48 \, \text{eV}}{k_B T} \right).
\end{equation} \\
After applying the ideal gas law and inserting Eq.~\eqref{eq:total_internal_energy}, the
pressure-internal energy relation turns into: 
\begin{equation}
\label{eq:energy_pressure_relation}
p = \frac{R \rho T}{\mu} = \frac{\rho \epsilon}
{\mu (\sum_i \epsilon_i  )} 
= (\gamma_{\text{eff}} - 1) \rho \epsilon,
\end{equation}
where $\gamma_{\text{eff}} = 1 + \frac{1}{\mu (\sum_i \epsilon_i  )}$ is the 
effective adiabatic index and $\mu$ is the mean molecular weight, given by  
\begin{equation}
\mu = 4 \left[ 2X \left( 1 + y + 2xy \right) + Y \right].
\end{equation}
Now, by using the relation from Eq.~\eqref{eq:energy_pressure_relation} 
an equation for the temperature can be derived:
\begin{equation}
T = \frac{\mu p}{R \rho}
= \frac{\mu(\rho, T) \left( \gamma_{\text{eff}}(\rho, T) - 1 \right)\rho \epsilon}{R \rho}
\end{equation}
which can be solved for a given internal energy and density as a root-finding problem:
\begin{equation}
\frac{\mu(\rho, T) \left( \gamma_{\text{eff}}(\rho, T) - 1 \right)\rho \epsilon}{R \rho} - T = 0.
\end{equation}

The sound speed is given by
\begin{equation}
c_s = \sqrt{\Gamma_1 \frac{p}{\rho}}
\end{equation}
where $\Gamma_1$ is the first adiabatic exponent, which is defined as 
\begin{equation}
\Gamma_1 = \frac{1}{C_V \left( T \right)} \left( \frac{p}{\rho T} \right) 
\chi_T^2 + \chi_\rho
\end{equation}
where $C_V \left( T \right)$ is the specific heat capacity at constant volume:
\begin{equation}
C_V \left( T \right) =  \left[ \frac{\partial \epsilon \left( T \right) }{ \partial T } \right]_V
\end{equation}
and the temperature and density exponents which are defined by 
\begin{equation}
\begin{split}
	\chi_T &= \left( \frac{\partial \ln P}{\partial \ln T} 
	\right)_\rho = 1 - \frac{\partial \ln \mu}{\partial \ln T} \\
	\chi_\rho &= \left( \frac{\partial \ln P}{\partial \ln \rho} 
	\right)_T = 1 - \frac{\partial \ln \mu}{\partial \ln \rho}.
\end{split}
\end{equation}

Since the computational effort to compute $\gamma_\mathrm{eff},\,\Gamma_1$
and $\mu$ for every cell at every time step is very high, we precompute 
them to create lookup tables. 
During the simulation, the values of $\gamma_\mathrm{eff},\,\Gamma_1$
and $\mu$ are interpolated from the lookup tables for given densities and
internal energies. How these tables
can be implemented is also explained in \citet{vaidya2015pvte}.

As our code is two-dimensional, we require the scale height to compute
the densities used for reading the adiabatic indices from the lookup table.
To compute the scale height, our method requires the adiabatic indices.
This results in a cyclic dependency.
We found that this is not an issue, as successive time-steps naturally act
as an iterative solver for this problem. We additionally always compute
the scale height twice, before and after updating the adiabatic indices 
and we perform this iteration twice per time step. 
We tested our implementation using the shock tube test. Because there
is no analytical solution for the shock tube test with non-constant adiabatic
indices, we compared our results against results generated with the \textsc{Pluto}
code. The test is shown in Fig.\,\ref{fig:shocktube} under the label 'PVTE' and
we find good agreement between our implementation and the implementation 
by \citet{vaidya2015pvte}.

%%%%%%%%%%%%%%%%%%%%%%%%%%%%%%%%%%%%%%%%%%%%

%%%%%%%%%%%%%%%%%%%%%%%%%%%%%%%%%%%%%%%%
\subsection{Viscosity}
%%%%%%%%%%%%%%%%%%%%%%%%%%%%%%%%%%%%%%%%
\label{sec:viscosity}

Viscosity is implemented as an operator splitting step, see Fig.~\ref{fig:flow} for its context.
The full viscous stress tensor reads \citep[see, e.g.][]{shu_physics_1992}
\begin{align}
\sigma_{ij} = 2\,\mu\,
    \left[
        \frac{1}{2} \left( \frac{\partial u_i}{\partial x_j} + \frac{\partial u_j}{\partial x_i} \right)
        - \frac{\delta_{ij}}{3} \nabla \vec{u} 
    \right] + \zeta \delta_{ij} \nabla \vec{u}\,,
\end{align}
where $i,j \in \{1,2\}$ indicate the spatial directions,
$\mu$ and $\zeta$ are the shear and bulk viscosity, respectively, and $\delta_{ij}$ is the Kronecker $\delta$.
The shear viscosity is given by $\mu = \nu \Sigma$ with the kinematic viscosity denoted by $\nu$.
In our case, $\zeta$ is neglected, although artificial viscosity (see Sect.~\ref{sec:artificial_viscosity}) reintroduces a bulk viscosity.
The kinematic viscosity is either given by a constant value or the $\alpha$-prescription \citep{shakura1973alpha} for which
\begin{align}
	\nu = \alpha c_\mathrm{s}^\mathrm{adb} H
\end{align}
with the adiabatic sound speed $c_\mathrm{s}^\mathrm{adb} = \sqrt{\gamma P/\Sigma}$ and the disk scale height $H$.

The relevant elements in polar coordinates of the viscous stress tensor are
\begin{align}
% r r - component
\sigma_{rr} &= 2 \nu \Sigma \left(\frac{\partial u_r}{\partial r} - \frac{1}{3} \nabla \cdot \vec{u} \right)\,, \\
% phi phi - component
\sigma_{\phi \phi} &= 2 \nu \Sigma \left(\frac{1}{r} \frac{\partial u_\phi}{\partial \phi} + \frac{u_r}{r} - \frac{1}{3} \nabla \cdot \vec{u} \right)\,, \\
% r phi - component
\sigma_{r \phi} &= \nu \Sigma \left(\frac{\partial u_\phi}{\partial r} + \frac{1}{r}\frac{\partial u_r}{\partial \phi} \right)\,, \\
% divergence of v
\nabla \cdot \vec{u} &= \frac{\partial u_r}{\partial r} + \frac{1}{r}\frac{\partial u_\phi}{\partial \phi} + \frac{u_r}{r}\,.
\end{align}

The momentum update is then performed according to \citet{kley1999mass_flow} (see also \citet{dangelo2002nested_grid}) as
\begin{align}
\Sigma \frac{\partial u_r}{\partial t} &= \frac{1}{r}\frac{\partial (r \sigma_{rr})}{\partial r} + \frac{1}{r}\frac{\partial \sigma_{r \phi}}{\partial \phi} - \frac{\sigma_{\phi \phi}}{r} \,. \label{eq:visocisty_momentum_update_r} \\
\Sigma \frac{\partial u_\phi}{\partial t} &= \frac{1}{r^2}\frac{\partial (r^2 \sigma_{r\phi})}{\partial r} + \frac{1}{r}\frac{\partial \sigma_{\phi \phi}}{\partial \phi} \,. \label{eq:visocisty_momentum_update_phi} \\
\end{align}
Finally, the energy update due to viscosity \citep{dangelo_thermohydrodynamics_2003} is given by
\begin{align}
\frac{\partial e}{\partial t} &= \frac{Q_\text{visc}}{\Delta t}  = \frac{1}{2 \nu \Sigma} 
\left[ \sigma_{rr}^2 + 2 \sigma_{r \phi}^2 + \sigma_{\phi \phi}^2 \right] + \frac{2 \nu \Sigma}{9} \left( \nabla \cdot \vec{u} \right)^2 \,,  \label{eq:viscous_heating}
\end{align}
where $Q_\text{visc}$ is to be used in the energy update in Eq.~\eqref{eqn:energy_update_implicit}.

%%%%%%%%%%%%%%%%%%%%%%%%%%%%%%%%%%%%%%%%
\subsubsection{Tscharnuter \& Winkler artificial viscosity}
%%%%%%%%%%%%%%%%%%%%%%%%%%%%%%%%%%%%%%%%
\label{sec:artificial_viscosity}
\label{sec:TW}
The role of artificial viscosity is to handle (discontinuous) shock fronts in finite-difference schemes.
This is achieved by smoothing the shock front over several grid cells by adding a bulk viscosity term.
\citet{tscharnuter1979artificial_viscosity} raised concerns about the formulation of artificial viscosity introduced by \citet{neumann1950shocks},
as it can produce artificial pressure even if there are no shocks \citep[e.g.][Sect.~6.1.4.]{bodenheimer2006numerical}.
\citet{tscharnuter1979artificial_viscosity} then proposed a tensor artificial viscosity, analogous to the viscous stress tensor, that is independent of the coordinate system and frame of reference.
For our implementation of this artificial viscosity, we follow \citet[][Appendix~B]{stone1992zeus1} who added two additional constraints on the artificial viscosity:
the artificial viscosity constant must be the same in all directions and the off-diagonal elements of the tensor must be zero to prevent artificial angular momentum transport.
Please note, that there is also an artificial viscosity described in the main text of \citet{stone1992zeus1}, sometimes referred to as the \textit{Stone and Norman} artificial viscosity, which does not have these properties and is only applicable in Cartesian coordinates.
Nonetheless, it is sometimes used in cylindrical and spherical coordinates.

In our case, we use the version suited for curve-linear coordinates and the artificial viscosity pressure tensor is given by
\begin{equation}
	\mathbf{Q} =
	  \begin{cases}
		l^2 \Sigma \, \left(\nabla \cdot \vec{u}\right)
			\left[\nabla\otimes \vec{u} - \frac{1}{3} \,\left(\nabla \cdot \vec{u}\right)\,\mathbf{I}\right] & \text{if\;} \nabla \cdot \mathbf{u} < 0\\
		0 & \text{otherwise}\\
	  \end{cases} \,,
  \end{equation}
where $l = q \,\Delta x$ is the distance over which shocks are smoothed with the dimensionless parameter $q$ near unity and the cells size $\Delta x$.
It is given by $\Delta x = \max (\Delta x_\mathrm{a}, \Delta x_\mathrm{b})$ where $a$ and $b$ indicate the grid of cell centers and interfaces, respectively. The contribution to the momentum equation is
\begin{equation}
	\Sigma \frac{\partial u_\phi}{\partial t} = \frac{\partial Q^\phi_{\; \phi}}{r\partial \phi}.
\end{equation}
\begin{equation}
	\Sigma \frac{\partial u_r}{\partial t} = \frac{\partial Q^r_{\; r}}{\partial r} + \frac{Q^r_{\;r}}{r} - \frac{Q^\phi_{\; \phi}}{r}.
\end{equation}
Finally, the shock heating caused by the artificial viscosity is given by
 \begin{equation}\label{eq:artificial_viscosity_heating}
	\frac{\partial e}{\partial t} = -l^2 \Sigma \, \left(\nabla \cdot \vec{u}\right)\frac{1}{3}\left[\left(\frac{\partial u_r}{\partial r}\right)^2 + \left(\frac{\partial u_\phi}{r \partial \phi} + \frac{u_r}{r}\right)^2 + \left(\frac{\partial u_\phi}{r \partial \phi} + \frac{u_r}{r} - \frac{\partial u_r}{\partial r}\right)^2\right].
\end{equation}
To ensure the stability of these updates, we use a CFL constraint analogous to the one in \citep[][, see their Sect.~4.6]{stone1992zeus1}
\begin{align}
\Delta t_\mathrm{art} = \frac{\Delta x^2}{4 l^2 \, \nabla \cdot \vec{u}} 
	= \frac{1}{4 q^2 \, \nabla \cdot \vec{u}} \,.
\end{align}

%%%%%%%%%%%%%%%%%%%%%%%%%%%%%%%%%%%%%%%%
\subsubsection{Local viscosity stabilizer}
%%%%%%%%%%%%%%%%%%%%%%%%%%%%%%%%%%%%%%%%
\label{sec:viscosity_stabilizer}

We found numerical instabilities in simulations of disks in close binary systems.
In these systems, the disk is truncated by tidal forces \citep[e.g.][]{artymowicz1994dynamics}.
At this truncation radius, the strong density gradients can cause numerical instabilities in the viscosity update step which drastically reduces the time step.

To prevent these instabilities, we designed a damping method that checks whether the viscosity update is too large and unstable and then reduces the update to a stable size. 
This method has the advantage that it is a local per-cell update that can be dropped into the existing code with only one modification to the update step. 
An alternative solution would be to implement a full implicit viscosity update step based on solving a linear system of equations which would have required substantial changes to our code in the viscosity update step.
Furthermore, this implicit update would be computationally more expensive whereas the overhead of the local damping method is negligible.
Because the instability is numerical and confined to only a small region, we argue that the damping method is a valid solution.

For our method, we interpret the viscosity update as a diffusion process. As we are only looking at a single cell,
we treat the velocities of the neighboring cells as constant. We then can write the velocity update due to viscosity in the form of
\begin{equation}
	\label{eq:explicit_visc_update}
	u^{t+\Delta t} = u^t + \Delta t \left( c_1 \cdot u^t + c_2 \right)\,,
\end{equation}
where in the nomenclature of Fig.~\ref{fig:flow}, $u^{t+\Delta t} = u_d$ and $u^t = u_c$.
The analytical solution to this equation is an exponential relaxation to the equilibrium velocity of $u_\mathrm{eq} =
  -\frac{c_2}{c_1}$.
When the explicit update overshoots the equilibrium velocity, the method becomes unstable.
One option to avoid the instability is to add $\Delta t \cdot c_1 > -1$ to the CFL criteria but this effectively freezes the simulation.
Instead, the code can now be configured to use
\begin{equation}
	\label{eq:implicit_visc_update}
 {u}_d = u_c + \Delta t \frac{c_1 \cdot u_c + c_2}{\mathrm{max}(0,\,\,1 + \Delta t\cdot c_1) - \Delta t \cdot c_1}\,.
\end{equation}
for the velocity update due to viscosity (see Sect.~\ref{sec:viscosity}).
For $\Delta t \cdot c_1 > -1$, the update reverts to \eqref{eq:explicit_visc_update}, 
while for $\Delta t \cdot c_1 < -1$, the new velocity is set to the equilibrium velocity $u_\mathrm{eq}$.

Compared to the other solution for too large time steps, which is to allow overshooting, we argue that an exponential decay to the equilibrium velocity is a physically more plausible choice.
However, one should keep in mind that this method violates angular momentum conservation.
In our tests, we found that only a few, low-density cells are unstable that have little influence on the whole simulation.
% rephrase until here

\begin{figure}
\begin{center}
\includegraphics[width=\linewidth]{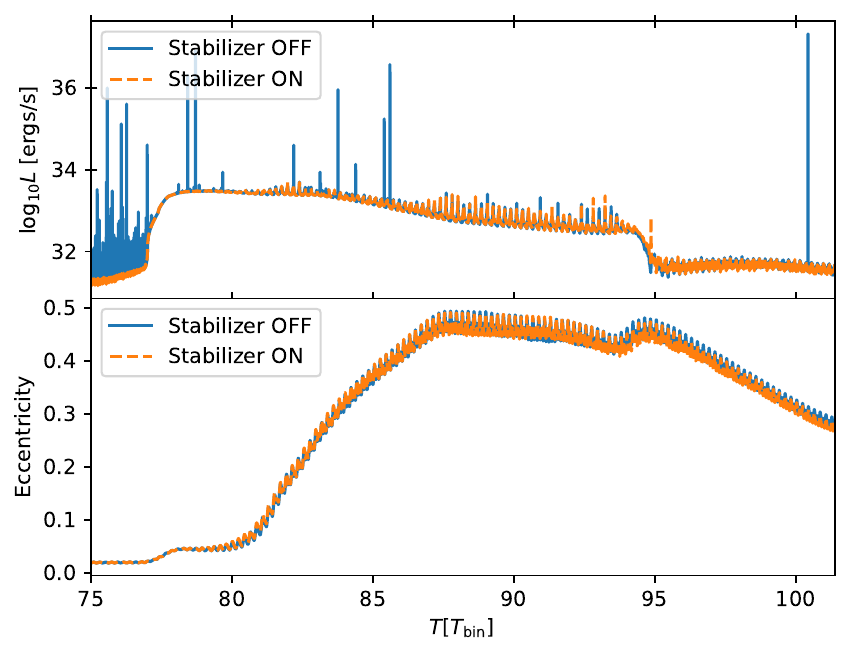}
\caption{\label{fig:viscosity_stabilizer} 
Luminosity and mass-weighted eccentricity during a super-outburst in a cataclysmic 
variable system with and without our viscosity stabilizer. The simulation
without the stabilizer required 3.8 times
more time-steps during the shown time frame.
}
\end{center}
\end{figure}

In Fig.\,\ref{fig:viscosity_stabilizer}, we show two simulations of a cataclysmic variable during a super-outburst, with the only difference being the stabilizing method turned on (blue line) or off (red line).
The top and bottom panels show the time evolution of luminosity and mass-weighted eccentricity, respectively.
The numerical instabilities that are dominating the luminosity (red line) are prevented by this method.
Yet, the overall time evolution of the eccentricity is similar.
Using the eccentricity as a proxy for the dynamical evolution of the disk, we can conclude that the stabilizing method only has a negligible impact on the physics.

\subsection{CFL criterion for heating and cooling}

A heating and cooling time step criterion was added similar to the cooling CFL criteria used in the \textsc{Pluto} code \citep{mignone2007pluto}. We found the additional CFL criteria beneficial to simulations of disks in close binaries that have truncated disks with strong density gradients (Wehner et al. 2024 in prep).
The time step must be smaller than
\begin{align}
	\Delta t_\mathrm{heatcool} = f \frac{e}{|Q_p - Q_-|}
\end{align}
where $f$ is a fudge factor to set the maximum energy change per time step.
We found that a factor of $f=10$ by trial-and-error to improve stability while not impacting the overall time step too much.

%%%%%%%%%%%%%%%%%%%%%%%%%%%%%%%%%%%%%%%%
\subsection{Scale-height dependent on all point masses}
%%%%%%%%%%%%%%%%%%%%%%%%%%%%%%%%%%%%%%%%
\label{sec:aspect_ratio}
The scale height is determined by the balance between vertical gravity and pressure forces. 
Realistically, the gravity of all N-body objects, including planets, should contribute to the gravitational force.
Including these contributions, the scale height can be computed as \citep{gunther2004}
\begin{equation}
	H = \left[\sum_i^{N_b} \frac{1}{H_i^2}\right]^{-1/2}\,,\quad H_i = \frac{c_{s}^\text{iso}}{\Omega_{\text{K}\,,i}}\,,
\end{equation}
where $c_{s}^\text{iso}$ is the isothermal sound speed and $\Omega_{\text{K}\,,i}$ is the Keplerian frequency with respect to the $i$-th N-body object.

For the computation of self-gravity presented in Sect.~\ref{sec:self-gravity}, we need a value for the aspect ratio.
In the case that the full N-body system influences the scale height, we compute the aspect ratio in an analog way
\begin{equation}
	\label{eq:nbody_h}
	h = \left[\sum_i^{N_b} \frac{1}{h_i^2}\right]^{-1/2}\,,\quad h_i = \frac{c_{s}^\text{iso}}{v_{\text{K}\,,i}}\,.
\end{equation}

\subsection{Particle system}
%%%%%%%%%%%%%%%%%%%%%%%%%%%%%%%%%%%%%%%%
\label{sec:dust}

\fargocpt can simulate solid particles of any size in a protoplanetary disk.
A collection of solid particles is represented as Lagrangian super-particles.
This means that a bunch of identical physical particles, with the same size and same density, are represented by one super-particle.
These super-particles are then evolved in the simulation according to the forces acting on them: a drag force due to the gas and gravity from the N-body objects and the disk if self-gravity of the gas is enabled.

These super-particles do not react back on the gas.
As such, the mass of each super-particle can be changed freely even after the simulation.
The number of super-particles determines how well the dynamics of solid particles are resolved or sampled in the simulation.
The first implementation of these Lagrangian super-particles in \fargocpt was used in \citet{picogna_how_2015}.

The particle system is synchronized to the N-body and hydro simulation once every iteration, i.e. at time $t^n$, and then integrated for a time equal to the time step resulting from the CFL criterion $\Delta t$, i.e. until $t^{n+1} = t^n + \Delta t$.
The code supports two different integrators for particle motion:
\begin{enumerate}
\item An implicit exponential midpoint integrator as described in \citet{mignone_particle_2019} Appendix~B.2.1 that combines two regimes of particle sizes ranging from micrometer-sized dust ($\text{St} \ll 1$) to meter-sized boulders ($\text{St} \geq 1$).
\item An explicit fifth order Runge-Kutta integrator with adaptive time stepping \citep{cash_variable_1990} to integrate very large particles with $\text{St} \gg 1$ such as planetesimals.
\end{enumerate}

In both integrators, the integration step can be performed in polar coordinates while the explicit integrator also supports Cartesian coordinates.
The polar formulation conserves angular momentum better in the case of circular particle orbits.
We found, however, that as soon as the orbits become eccentric, this advantage disappears.

%%%%%%%%%%%%%%%%%%%%%%%%%%%%%%%%%%%%%%%%
\subsubsection{Drag force}
%%%%%%%%%%%%%%%%%%%%%%%%%%%%%%%%%%%%%%%%

The friction force for the gas drag is calculated as a smooth interpolation between the Epstein and Stokes regimes.
We use a model that follows \citet{picognaParticleAccretionPlanets2018b} but with the Epstein drag calculated according to \citet{woitkeDustBrownDwarfs2003a}.

The drag model depends on three dimensionless numbers: Knudsen number $\text{Kn}$, Mach number $\text{Ma}$, and Reynolds number $\text{Re}$.
Those are given by
\begin{align}
	\text{Kn} &= \frac{l}{2a}\,, \\
	\text{Ma} &= \frac{v_\text{rel}}{c_s}\,, \\
	\text{Re} &= \frac{2 a \rho_s v_\text{rel}}{\nu}\,,
\end{align}
where $a$ is the radius of the solid particle, $l$ is the mean free path of the gas molecules, $c_s$ is the sound speed, $\rho_s$ is the material density of the particle, $v_\text{rel}$ is the relative velocity between the particle and the gas, and $\nu$ is the kinematic viscosity coefficient of the gas.
Note that the latter is different from the kinematic viscosity coefficient used to model turbulent accretion with the $\alpha$ model.

The mean free path of the gas molecules is given by \citep{haghighipourPressureGradientsRapid2003}
\begin{align}
	l = \frac{m_0}{\pi a_0 \rho_g} = 4.72\times 10^{-9}\,\text{cm}\,\frac{\text{g}/\text{cm}^3}{\rho_g}\,.
\end{align}
with $a_0 = 1.5 \times 10^{-8}\,\text{cm}$, assuming the gas
consists primarily of H$_2$ molecules, and the mass volume density of the gas $\rho_g$.
The kinematic viscosity coefficient is given by
\begin{align}
	\nu = \frac{1}{3} m_0 \frac{v_\text{thermal}}{\sigma}\,,
\end{align}
with the thermal velocity $v_\text{thermal} = \sqrt{8/\pi} \, c_s^\text{iso}$ and the collisional cross-section between the gas molecules $\sigma = \pi a_0^2$.

The drag force $\vec{F}_\text{drag}$, a particle in the disk is subjected to, is related to the stopping time $t_\text{stop}$ via
\begin{align}
\vec{F}_\mathrm{drag} = - \frac{m_\mathrm{s}}{t_\mathrm{stop}} \vec{v}_\mathrm{rel}\,,
\end{align}
with the mass of a solid particle $m_\mathrm{s}$.
The stopping time is given by
\begin{align}\label{eqn:stopping_time}
t_\text{stop} = \frac{4}{3} \frac{l\, \rho_\text{s}}{\rho_g C_\text{d} c_s \,\text{Kn}}\,.
\end{align}
The total drag coefficient results from a quadratic interpolation between the free molecular flow (Epstein) and viscous regime (Stokes) and is given by \citep[][see their Eq.~(18)]{woitkeDustBrownDwarfs2003a}
\begin{align}
	C_\text{d} = \frac{9 \, \text{Kn}^2 \, C_\text{E} + C_\text{S}}{\left( 3\,\text{Kn} + 1 \right)^2}\,.
\end{align}
The Epstein drag coefficient is a smooth interpolation of the sub- and supersonic regimes of the Stokes drag \citep[][see their Eq.\,(13)]{woitkeDustBrownDwarfs2003a}:
\begin{align}
 C_\text{E} = 2 \sqrt{\text{Ma}^2 + \frac{128}{9\pi}}\,.
\end{align}
The Stokes drag coefficient is given by \citep[Eq.~(15) in][]{woitkeDustBrownDwarfs2003a}:
\begin{align}
C_\text{S} = 
\begin{cases} 
\frac{24\,\text{Ma}}{\text{Re}} + 3.6\,\text{Ma}\,\text{Re}^{-0.313} & ,\,\text{Re} \leq 500 \\
9.5\times10^{-5}\,\text{Ma}\,\text{Re}^{1.397} & ,\,500 < \text{Re} \leq 1500\\
2.61\,\text{Ma} & ,\,1500 < \text{Re}
\end{cases}
\end{align}
Please note that the expressions here differ from the ones in \citet{picognaParticleAccretionPlanets2018b} and \citet{woitkeDustBrownDwarfs2003a} because ours are formulated for computing $t_\text{stop}$ in Eq.~\eqref{eqn:stopping_time} while the others are formulated for computing $\vec{F}_\text{drag}$ directly.
In Appendix~\ref{app:dust_drift_test} we present a test of the radial dust drift velocity resulting from the gas drag.

%%%%%%%%%%%%%%%%%%%%%%%%%%%%%%%%%%%%%%%%
\subsubsection{Dust diffusion}
%%%%%%%%%%%%%%%%%%%%%%%%%%%%%%%%%%%%%%%%
\label{sec:dust_diffusion}

Simulating dust with Stokes numbers around unity requires treating dust diffusion due to unresolved small-scale turbulent motion of the gas. Otherwise, this dust accumulates nonphysically in a single point at a pressure maximum, such as the centers of large-scale vortices.
Our code includes dust diffusion modeled with stochastic kicks in analogy to \citet{charnoz_three-dimensional_2011}.
In this model, the Lagrangian super-particles receive a kick at every time step.
The kick is only applied in the radial direction such that
\begin{align}
r^\mathrm{new} = r^\mathrm{old} + \delta r
\end{align}
and the azimuthal velocity is corrected to conserve angular momentum
\begin{align}
\dot{\phi}^\mathrm{new} = \dot{\phi}^\mathrm{old} \frac{r^\mathrm{old}}{r^\mathrm{new}}\,.
\end{align}
The correction of the azimuthal velocity is required to avoid a nonphysical drift due to the changed angular momentum.

The kick strength is 
\begin{align}\label{eqn:dust_diffusion_kick_strength}
\delta r &= \langle \delta r \rangle \Delta t \Omega_\text{K} + W \sigma + \delta_\text{2D}
\end{align}
where we the symbols carry the same meaning as in \citet{charnoz_three-dimensional_2011} (see their Eq.~(17)).
$\langle \delta r \rangle = \Delta t\,D_d / \rho_g \,\partial \rho_g/\partial r$ is the mean 
and $\sigma^2 = 2D_d \Delta t$ is the variance of a random Gaussian variable, 
$W$ is a standard normal random variable, and $\delta_\text{2D}$ is an additional displacement to take into account the second dimension, i.e. the diffusion in the $\phi$ direction.
The dust diffusion coefficient is $D_d = \nu/\mathrm{Sc} = \alpha c_s H/\mathrm{Sc}$ with the Schmidt number $\mathrm{Sc} = (1 + \mathrm{St}^2)^2/(1 + 4 \mathrm{St}^2)$ \citep{youdinParticleStirringTurbulent2007}.
The factor $\delta t \Omega_\text{K}$ takes into account the time correlation in the kicks due to the gas turbulence \citep[see Sect.~2.7 in][]{charnoz_three-dimensional_2011}.

The 2D correction
\begin{align}\label{eqn:dust_diffusion_2d_correction}
\delta_\text{2D} = \sqrt{r^2 + (W\sigma)^2} -r
\end{align}
can be derived by a geometrical argument taking into account the diffusive spread in the azimuthal direction.
Consider a dust particle on a circular orbit that is displaced by a turbulent kick in the azimuthal direction. 
Its final location will be radially further out from where it started, independent of whether it was kicked in the direction of the orbit or opposite to it.
This positive radial change is taken into account by $\delta_\text{2D}$.

To save computational costs, we use the same standard random number $W$ in Eq.~\eqref{eqn:dust_diffusion_kick_strength} and in Eq.~\eqref{eqn:dust_diffusion_2d_correction}.
This introduces correlations between two contributions to the kick.
However, the factor $W\sigma$ is a small number compared to the relevant length scales and the 2D correction is quadratic in $W\sigma$ which can be seen by a Taylor expansion of Eq.~\eqref{eqn:dust_diffusion_2d_correction}: $\delta_\text{2D} \approx \frac{1}{2} \left( W\sigma \right)^2 / r$.
Additionally, the direction of the radial kick and the 2D correction are not correlated, because the latter is always positive.
Thus the effect of the correlations should be negligible.

To our knowledge, this correction has not been used before when kicks were only applied in the radial direction. However, we found it to be necessary to match the solutions of the radial advection-diffusion equation, see Appendix~\ref{app:dust_diffusion_test}.

For generating the random numbers for these steps, we added the small non-cryptographic random generator by Bob Jenkins (JSF)
which is fast and well-tested. The implementation of the generator as well as tests are shown in \citet{neill2018jsf}.

A test of this procedure is presented in Appendix~\ref{fig:dust_diffusion_test} in the form of a comparison of simulations of the spread of a thin dust ring simulated in \fargocpt against a solution of the radial advection-diffusion equation.
We find excellent agreement between the two approaches.

%%%%%%%%%%%%%%%%%%%%%%%%%%%%%%%%%%%%%%%%
\subsection{Accretion onto point masses}
%%%%%%%%%%%%%%%%%%%%%%%%%%%%%%%%%%%%%%%%
\label{sec:planet_accretion}

When accretion is enabled, mass is removed from the disk and added to the N-body object.
The mass is removed from the vicinity of the N-body object similar to \citet{kley1999mass_flow} with a fixed half-emptying time.
The momentum of the accreted mass is added to the N-body object.
The radius around the N-body object from which mass is removed, $R_\text{acc}$, is given by a fraction of the Roche lobe radius $R_\mathrm{Roche}$, which we calculate as described in Sect.~\ref{sec:gravitational_smoothing}.
The accretion radius is given by
\begin{align}
	R_\mathrm{acc} = c \cdot R_\mathrm{Roche}\,.
\end{align}
For a cell with mass $m_i$ and distance $d_i$ to the accreting object, the rate of mass removal is given by
\begin{align}
	\dot{m_i} = - f_i m_i \frac{\log(2)}{T_\text{P}}\,,
\end{align}
where $f_i$ is the accretion fraction for cell $i$.
Close to the planet, it is higher and decreases with distance.
We use a simple two-step function such that
\begin{align}
	f_i =
	\begin{cases}
		 2 f_\text{acc}  &\text{if  } R_\mathrm{acc}/2 < d_i \leq R_\mathrm{acc} \\
		 \ \ f_\text{acc} &\text{if  } d_i > R_\mathrm{acc}/2 \\
		 \ \ 0 &\text{if  } d_i > R_\mathrm{acc}
	\end{cases}\,
\end{align}
with the accretion parameter $f_\text{acc}$ which can be chosen individually for each N-body object.
$f_i$ is additionally limited by the mass in the cell $m_i$ such that $f_i = \min(f_i, (m_i - A_i \Sigma_\text{floor}) T_\text{P} /(\log(2)\Delta t))$, where $A_i$ is the area of cell $i$ and $\Sigma_\text{floor}$ is the density floor.
This scheme effectively takes away mass from the cells with a half-emptying time of $T_\text{P} / f_i$.
Summation over the cells in the vicinity of the body yields the accretion rate and momentum transfer onto it:
\begin{align}
\dot{M} &= - \sum_{i\in\mathcal{V}} \dot{m_i}\,, \label{eq:accretion_rate_simple} \\
\dot{\vec{P}} &= - \sum_{i\in\mathcal{V}} \dot{m_i} \vec{u_i}\,. \label{eq:accretion_momentum_simple}
\end{align}
where $\mathcal{V}$ is the set of indices of cells that are located within $R_\text{acc}$ of the accreting object.

For accretion onto binary stars from a circumbinary disk, we can use a more sophisticated model of accretion.
Assume that, within the cavity of the circumbinary disk, both stars are surrounded by their own disks which are tidally truncated, keeping them small.
Assume further that within these disks the accretion happens according to a simple 1D viscous accretion-disk model and that the stars accrete at the same rate as the mass flows through these disks.
Then, the accretion rate onto the stars is given by \citep{lynden-bellEvolutionViscousDiscs1974}
\begin{align}\label{eq:viscous_accretion_1D}
	\dot{M}_\mathrm{acc} = 3 \pi \nu \Sigma s\,.
\end{align}
The free parameter $s$ accounts for the increase in accretion close to an object due to gas friction at the boundary layer.
The idea is analogous to the viscous inflow boundary condition (Sect.~\ref{sec:boundary_conditions_vr}).
For our purposes, we assume $\nu$ and $\Sigma$ to be constant and equal to the average values within $R_\text{acc}$.

With this model of accretion, we can modify the scheme above in such a way that the rate of mass removed from the disk in the vicinity of the object is equal to the 1D viscous accretion rate given by Eq.~\eqref{eq:viscous_accretion_1D}.
To achieve this, we choose $f_i$ such, that
\begin{align}
\sum_{i\in\mathcal{V}} \dot{m}_i = \sum_{i\in\mathcal{V}} f_i m_i = \Delta t \dot{M}_\mathrm{acc}\,.
\end{align}
Again, we want $f(d)$ to vary with the distance $d$ to the accreting object and we want $f(d)$ to be zero for $d > R_\mathrm{acc}$.
Now, instead of a step function, we choose a smooth power law function for $f(d)$ such that
\begin{align}
	f(d) = f_0 \left(1 - \left(\frac{d}{R_\mathrm{acc}}\right)^q\right)\,,
\end{align}
with the understanding that  $f(d > R_\text{acc}) = 0$.
With this choice, the physical model from above, and assuming $q>0$, we find
\begin{align}
	f_0 = 3 \frac{q+2}{q} s \frac{\nu}{R_\text{acc}}\,.
\end{align}
In the code, we use $q=1$.
This enables the physically motivated computation of an accretion rate onto a secondary star which is located within the computational domain.

%%%%%%%%%%%%%%%%%%%%%%%%%%%%%%%%%%%%%%%%
\subsection{Boundary conditions}
%%%%%%%%%%%%%%%%%%%%%%%%%%%%%%%%%%%%%%%%
\label{sec:boundary_conditions}
This section describes the boundary conditions (BCs) that are implemented in the code.
We start by stating the most basic BCs and then describe the more complex ones.
In the code, BCs are applied by setting the value of $\Sigma$, $e$, $u_\phi$ in the center of the ghost cells and $u_r$ on its two interfaces.

\begin{figure}
	\begin{center}
	\includegraphics[width=0.8\linewidth]{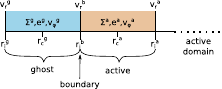}
	\caption{\label{fig:boundary_conditions} Schematic of the location of the quantities for the boundary conditions. The first active cell is shown in beige and the ghost cells are in blue. The ghost cells are not updated by the source terms and transport steps. Their values are used to apply the boundary conditions.}
	\end{center}
\end{figure}

The nomenclature used in this section is as follows.
The superscripts $a$ and $g$ denote the last cell in the active domain and the ghost cell, respectively.
The location of the cell centers are $r^a$ and $r^g$.
Additionally, the superscript $b$ denotes the boundary at $R_\text{min}$ and $R_\text{max}$.
The locations of the interfaces are $r_i^a$ for the interface between the first and second active cell, $r_i^b$ for the boundary interface, and $r_i^g$ for the interface of the ghost cell facing away from the active domain.
Values in the center of the ghost cells are $\Sigma^g$, $e^g$ and $u_\phi^g$.
The two values on the ghost cell interfaces are $u_r^b$, at the interface between the active domain and the ghost cell and $u_r^g$.
The values in the first cell of the active domain are $\Sigma^a$, $e^a$, $u_\phi^a$ and the radial velocity at the interface between the first and second active cell is $u_r^a$.
See Fig.~\ref{fig:boundary_conditions} for a schematic of the location of these quantities.

In the code, BCs can either be set for each variable individually or for all variables at once.
We call the latter case a composite BC.
Here, we start by introducing the individual BCs.

%%%%%%%%%%%%%%%%%%%%%%%%%%%%%%%%%%%%%%%%
\subsubsection*{Boundary conditions applicable to all variables}
%%%%%%%%%%%%%%%%%%%%%%%%%%%%%%%%%%%%%%%%
\label{sec:boundary_conditions_basic}

Several BCs are available for each variable.
Those are summarized in Table~\ref{tab:boundary_conditions_all}.

The \emph{zero-gradient} BC enforces a radial derivative of zero at the boundary.
It copies the value from the last active cell into the ghost cell for cell-centered variables and from the last active cell interface into the ghost cell interface for interface variables.

The \emph{disk model} BC is used to set the boundaries according to a specific disk model.
The code currently supports a power law disk as a model.
The functions $x_\text{dm}(r)$ specify the disk model for each variable $x$.
An example is the surface density $\Sigma_\text{dm}(r) = \Sigma_0 (r/r_0)^{-p}$.

Another choice to set the boundaries to a specific model is the \emph{reference} BC.
In this case, the reference values $x_\text{ref}(r)$ are loaded from a special simulation snapshot.
The same snapshot is used for the damping zones (see Sect.~\ref{sec:damping_zones}).
This snapshot is generated at the beginning of the simulation and contains the initial conditions by default.
To specify a custom model, the user can run the simulation for zero time steps (using the \verb|-N 0| command line flag) and then replace the reference values with the desired arbitrary model using an external tool.
Then, the simulation can be continued from snapshot number 0.

\begin{table}[h]
	\begin{center}
		\caption{Boundary conditions available all variables $x \in \{\Sigma, e, u_r, u_\phi\}$.\label{tab:boundary_conditions_all}}
		\begin{tabular}{ |c|c| } 
		 \hline
		 Condition & Formula \\ 
		 \hline
		 zero-gradient & $x^{g,b} = x^a$ \\ 
		 disk model & $x^{g,b} = x_\text{dm} (r^{g,b})$ \\ 
		 reference & $x^{g,b} = x_\text{ref}(r^{g,b})$ \\ 
		 \hline
		\end{tabular}
		\tablefoot{The superscript $g,b$ indicates that the same formula is used for variables in the center of the ghost cell and on the interfaces. The radii are understood to be at the respective locations.}
		\end{center}
\end{table}

%%%%%%%%%%%%%%%%%%%%%%%%%%%%%%%%%%%%%%%%
\subsubsection{Radial velocity boundary conditions}
%%%%%%%%%%%%%%%%%%%%%%%%%%%%%%%%%%%%%%%%
\label{sec:boundary_conditions_vr}

Several BCs apply only to the radial velocity $u_r$, see Table~\ref{tab:boundary_conditions_vr}.
They are mostly connected to the flow through the boundary.

By mirroring the flow, we can simulate a \emph{reflective} boundary.
This is achieved by setting the radial velocity to zero on the boundary and to the negative value of the radial velocity in the last active cell at the other interface of the ghost cell.
This keeps mass from flowing out of the domain but also reflects waves back into the domain and can cause wave interference patterns and instabilities in the worst case.
Usually, the reflective boundary is used together with damping zones (see Sect.~\ref{sec:damping_zones}) to prevent these downsides.

The opposite behavior is achieved by the \emph{outflow} BC.
It lets mass flow out of the domain but does not allow mass to flow into the domain.
This is achieved by using a zero-gradient condition if the velocity vector is pointing outwards at the boundary and by setting the radial velocity to zero otherwise.
Pointing outwards means that the radial velocity is positive at the outer boundary and negative at the inner boundary.
Despite being an outflow condition for mass, this prescription can still reflect waves back into the domain, but it is less of an issue as the boundary tends to form an empty area between the disk and the boundary.

At the inner boundary, it can be advantageous to more closely control the flow through the boundary to model a certain accretion rate onto the star.
There are currently two options to influence this flow.
The first option is the \emph{viscous outflow} BC.
This BC assumes a steady-state accretion disk at the inner boundary.
Then, analogous to the viscous accretion model presented in Sect.~\ref{sec:planet_accretion}, the accretion rate through the boundary is given by \citep{lynden-bellEvolutionViscousDiscs1974}
\begin{align}\label{eq:mdot_viscous_accretion_1D}
	\dot{M} = 3 \pi \nu \Sigma s\,,
\end{align}
with the free parameter $s$ for which 5 is a suitable value for accretion onto a solar-type star \citep{pierens2008constraints}.
The BC is implemented by setting the radial velocity at the boundary $u_r^{g/b} =  v_\text{visc}(r_i^{g/b})$ with the viscous inflow speed
\begin{align}\label{eq:viscous_inflow_speed}
	v_\text{visc} = - \frac{3}{2} \frac{s \nu}{r} \,.
\end{align}

Another choice is to set the radial velocity to a fraction of the Keplerian velocity at the inner boundary.
This is done by the \emph{Keplerian} BC for the radial direction.
It sets the radial velocity to the negative of the fraction of the Keplerian velocity at the inner boundary.
This BC can also be applied at the outer boundary.

\begin{table}[h]
\caption{Boundary conditions for the radial velocity $u_r$. \label{tab:boundary_conditions_vr}}
\begin{center}
\begin{tabular}{ |c|c| } 
 \hline
 Condition & Formula \\ 
 \hline
 reflective & $u_r^g = - u_r^a$, $u_r^b = 0$ \\ 
 outflow & $\begin{cases} u_r^{g/b} = u_r^a & \text{if } \vec{u}^a \text{ points outwards} \\ u_r^{g/b} = 0 &\text{otherwise}\end{cases}$ \\ 
 viscous outflow & $u_r^{g/b} = - c \,v_\text{visc} (r_i^{g/b})$, at $R_\text{min}$ \\ 
 Keplerian & $u_r^{g/b} = c\,v_\text{K}(r_i^b)$ \\ 
 \hline
\end{tabular}
\tablefoot{The factors $c$ are different for each boundary condition and can be selected by the user in the config file. The boundary conditions from Table~\ref{tab:boundary_conditions_all} are also available for $u_r$.}
\end{center}
\end{table}

%%%%%%%%%%%%%%%%%%%%%%%%%%%%%%%%%%%%%%%%
\subsubsection{Azimuthal velocity boundary conditions}
%%%%%%%%%%%%%%%%%%%%%%%%%%%%%%%%%%%%%%%%
\label{sec:boundary_conditions_vphi}

\begin{table}[h]
	\begin{center}
	\caption{Boundary conditions for the azimuthal velocity $u_\phi$.}
	\label{tab:boundary_conditions_vphi}
	\begin{tabular}{ |c|c| } 
	 \hline
	 Condition & Formula \\ 
	 \hline
	 Keplerian & $u_\phi^g = c\,v_\text{K}(r_c^g) - r\Omega_{F}$ \\ 
	 zero-shear & $u_\phi^g = r_c^g \,\Omega^a$ with $\Omega^a = u_\phi^a / r_c^a$ \\ 
	 balanced & $u_\phi^g = \sqrt{v_\text{K}^2 \left(\mathcal{S} + \mathcal{P} + \mathcal{Q} \right) - r g_r } - r\Omega_{F}$ \\ 
	 \hline
	\end{tabular}
	\tablefoot{The boundary conditions from Table~\ref{tab:boundary_conditions_all} are also available for $u_\phi$.}
	\end{center}
	\end{table}

Table~\ref{tab:boundary_conditions_vphi} summarizes the BCs that are available for the azimuthal velocity.
The most basic condition is the \emph{Keplerian} BC which sets the azimuthal velocity to a fraction of the Keplerian velocity at the boundary, corrected for the frame rotation.
The fraction can be chosen to be sub- or super-Keplerian to reflect additional pressure support or other effects.
It can be used to model boundary layers where the disk connects to the star.
Then the fraction is chosen such that $u_\phi$ is the surface rotation velocity of the star.

Depending on the flow close to the boundary, the Keplerian or any of the basic BCs might lead to shear at the boundary.
This results in torques at the boundary which might be nonphysical or undesirable.
To avoid this, we can use the \emph{zero-shear} BC which removes the shear at the boundary by scaling the azimuthal velocity in the last active cell to the ghost cell such that both have the same angular velocity.

Finally, the code supports the \emph{balanced} BC which sets the azimuthal velocity such that the centrifugal force is in equilibrium with all other forces.
This is especially useful for equilibrium disk models, see also Appendix~\ref{sec:app_equilibrium_disk}.
In addition to the gravity of the central object, this takes into account pressure, the quadrupole moment from a central binary, smoothing, and self-gravity.
It reads
\begin{align}\label{eq:equilibrium_boundary_vaz}
	u_\phi^g &= \sqrt{v_\text{K}^2 \left(\mathcal{S} + \mathcal{P} + \mathcal{Q} \right) - r g_r } - r\Omega_{F}\,,
\end{align}
where $g_r$ is the radial component of the self-gravity acceleration (see Sect.~\ref{sec:self-gravity}) and $\mathcal{P}$, $\mathcal{S}$, and $\mathcal{Q}$ represent pressure, smoothing, and quadrupole moment, respectively, and the last term accounts for the rotating frame.
This equation can be derived (see Appendix~\ref{sec:app_equilibrium_disk}) from the radial force balance starting from the radial momentum conservation Eq.~\eqref{eq:momentum_r_eq_polar}.

Usually, the term $\mathcal{S}$ equals $1$ in other instances of this formula in the literature.
However, when the gravitational potential (Eq.~\eqref{eq:potential_2}) is differentiated to calculate the external forces acting on the disk, additional terms that depend on the smoothing length appear, because the smoothing length depends on the location in the disk.
These terms are accounted for in the centrifugal balance by
\begin{align}\label{eq:smoothing_support}
	\mathcal{S} = \frac{1 + (h \alpha_\text{sm})^2 \left(1+\frac{\mathrm{d\,log}\,h}{\mathrm{d\,log}\,r}\right)}{\left( 1 + (h \alpha_\text{sm})^2 \right)^{3/2}}\,,
\end{align}
where $\alpha_\text{sm}$ is the smoothing parameter (see Sect.~\ref{sec:gravitational_smoothing}).
See Appendix~\ref{sec:app_equilibrium_disk} for more detail.
The pressure term is as usual given by \citep[e.g.][Eq.~(3.4)]{baruteauPredictiveScenariosPlanetary2008a}
\begin{align}
	\mathcal{P} = h^2 \frac{\mathrm{d\,log\,P}}{\mathrm{d\,log}\,r} = h^2 \left(2\frac{\mathrm{d\,log}\,h}{\mathrm{d\,log}\,r} + \frac{\mathrm{d\,log}\,\Sigma}{\mathrm{d\,log}\,r} - 1\right)\,,
\end{align}
where where the last equality follows from $P=c_{s,\text{iso}}^2 \Sigma$ and $c_{s,\text{iso}} = h\,v_\text{K}$.
For simulations of circumbinary disks, it can be advantageous to account for the quadrupole term of the gravitation potential of the binary.
The inclusion of this term can be turned on by the user.
The term reads \citep[][Eq.~(23)]{munoz2019quadropole}
\begin{align}\label{eq:quadrupole_term}
	\mathcal{Q} = \frac{3 Q}{r^2}\,,
\end{align}
with the quadrupole moment of the binary \citep[][Eq.~(24)]{munoz2019quadropole}
\begin{align}
Q = \frac{a_b^2}{4} \frac{q_b}{(1+q_b)^2} \left(1 + \frac{3}{2} e_b^2\right)\,. \label{eq:quadrupole_moment}
\end{align}
Here, $a_b$ is the binary separation, $q_b$ is the binary mass ratio $q_b$, and $e_b$ is the binary eccentricity.

%%%%%%%%%%%%%%%%%%%%%%%%%%%%%%%%%%%%%%%%
\subsubsection{Radiative transfer boundary conditions}
%%%%%%%%%%%%%%%%%%%%%%%%%%%%%%%%%%%%%%%%
\label{sec:boundary_conditions_RT}

The radiative transfer module solves a diffusion equation.
In the azimuthal boundary, the periodicity is built in, but on the radial boundaries, a BC is required.
If the module is enabled, the user must specify such conditions.
The BC is applied right before the SOR solver is called.

There are three options available.
First, a \emph{zero-gradient} conditions for the diffusion coefficient $K$ (see Eq.~\eqref{eq:fld_equation}).
Second, a \emph{zero-flux} condition which sets the diffusion coefficient to zero at the boundary ($r_i^b$).
Third, an \emph{outflow} condition which allows flux leaving the domain, implemented by setting $T=T_\text{floor}$ in the ghost cell.

%%%%%%%%%%%%%%%%%%%%%%%%%%%%%%%%%%%%%%%%
\subsubsection{Composite boundary conditions}
%%%%%%%%%%%%%%%%%%%%%%%%%%%%%%%%%%%%%%%%
\label{sec:boundary_conditions_composite}

The BCs presented above apply to single variables.
This subsection introduces the composite BCs, collections of BCs which apply to the variables, $\Sigma$, $e$, and $u_r$.
The BC for $u_\phi$ still needs to be set individually, except for the \emph{reference} case.

The \emph{outflow}, \emph{reflecting}, and \emph{zero-gradient} composite BCs are simply shorthands to set the individual BCs as described in Table~\ref{tab:special-boundaries}.
These BCs reflect what is commonly used in other codes.

The \emph{center of mass} BC is intended for the special case of simulating
infinite circumbinary disks in the center of one of the stars. It enforces
the initial power law profile for density and temperature,
the equilibrium for azimuthal velocity according to Eq.~\eqref{eq:equilibrium_boundary_vaz}, 
and the viscous speed for the radial velocity at the boundary and the outer damping region
with respect to the center of mass of the N-body system instead of the coordinate center.
In a second step, the gas velocities are transformed from polar to cartesian coordinates, 
shifted by the velocity of the coordinate center and transformed back to polar coordinates
in the frame of the central object.
\begin{table}[h]
\begin{center}
\caption{Composite boundaries and their descriptions}
\label{tab:special-boundaries}
\begin{tabular}{ |p{0.3\linewidth}|p{0.5\linewidth}| } 
 \hline
 Condition & Description \\ 
 \hline
 \centering center of mass & Calculate a disk model centered around the center of mass and shift it to the primary frame. \\ 
 \hline
 \centering Roche lobe overflow & Simulates mass overflow through the L1 point between binary stars. \\ 
 \hline
 \centering outflow & \{$\Sigma$, $e$\} = zero-gradient, $v_r$ = outflow \\
 \hline
 \centering reflecting & \{$\Sigma$, $e$\} = zero-gradient, $v_r$ = reflecting \\
 \hline
 \centering reference & \{$\Sigma$, $e$, $u_r$, $u_\phi$\} = reference \\
 \hline
 \centering custom & Template for the user to modify. \\
 \hline
\end{tabular}
\end{center}
\end{table}

For simulations of cataclysmic variables, we implemented a \emph{Roche Lobe overflow} BC that models the mass flow through the L1 point between binary stars when one of the stars is overflowing its Roche lobe.
While the function is implemented generally, it only works as intended if the primary is at the coordinate center and the outer edge of the domain has the size of the Lagrangian $L_1$ point.

We compute the width of the mass stream using the approximate function in \cite{BrianWarner2003} Sect. 2.4.1 which is a simplified version of the isothermal model for a Roche-lobe overflow from \cite{meyer1983}:
\begin{equation}
	W \approx \sqrt{\frac{2.4\cdot 10^{13}}{\pi} \left( \frac{T_s}{\mathrm{K}} \right) T^2_\mathrm{orb}(h) \,\mathrm{cm^2}}.
\end{equation}
We then select the cell in the outer boundary, which is closest to the secondary and smooth the mass stream with a Gaussian profile over 3 times its width.
We found that the initial stream width and temperature are of little importance as it quickly reaches a new equilibrium upon entering the simulation domain and that the width is mostly determined by the grid resolution.

\begin{figure}
\begin{center}
\includegraphics[width=\linewidth]{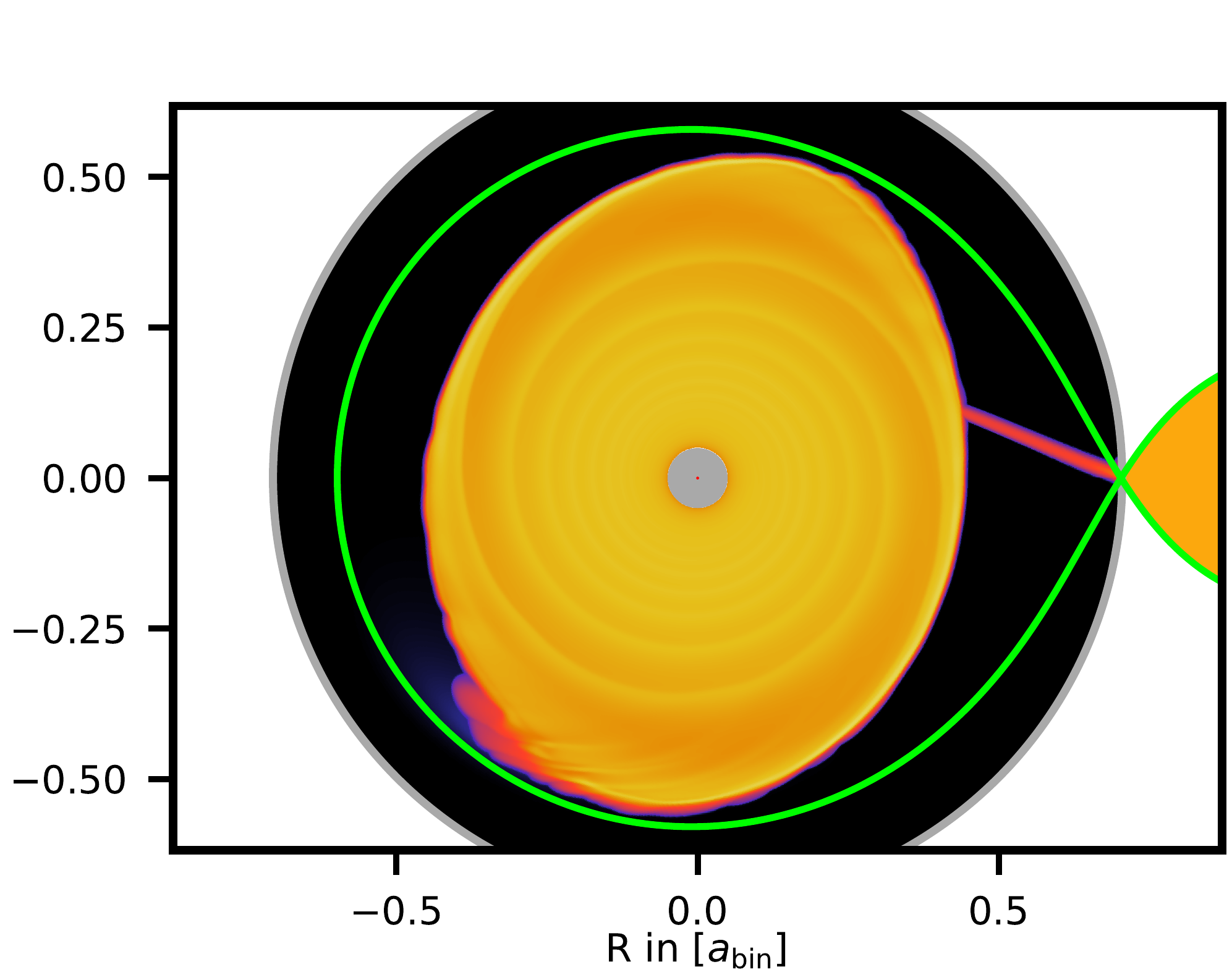}
\caption{\label{fig:roche_lobe_overflow}
Logarithmic density plot of a simulation with the \emph{Roche Lobe overflow} boundary condition showing the stream of material emerging from the $L_1$ point.
The green line indicates the Roche lobe of the binary system and the grey circle is the outer boundary of the simulation domain.
Outside the small region around the Lagrangian $L_1$ point the boundary condition is set to \emph{outflow}.}
\end{center}
\end{figure}

The initial radial velocity of the stream is computed, the same as in \cite{kley2008binary}, as a small fraction binary's orbital frequency:
\begin{equation}
	u_{r, \text{stream}} = -2 \cdot 10 ^{-3} \Omega_\mathrm{bin} r_\mathrm{cell}
\end{equation}
though, again, the exact value of $u_{r, \text{stream}}$ does not make a difference as long it is small compared to the orbital velocity.
For this localized infall condition, the described boundary values are set inside the transport step to ensure the chosen mass flux.
Please note, that the Roche lobe overflow condition is only applied to the cells within the width of the stream.
The rest of the ghost cells are unaffected.
Thus, when using the Roche lobe overflow BC, additional individual BCs have to be specified for all variables.
A demonstration of the \emph{Roche Lobe overflow} BC is presented
in Fig.~\ref{fig:roche_lobe_overflow}. Note that is up to the user to ensure that the outer boundary is located at the Lagrangian $L_1$ point.
%
%%%%%%%%%%%%%%%%%%%%%%%%%%%%%%%%%%%%%%%%
\subsubsection{Damping zones}
%%%%%%%%%%%%%%%%%%%%%%%%%%%%%%%%%%%%%%%%
\label{sec:damping_zones}

\fargocpt supports damping zones (DZ) at the inner and outer boundary as presented in \citet{deval-borroComparativeStudyDiscplanet2006a}.
The DZs are used to damp waves that are reflected at the boundaries.
We choose to use the words \emph{damping zone} instead of \emph{damping boundary condition} because they are, strictly speaking, not a boundary condition in the context of partial differential equations.
They behave more akin to a heat bath in thermodynamics.

The DZs are implemented as an exponential relaxation of any of the quantities $X \in \{\Sigma, e, u_r, u_\phi\}$.
The damping is given by
\begin{equation}
	\label{eq:damping_derivative}
	\frac{\partial X}{\partial t} = - \frac{X - X_\mathrm{ref}(\vec{r})}{\tau_\mathrm{damp}} a(r)\,,
\end{equation}
with the damping timescale $\tau_\mathrm{damp}$ and the damping function $a(r)$.

The reference value $X_\mathrm{ref}$ is saved as a separate snapshot at the beginning of the simulation and by default contains the initial values.
It can also be manually changed by the user to support arbitrary damping fields in 2D.
Other choices for $X_\mathrm{ref}$ include 0,
the azimuthal average of the ring, or for the case of inner radial velocity, the viscous speed according to Eq.~\eqref{eq:viscous_inflow_speed}.
This relaxation step is applied at the end of each time step as the last update of the hydro system, see Fig.~\ref{fig:flow}.

The regions of the DZs are defined by two separate parameters $c_\mathrm{di}$ and $c_\mathrm{do}$ for the inner and outer zone, respectively.
The inner damping zone reaches from $R_\text{min}$ to $R_\text{di} = c_\text{di} R_\text{min}$ and the outer damping zone reaches from $R_\text{do} = c_\text{do} R_\text{max}$ to $R_\text{max}$.

The damping timescale $\tau_\mathrm{damp}$ is given by
\begin{align}\label{eq:damping_timescale}
\tau = \beta \Omega_\text{K}(R_\text{ref})\,,
\end{align}
where $\beta$ is a free parameter to configure the damping strength.
The reference radius $R_\text{ref}$ can be chosen by the user for each zone separately and defaults to inner and outer boundary radius.

The damping function $a(r)$ is used to smoothly reduce the damping from one at the boundary to zero towards the active computational domain.
We use a second-order polynomial with the condition that its first derivative vanishes at the transition from the damping zone to the active domain located at $R_\text{di/do}$.
This is fulfilled by the choice
\begin{align}\label{eq:damping_function}
	a(r) = \begin{cases}
		\left(\frac{R_\text{di} - r}{R_\text{di} - R_\text{min}}\right)^2 & \text{if } R_\text{min} \leq r \leq R_\text{di} \\
		\left(\frac{r - R_\text{do} }{R_\text{max} - R_\text{do}}\right)^2 & \text{if } R_\text{do} \leq r \leq R_\text{max} \\
		0 & \text{otherwise}
	\end{cases}\,.
\end{align}
The update step is discretized using the analytical solution of Eq.~\eqref{eq:damping_derivative} to avoid overshooting and is given by
\begin{align}
X^{t+\Delta t} = X_\text{ref} + (X^t - X_\text{ref}) \exp\left( - \frac{\Delta t a(r)}{\tau_\text{damp}} \right)
\end{align}

The DZ can be enabled for each quantity and in each zone individually.
Please note, that the DZ creates and removes mass, energy and momentum in the simulation domain and, thus, breaks conservation of these quantities.
The change of mass due to the DZ can be monitored.

%%%%%%%%%%%%%%%%%%%%%%%%%%%%%%%%%%%%%%%%
%%%%%%%%%%%%%%%%%%%%%%%%%%%%%%%%%%%%%%%%
%%%%%%%%%%%%%%%%%%%%%%%%%%%%%%%%%%%%%%%%
\section{Software features}
%%%%%%%%%%%%%%%%%%%%%%%%%%%%%%%%%%%%%%%%
%%%%%%%%%%%%%%%%%%%%%%%%%%%%%%%%%%%%%%%%
%%%%%%%%%%%%%%%%%%%%%%%%%%%%%%%%%%%%%%%%
\label{sec:software}

%%%%%%%%%%%%%%%%%%%%%%%%%%%%%%%%%%%%%%%%
\subsection{Hybrid parallelization}
%%%%%%%%%%%%%%%%%%%%%%%%%%%%%%%%%%%%%%%%

\fargocpt has been parallelized using a hybrid MPI + OpenMP approach.
This is currently the most efficient way to make use of the underlying CPU and memory structure of most supercomputers.
A scaling test can be found in Appendix~\ref{sec:app_scaling_test} which shows that the code reasonably scales at least up to 500 CPU cores. 

Modern computers often have multiple NUMA (non-uniform memory access) nodes per processor package.
As a result, a specific CPU core can access a part of the system memory with lower latency than the rest of the memory.
Because hydrodynamics simulations rely heavily on memory access, it is important to instruct the operating system to take this into account.

This is automatically attempted by default when using the launcher (see Sect.~\ref{sec:usability}), but can be done manually and the automatic mapping should be checked.
In practice, one MPI process is launched per NUMA node and OpenMP threads are launched as many as there are cores per NUMA node.
It is usually worthwhile to tune the execution settings (number of processes and threads) to the architecture of the computer used to run the simulation.
To find out the number of NUMA nodes and cores per NUMA node on a Linux system, one can use the \textbf{lstopo} or \textbf{lscpu} utilities.
For consumer PCs, one might need to enable a setting in the bios for the operating system to be aware of the NUMA topology.

%%%%%%%%%%%%%%%%%%%%%%%%%%%%%%%%%%%%%%%%
\subsection{Interactivity via signals}
%%%%%%%%%%%%%%%%%%%%%%%%%%%%%%%%%%%%%%%%

Planet-disk interaction simulations with any of the available codes typically run in a non-interactive mode, and only provide feedback through logs and monitor files once a predefined time step has been covered by the simulation.
For example, in the case that the CFL time step tends to zero, this can lead to simulations freezing without any indication of the cause.

To address issues such as frozen simulations, we have introduced UNIX signal handling with three signals.
One can interact with the simulation using the \textbf{kill} command, for example, using \textbf{kill -SIGUSR1 <fargo pid>}.

\texttt{SIGUSR1} allows users to request a status report, including current simulation time and details on CFL time step constraints, aiding in identifying freeze causes. \texttt{SIGUSR2} enables immediate stack trace printing, useful in development for locating algorithmic issues or runtime issues, for instance, when a cluster filesystem hangs. 
Finally, \texttt{SIGTERM} allows for a graceful shutdown, saving a snapshot for later resumption, beneficial in cluster environments or when simulations must be paused due to resource sharing.

%%%%%%%%%%%%%%%%%%%%%%%%%%%%%%%%%%%%%%%%
\subsection{Usability}
%%%%%%%%%%%%%%%%%%%%%%%%%%%%%%%%%%%%%%%%
\label{sec:usability}

We aimed to make using the code as straightforward as possible to encourage students to perform numerical experiments.
This includes changes to the command line interface as well as the restructuring of the config files discussed later in Sect.~\ref{sec:config_files}.

The command line interface traditionally included the selection of a mode, either to start or restart a simulation and the specification of the path to a config file.
A common task is to restart an existing simulation to extend it after its initially targeted time was reached or after the allocated wall time on a shared computing resource has elapsed.

Traditionally, this required specifying the output number of the snapshot that was last written to disk.
In our experience, this manual task can be error-prone, sometimes resulting in the loss of parts of the simulation data.
We added the \textit{automatic} mode which starts the simulation if no simulation output has been written yet and restarts the simulation from the latest snapshot present.
This has proven to essentially eliminate errors during routine restarting of simulations, saving duplicate spending of computational resources.

Furthermore, the compiled executable is launched through a Python launcher to handle the setup for parallel execution.
This launcher guesses an appropriate selection of the OpenMP thread and MPI process numbers under several different workload managers and MPI implementations.
Again, this is aimed to enable straightforward access to the code without specific knowledge about OpenMP and MPI, which can initially be a substantial hurdle.
Experienced users can skip this launcher, write their own or manually specify the parameters, to achieve a potentially more optimal configuration.

Installing all required libraries and compiling the code can be a substantial hurdle, especially for someone not familiar with Linux operating systems.
To this end, we made \fargocpt seamlessly available inside the virtual computers provided by GitHub called \emph{codespaces} \footnote{\url{https://github.com/features/codespaces}}.
Using this service, one can start a virtual machine with the code ready to go from within the \fargocpt GitHub repository \footnote{\url{https://github.com/rometsch/fargocpt}} with the click of a button.
Then one can get familiar with or use the code with the integrated Jupyter notebooks, all in the browser and without needing to use the command line or compile the code.
Although these codespaces have runtime restrictions, the freely available resources are enough, at the time of writing, for learning to use the code, prototyping setups and even low-resolution scientific simulations.
When this free service should no longer be available, the \fargocpt code can still be used in a similar fashion using the \texttt{docker} container provided in the GitHub repository.

%%%%%%%%%%%%%%%%%%%%%%%%%%%%%%%%%%%%%%%%
\subsection{Python interface}
%%%%%%%%%%%%%%%%%%%%%%%%%%%%%%%%%%%%%%%%
\label{sec:python_interface}

We created a Python module that comes with the code to facilitate starting the code from within a Jupyter Notebook or Python scripts, loading data from output directories and creating an interactive overview plot.
It comes with a command line interface that can be used to start simulations with an automatically determined suitable CPU allocation and to inspect output data.

The GitHub repository of the code includes multiple Jupyter notebooks with example cases that illustrate how to build and use the code.
Additionally, there are examples of how to load and visualize the output data and some common simulation scenarios.

%%%%%%%%%%%%%%%%%%%%%%%%%%%%%%%%%%%%%%%%
\subsection{C++ port and code restructuring}
%%%%%%%%%%%%%%%%%%%%%%%%%%%%%%%%%%%%%%%%

FargoADSG \citep{baruteauPredictiveScenariosPlanetary2008a} was converted from a C code to a C++ object-oriented by \citet{mullerTreatingGravityThindisk2012}.
Notably, the data structure was converted to be object-oriented, a data class holding a number of other structures for the storage of physical quantities defined on a polar grid.
Functionalities of these classes include the tracking of units with pre- and post-hooks for input and output.

Later, functionally separate parts of the code were split up (into C++ namespaces).
Examples include software aspects such as the parsing and storage of parameters, units and output, and physics modules such as the N-body system, boundary conditions or radiative transport.
This was done to highlight the structure of the code to make it easier to maintain and extend, and to make it easier to understand for new users.

%%%%%%%%%%%%%%%%%%%%%%%%%%%%%%%%%%%%%%%%
\subsection{Config files}
%%%%%%%%%%%%%%%%%%%%%%%%%%%%%%%%%%%%%%%%
\label{sec:config_files}

Config files have been changed from a custom flavor of the \texttt{ini} file format to the well-defined and documented \texttt{yaml} format.
The advantage of this is twofold.
First, the config files can be easily processed and generated using scripts, for example, with any of the \texttt{yaml} packages in the Python ecosystem.
This can help avoid errors in preparing simulations for parameter studies.
Along the same lines, the parsing of the config files in the actual C++ simulation code is off-loaded to an existing and tested library which can handle and report syntax errors in the config files and handle type conversions.
The latter two are examples in which the authors experienced time-consuming errors with other codes in the past.

Second, \texttt{yaml} supports structured data which enables setting the parameters and initial conditions of the N-body objects within the config file.
This change removes the column-based and unchecked planet file in favor of a structured entry in the main configuration file.

An aspect that sets \fargocpt apart from other versions of the code is that the config file is the only place that is changed to configure the simulations.
No compile time parameters are used.
In our opinion, this makes the code easier to use for first-time users.

%%%%%%%%%%%%%%%%%%%%%%%%%%%%%%%%%%%%%%%%
\subsection{Output format}
%%%%%%%%%%%%%%%%%%%%%%%%%%%%%%%%%%%%%%%%
\label{sec:output_format}

Originally, all simulation output files were stored in one single output folder for all snapshots.
Depending on the simulation, several thousand snapshots can be written to disk resulting in several thousand to several tens of thousands of files in a single folder.
This makes working with these simulations cumbersome because simply listing the contents of this output directory can take multiple tens of seconds on cluster file systems which are optimized for parallel throughput rather than metadata access.

Furthermore, extracting a single snapshot from a simulation directory included the error-prone manual extraction of various state variables from different text files.

In \fargo-like codes, time series data is usually written to tab-separated-value text files.
However, the contents of the columns are often not described, neither the content of a column nor its unit.
This lack of description of the data is made worse by the fact that the structure of these files can change between different versions of the same code or be modified by the user to track any quantity of interest that is not already present in the standard output leading to confusion and necessitating the knowledge of which particular version of the code was used to generate a particular set out output data.
While we generally encourage saving which exact version of the code was used (e.g., the git commit id), even storing the source code of the simulation with the output data, reading the values from a text file should not necessitate the study of the accompanying source code.

In addition, the input files for the simulation were required to infer the unit system used in the simulation because the base units of length, mass and time can be set in the configuration file.
Hence, the outputs were not self-descriptive.

We remedied these issues with the following changes.
The output directory was restructured such that each snapshot was saved into a separate directory within the \texttt{snapshot} sub-directory.
Such a directory contains all binary data about density, energy and velocity fields, all scalar quantities such as the time, the rotation angle of the frame and intermediate integration variables and all information about the state of the N-body and particle systems.
Additionally, a copy of the input file is saved for each snapshot to save the history of parameters should they change during a restart.

Additionally, yaml info files are written to the output directory containing information about the units used in the simulation and the quantities that were written as parts of the snapshots to unambiguously specify the physical quantities.
This grounds these code units on a physical scale.
Rescaling of the results can still be done should the set of physical assumptions allow it.

The files containing time series data are now collected in one single sub-directory called \texttt{monitor}.
Each such text file includes a header that describes the content of each column in the text file and specifies its unit in a string readable by the \texttt{astropy} Python library in an automated fashion.

Furthermore, the output directory now contains an empty text file with the name \texttt{fargocpt\_output\_vx\_y}, where \texttt{x\_y} specifies the major and minor version of the code.
This helps with the automation of postprocessing when multiple versions of the same code or even different codes are used in a single project.

%%%%%%%%%%%%%%%%%%%%%%%%%%%%%%%%%%%%%%%%
\subsection{Unit system}
%%%%%%%%%%%%%%%%%%%%%%%%%%%%%%%%%%%%%%%%

Because we carry out numerical simulations of physical processes, we necessarily need to choose a system of units.
This unit system, often called \emph{code units}, can be specified using two or four parameters in the config file.
The base length $L_0$ and the base mass $M_0$ are required and the base time $T_0$ and the base temperature $\Theta_0$ are optional.
If the base time $T_0$ is not chosen explicitly, it is computed such that $\mathrm{G}=1$ in code units.
This is equivalent to one Keplerian orbit at distance $L_0$ around an object of mass $M_0$ having a period of $2\pi$, thus,
\begin{align}\label{eqn:definition_base_time}
	T_0 = \sqrt{ \frac{L_0^3}{ \mathrm{G} M_0} } = \frac{1}{\Omega_K\big|_{M_0, L_0}}\,
\end{align}
with the gravitational constant $\mathrm{G}$ and the Keplerian angular velocity $\Omega_K = \sqrt{\mathrm{G}M/R^3}$.
If the temperature unit is not specified, it is calculated such that the specific gas constant $R = \frac{\text{k}_\text{B}}{\mu}$ is unity in code units, thus
\begin{align}
	\Theta_0 = \frac{\text{G}\,\mu\,M_0}{\text{k}_\text{B}\,\,L_0}\,,
\end{align}
with the mean molecular weight $\mu$ and the Boltzmann constant $\text{k}_\text{B}$.
Internally, all calculations are carried out and all output is written in the code unit system $\mathcal{U} = (L_0, M_0, T_0, \Theta_0)$.

In the config file, the user can specify physical parameters either as a number without units, in which case they are interpreted to be in code units, or they can be specified with a number and a unit symbol. This number and unit are then automatically converted to code units.
As an example, consider a unit system with $L_0 = 1\,\mathrm{au}$ and $M_0 = 1\,\mathrm{M}_\odot$. 
The reference surface density, $\Sigma_0$, which has the key \verb|Sigma0| in the config file can then either be specified in code units using
\begin{lstlisting}[language=Python]
Sigma0: 1e-5
\end{lstlisting}
in which case, $\Sigma_0 = 10^{-5}\,\frac{M_0}{L_0^2} = 10^{-5} \frac{\mathrm{M}_\odot}{\mathrm{au}^2} \approx 88.85\,\frac{\mathrm{g}}{\mathrm{cm}^2}$.
Alternatively, the same could be specified using:
\begin{lstlisting}[language=Python]
Sigma0: 88.85 g/cm2
\end{lstlisting}
The former version is more informative in simulations of a scale-free problem while the latter version is more informative in simulations aimed at simulating existing protoplanetary disk systems.
This flexibility in terms of units combines the usefulness of a unit system adapted to the physical problem at hand (code units) with the necessity to specify certain physical parameters in physical units, such as parameters inferred from observations or experiments.
We hope that this feature helps to avoid common conversion errors in setting up simulations.

The implementation of the unit parsing is based on the \codestyle{C++} units runtime library \citep{LLNLunits}. All conversions are performed in the initialization step, so variables during the actual simulations do not carry any units.

Unit symbols supported in the config file are specified by the \citet{LLNLunits} library\footnote{See \url{https://units.readthedocs.io/en/latest/user-guide/from_string.html} for more details.}, they include all SI units with prefixes, and the convenience units \verb|au|, \verb|solMass|, \verb|solRadius|, \verb|jupiterMass|, \verb|jupiterRadius|, \verb|earthMass|, \verb|earthRadius|.
Usually, most unit strings that work with the Python \texttt{astropy} package also work here.
Furthermore, combinations of powers of units are supported, as in the example above.

In addition to the definition of units, \citep{LLNLunits} is also the source for the definition of physical constant\footnote{\url{https://units.readthedocs.io/en/latest/user-guide/Physical_constants.html}}, namely the gravitational constant $G$, the Boltzmann constant $k_\mathrm{B}$, the atomic mass units $\mathrm{u}$, the Planck constant $\mathrm{h}$, and the speed of light $\mathrm{c}$. These are based on the 2019 redefinition of the SI units\footnote{\url{https://www.nist.gov/si-redefinition/meet-constants}} and the NIST 2018 CODATA\footnote{\url{https://physics.nist.gov/cuu/Constants/Table/allascii.txt}} physical constants table.

%%%%%%%%%%%%%%%%%%%%%%%%%%%%%%%%%%%%%%%%
\subsection{Test suite}
%%%%%%%%%%%%%%%%%%%%%%%%%%%%%%%%%%%%%%%%
\label{sec:testsuite}

Having a test suite is crucial to illustrate that the code is working as intended and that the various physical modules of the code actually provide approximations to the underlying equations.
This is an essential part of any simulation code.

\fargocpt comes with an automatic test suite.
The test suite can be run by executing the \texttt{run\_tests.sh} script within the \texttt{tests} directory.
This automatically executes multiple test cases and compares the results to reference data or theoretical expectations.
The result of the test is then either \texttt{passed} or \texttt{failed}.
Each test case therefore includes threshold values that define the maximum allowed deviation from the reference data.
The test suite is designed to be run on a local computer in a matter of minutes and does not require a supercomputer.
This makes testing the code base relatively easy and cheap.

Currently, the automatic tests include tests for most of the major physical modules.
The tests are
\begin{itemize}
	\item Steady state accretion disk (gas mass flow) (App.~\ref{app:steady_state_accretion_test})
	\item Shocktube (ideal and caloric equation of state) (App.~\ref{sec:shock_tube})
	\item Viscous spreading ring (App.~\ref{sec:spreading_ring})
	\item Viscous heating-cooling equilibrium temperature (App.~\ref{sec:heating_cooling_test})
	\item Irradiation-cooling equilibrium temperature
	\item Cold disk with the ideal equation of state with and without planet
	\item N-body integration: Kepler orbits
	\item Dust diffusion (App.~\ref{app:dust_diffusion_test})
	\item Dust drift (App.~\ref{app:dust_drift_test})
	\item Type I migration planet torque (App.~\ref{app:planet_torque_test})
	\item Flux-Limited Diffusion 1D (App.~\ref{appendix:FLD_test})
	\item Flux-Limited Diffusion 2D - direct test of the diffusion equation solver (App.~\ref{appendix:sor_solver_test})
	\item Self-gravity solver (App.~\ref{appendix:sg_solver_test})
	\item Planet orbiting a disk
\end{itemize}

See the referenced appendices for more details.

%%%%%%%%%%%%%%%%%%%%%%%%%%%%%%%%%%%%%%%%
%%%%%%%%%%%%%%%%%%%%%%%%%%%%%%%%%%%%%%%%
%%%%%%%%%%%%%%%%%%%%%%%%%%%%%%%%%%%%%%%%
\section{Discussion}
%%%%%%%%%%%%%%%%%%%%%%%%%%%%%%%%%%%%%%%%
%%%%%%%%%%%%%%%%%%%%%%%%%%%%%%%%%%%%%%%%
%%%%%%%%%%%%%%%%%%%%%%%%%%%%%%%%%%%%%%%%
\label{sec:discussion}

%%%%%%%%%%%%%%%%%%%%%%%%%%%%%%%%%%%%%%%%
\subsection{Leap-Frog like scheme}
%%%%%%%%%%%%%%%%%%%%%%%%%%%%%%%%%%%%%%%%

In addition to the integration schedule presented in Fig.~\ref{fig:flow}, we also implemented a Leap-Frog-like schedule presented in Appendix~\ref{appendix:leap_frog_scheme}.
This scheme performs the source term step twice with half of the time step size and the transport step once with the full time step size.
The transport step is performed in between the two source term steps.

This allows for larger simulation time steps by relaxing the CFL criteria of
the source step but it also becomes more expensive because the source term is evaluated twice.
This is prohibitive when self-gravity or radiative transport is enabled.
Without these enabled, the scheme still runs around $7\%$ slower than the Euler scheme.

While in theory, a Leap-Frog scheme has higher accuracy than an Euler scheme,
and it is advantageous to conduct the transport step less often and
with larger time steps due to numerical diffusion, we found that the benefits are negligible for typical planet disk interaction simulations.
We found the Leap-Frog scheme only to be beneficial for simulations where the source terms dominate the numerical errors and not the transport step, for instance, in test simulations of an equilibrium disk.
An example of such a simulation is the heating and cooling test presented in Appendix~\ref{sec:heating_cooling_test}.
In this model, the transport step is effectively only an advection along the azimuthal direction, which is trivial in the polar coordinate system.

Because of these reasons, we see no benefit in using the Leap-Frog scheme as implemented in the current version of the code over the default scheme for typical planet-disk interaction simulations.

The operator splitting scheme we use is formally only accurate up to first-order in time (worst case).
Therefore, we suspect that more substantial improvements to accuracy can be made by implementing higher-order time stepping such as a second-order Runge-Kutta scheme.
This is left for future work.

%%%%%%%%%%%%%%%%%%%%%%%%%%%%%%%%%%%%%%%%
\subsection{Why do we need another FARGO code?}
%%%%%%%%%%%%%%%%%%%%%%%%%%%%%%%%%%%%%%%%

\fargocpt is another addition to the family of \fargo codes.
This raises the question of why the community needs another \fargo code.
Or put differently, why do we not just use \fargothreed or \fargoca codes which even support 3D simulations?

Our answer to this question is twofold.
First, to the knowledge of the authors, the collection of physics modules implemented in \fargocpt is unique compared to other publicly available versions of codes for the study of planet-disk interaction.
At present, the \fargocpt code includes different equations of state, viscous heating and irradiation, local $\beta$-cooling, cooling through the disk surfaces, midplane energy transport, self-gravity, high-order N-body integration, accretion onto N-body objects, and a particle module which includes gas drag laws for a wide range of particle sizes and dust diffusion, all while making use of the \fargo speedup.
For example, the public version of \fargothreed does not support self-gravity, while \fargoadsg and Athena++ do not support radiation transport.
Both are processes that are important in current problems of planet-disk interaction \citep[e.g.][]{ziamprasModellingPlanetinducedGaps2023} and protoplanetary disks \citep[e.g.][]{rendonrestrepoMorphologyDynamicalStability2022}.
\fargocpt also includes many of the typically used effective models for planet-disk interaction, such as $\beta$-cooling.
While these effective models can be implemented in other codes with relative ease by an experienced programmer, these implementations still have to be manually validated by the user.
Hence, having a tested implementation is nearly always preferable \citep{wilson_best_2012}, even if they are only taken as a starting point and modified for a specific problem.

Second, we believe that the code presented here is easier to use, understand and modify than other versions of \fargo.
This is especially important for students who are new to the field and want to perform numerical experiments.
We anticipate that this will lead to a more efficient learning process and a more efficient use of the time of the students.
Additionally, features such as the support for physical units in the config files and a self-contained output reduce the chance of human error in the simulation workflow.
This leads to a more efficient use of the attention of the user to the physical model behind the simulation and the scientific problem at hand.

Furthermore, we are not aware of any other hydrodynamics code for the study of planet-disk interactions that can be used in the browser.
This property can make \fargocpt a valuable tool for teaching and learning about planet-disk interactions.
Indeed, it was already successfully used to teach hydrodynamics and planet migration at the SPP 1992 summer school on planet formation in Rauenberg, Germany, in August of 2023.

%%%%%%%%%%%%%%%%%%%%%%%%%%%%%%%%%%%%%%%%
\subsection{Future development}
%%%%%%%%%%%%%%%%%%%%%%%%%%%%%%%%%%%%%%%%
\label{sec:future_development}

As with any software project, many possible improvements can be made to the code base.
Here, we outline some of the potential improvements that we believe might be worthwhile to be incorporated into future iterations of the code:

\noindent\textbf{More general self-gravity solver:}
One of the foundational assumptions of our current self-gravity module is that the aspect ratio needs to be assumed constant for the Fourier Method to work.
This limits the accuracy of simulations with the combination of self-gravity and radiation physics, the latter of which generally leads to a non-uniform aspect ratio.
Removing this limitation would, for example, allow for more accurate studies of gravitational collapse within the disk, which includes a balance between pressure, built up by compression heating and reduced by radiation transport, and self-gravity.
This could be achieved using tree-based or multigrid methods.

\noindent\textbf{Higher-order time-stepping:}
By integrating higher-order time-stepping techniques, we can potentially achieve better temporal resolution and improved simulation stability, thus ensuring more accurate representations of physical systems over time.
The groundwork for such a change has already been laid in the Leap-Frog-like scheme presented in Appendix~\ref{appendix:leap_frog_scheme} and a second-order Runge-Kutta scheme could be implemented in a similar fashion.

\noindent\textbf{Matrix solvers for heating, cooling and viscosity:}
Currently, the heating and cooling terms, as well as the viscosity update rely on either a simple implicit but local update step and the viscosity update uses a simple explicit update step (see Sect.~\ref{fig:viscosity_stabilizer} for resulting issues).
These updates can be replaced by fully implicit updates which rely on a matrix or linear system solver to increase accuracy and stability (see also Sect.~\ref{sec:viscosity_stabilizer}).
To this end, the SOR linear system solver used in the flux-limited diffusion model could be adapted to the heating, cooling and viscosity update steps.

\noindent\textbf{Irradiation Using Ray-tracing}
For studying scenarios like accretion onto planets, where the impact of radiation sources can dominate the local evolution, introducing a ray-tracing mechanism for irradiation can be beneficial. 
Currently, the irradiation is computed using a simple distance-based approximation, formally only valid for a single star in the center of a flaring disk.
While being computationally expensive, ray tracing can provide a more precise depiction of the dynamics of the disk when it is heated by accreting planets including influences of shadows.

%%%%%%%%%%%%%%%%%%%%%%%%%%%%%%%%%%%%%%%%
\begin{acknowledgements}
	TR and LJ would like to express their gratitude to Alex Ziampras for the numerous insightful and productive discussions.
	TR, GP, WK and CD acknowledge funding from the Deutsche Forschungsgemeinschaft (DFG) research group FOR 2634 ''Planet Formation Witnesses and Probes: Transition Disks''
	under grant DU 414/22-1, and KL 650/29-1, 650/29-2, 650/30-1, and 650/30-2.
	SRR acknowledges funding from the European Union (ERC, Epoch-of-Taurus, 101043302). Views and opinions expressed are however those of the authors only and do not necessarily reflect those of the European Union or the European Research Council. Neither the European Union nor the granting authority can be held responsible for them.
	The authors acknowledge support by the High Performance and Cloud Computing Group at the Zentrum f\"ur Datenverarbeitung of the University of T\"ubingen, the state of Baden-W\"urttemberg through bwHPC and the German Research Foundation (DFG) through grant INST\,37/935-z1\,FUGG.
	Plots in this paper were made using the Python library \texttt{matplotlib} \citep{hunter_matplotlib_2007}.
\end{acknowledgements}

\bibliographystyle{aa}
\bibliography{paper}

\begin{thebibliography}{109}
\expandafter\ifx\csname natexlab\endcsname\relax\def\natexlab#1{#1}\fi

\bibitem[{Anderson {et~al.}(2020)Anderson, Tannehill, \&
  Pletcher}]{anderson_computational_2020}
Anderson, D.~A., Tannehill, J.~C., \& Pletcher, R.~H. 2020, Computational fluid
  mechanics and heat transfer, fourth edition edn., Computational and physical
  processes in mechanics and thermal sciences (Boca Raton, FL: CRC Press)

\bibitem[{{Artymowicz} \& {Lubow}(1994)}]{artymowicz1994dynamics}
{Artymowicz}, P. \& {Lubow}, S.~H. 1994, \apj, 421, 651

\bibitem[{Baruteau(2008)}]{baruteauPredictiveScenariosPlanetary2008a}
Baruteau, C. 2008, PhD thesis, Observatoire de Paris

\bibitem[{Baruteau {et~al.}(2014)Baruteau, Crida, Paardekooper, Masset, Guilet,
  Bitsch, Nelson, Kley, \& Papaloizou}]{baruteau_planet-disk_2014}
Baruteau, C., Crida, A., Paardekooper, S.~J., {et~al.} 2014, Planet-{Disk}
  {Interactions} and {Early} {Evolution} of {Planetary} {Systems} (eprint:
  arXiv:1312.4293: University of Arizona Press)

\bibitem[{Baruteau \&
  Masset(2008{\natexlab{a}})}]{baruteauCorotationTorqueRadiatively2008a}
Baruteau, C. \& Masset, F. 2008{\natexlab{a}}, \apj, 672, 1054

\bibitem[{Baruteau \&
  Masset(2008{\natexlab{b}})}]{baruteauTypePlanetaryMigration2008}
Baruteau, C. \& Masset, F. 2008{\natexlab{b}}, \apj, 678, 483

\bibitem[{Baruteau \& Zhu(2016)}]{baruteauGasDustHydrodynamical2016a}
Baruteau, C. \& Zhu, Z. 2016, \mnras, 458, 3927

\bibitem[{Bell \& Lin(1994)}]{bellUsingFUOrionis1994}
Bell, K.~R. \& Lin, D. N.~C. 1994, \apj, 427, 987

\bibitem[{Benítez-Llambay \&
  Masset(2016)}]{benitez-llambayFARGO3DNewGPUoriented2016}
Benítez-Llambay, P. \& Masset, F. 2016, \apjs, 223, 11

\bibitem[{Bertin \& Lodato(1999)}]{bertin_class_1999}
Bertin, G. \& Lodato, G. 1999, \aap, 350, 694

\bibitem[{Binney \& Tremaine(1987)}]{binneyGalacticDynamics1987}
Binney, J. \& Tremaine, S. 1987, Galactic dynamics (Princeton University Press)

\bibitem[{Bodenheimer {et~al.}(2006)Bodenheimer, Laughlin, Rozyczka, Plewa,
  Yorke, \& Yorke}]{bodenheimer2006numerical}
Bodenheimer, P., Laughlin, G., Rozyczka, M., {et~al.} 2006, Numerical Methods
  in Astrophysics: An Introduction, Series in Astronomy and Astrophysics
  (Taylor \& Francis)

\bibitem[{Cash \& Karp(1990)}]{cash_variable_1990}
Cash, J.~R. \& Karp, A.~H. 1990, ACM Transactions on Mathematical Software, 16,
  201

\bibitem[{Charnoz {et~al.}(2011)Charnoz, Fouchet, Aleon, \&
  Moreira}]{charnoz_three-dimensional_2011}
Charnoz, S., Fouchet, L., Aleon, J., \& Moreira, M. 2011, \apj, 737, 33

\bibitem[{{Chiang} \& {Goldreich}(1997)}]{chiang1997spectral_energy}
{Chiang}, E.~I. \& {Goldreich}, P. 1997, \apj, 490, 368

\bibitem[{Chrenko {et~al.}(2017)Chrenko, Brož, \&
  Lambrechts}]{chrenkoEccentricityExcitationMerging2017}
Chrenko, O., Brož, M., \& Lambrechts, M. 2017, \aap, 606, A114

\bibitem[{Commerçon {et~al.}(2011)Commerçon, Teyssier, Audit, Hennebelle, \&
  Chabrier}]{commerconRadiationHydrodynamicsAdaptive2011}
Commerçon, B., Teyssier, R., Audit, E., Hennebelle, P., \& Chabrier, G. 2011,
  \aap, 529, A35

\bibitem[{{Cox} \& {Stewart}(1969)}]{cox1969opacities}
{Cox}, A.~N. \& {Stewart}, J.~N. 1969, Nauchnye Informatsii, 15, 1

\bibitem[{Crida {et~al.}(2007)Crida, Morbidelli, \&
  Masset}]{cridaSimulatingPlanetMigration2007b}
Crida, A., Morbidelli, A., \& Masset, F. 2007, \aap, 461, 1173

\bibitem[{{D'Angelo} \& {Bodenheimer}(2013)}]{dangelo2013}
{D'Angelo}, G. \& {Bodenheimer}, P. 2013, \apj, 778, 77

\bibitem[{{D'Angelo} {et~al.}(2002){D'Angelo}, {Henning}, \&
  {Kley}}]{dangelo2002nested_grid}
{D'Angelo}, G., {Henning}, T., \& {Kley}, W. 2002, \aap, 385, 647

\bibitem[{D'Angelo {et~al.}(2003)D'Angelo, Henning, \&
  Kley}]{dangelo_thermohydrodynamics_2003}
D'Angelo, G., Henning, T., \& Kley, W. 2003, \apj, 599, 548

\bibitem[{{D'Angelo} \& {Marzari}(2012)}]{dangelo2012outward_migration}
{D'Angelo}, G. \& {Marzari}, F. 2012, \apj, 757, 50

\bibitem[{de~Val-Borro {et~al.}(2006)de~Val-Borro, Edgar, Artymowicz,
  Ciecielag, Cresswell, D'Angelo, Delgado-Donate, Dirksen, Fromang,
  Gawryszczak, Klahr, Kley, Lyra, Masset, Mellema, Nelson, Paardekooper,
  Peplinski, Pierens, Plewa, Rice, Schäfer, \&
  Speith}]{deval-borroComparativeStudyDiscplanet2006a}
de~Val-Borro, M., Edgar, R.~G., Artymowicz, P., {et~al.} 2006, \mnras, 370, 529

\bibitem[{Frigo \& Johnson(2005)}]{FFTW05}
Frigo, M. \& Johnson, S.~G. 2005, Proceedings of the IEEE, 93, 216

\bibitem[{Gammie(2001)}]{gammieNonlinearOutcomeGravitational2001}
Gammie, C.~F. 2001, \apj, 553, 174

\bibitem[{Geiser {et~al.}(2017)Geiser, Hueso, \& Martinez}]{geiser2017new}
Geiser, J., Hueso, J.~L., \& Martinez, E. 2017, Journal of Computational and
  Applied Mathematics, 309, 359

\bibitem[{{G{\"u}nther} {et~al.}(2004){G{\"u}nther}, {Sch{\"a}fer}, \&
  {Kley}}]{gunther2004}
{G{\"u}nther}, R., {Sch{\"a}fer}, C., \& {Kley}, W. 2004, \aap, 423, 559

\bibitem[{Haghighipour \& Boss(2003)}]{haghighipourPressureGradientsRapid2003}
Haghighipour, N. \& Boss, A.~P. 2003, \apj, 583, 996

\bibitem[{{Hawley} {et~al.}(1984){Hawley}, {Smarr}, \&
  {Wilson}}]{hawley1984sod}
{Hawley}, J.~F., {Smarr}, L.~L., \& {Wilson}, J.~R. 1984, \apj, 277, 296

\bibitem[{Hayashi(1981)}]{hayashiStructureSolarNebula1981}
Hayashi, C. 1981, Progress of Theoretical Physics Supplement, 70, 35

\bibitem[{Hubeny(1990)}]{hubenyVerticalStructureAccretion1990}
Hubeny, I. 1990, \apj, 351, 632

\bibitem[{Hunter(2007)}]{hunter_matplotlib_2007}
Hunter, J.~D. 2007, Computing In Science \& Engineering, 9, 90

\bibitem[{{Ichikawa} \& {Osaki}(1992)}]{ichikawa1992dwarfnova}
{Ichikawa}, S. \& {Osaki}, Y. 1992, \pasj, 44, 15

\bibitem[{Jordan {et~al.}(2021)Jordan, Kley, Picogna, \&
  Marzari}]{jordanDisksCloseBinary2021}
Jordan, L.~M., Kley, W., Picogna, G., \& Marzari, F. 2021, \aap, 654, A54

\bibitem[{Joseph {et~al.}(2023)Joseph, Ziampras, Jordan, Turpin, \&
  Nelson}]{joseph_measuring_2023}
Joseph, J., Ziampras, A., Jordan, L., Turpin, G.~A., \& Nelson, R.~P. 2023,
  \aap, 678, A134

\bibitem[{{Kimura} {et~al.}(2020){Kimura}, {Osaki}, {Kato}, \&
  {Mineshige}}]{kimura2020tilt}
{Kimura}, M., {Osaki}, Y., {Kato}, T., \& {Mineshige}, S. 2020, \pasj, 72, 22

\bibitem[{{Klahr} \& {Kley}(2006)}]{klahr2006}
{Klahr}, H. \& {Kley}, W. 2006, \aap, 445, 747

\bibitem[{Kley(1989)}]{kleyRadiationHydrodynamicsBoundary1989}
Kley, W. 1989, \aap, 208, 98

\bibitem[{{Kley}(1998)}]{kley1998coriolis}
{Kley}, W. 1998, \aap, 338, L37

\bibitem[{{Kley}(1999)}]{kley1999mass_flow}
{Kley}, W. 1999, \mnras, 303, 696

\bibitem[{Kley \& Crida(2008)}]{kley_migration_2008}
Kley, W. \& Crida, A. 2008, \aap, 487, L9

\bibitem[{Kley \& Nelson(2012)}]{kley_planet-disk_2012}
Kley, W. \& Nelson, R.~P. 2012, \araa, 50, 211

\bibitem[{{Kley} {et~al.}(2008){Kley}, {Papaloizou}, \&
  {Ogilvie}}]{kley2008binary}
{Kley}, W., {Papaloizou}, J.~C.~B., \& {Ogilvie}, G.~I. 2008, \aap, 487, 671

\bibitem[{Kolb {et~al.}(2013)Kolb, Stute, Kley, \&
  Mignone}]{kolbRadiationHydrodynamicsIntegrated2013b}
Kolb, S.~M., Stute, M., Kley, W., \& Mignone, A. 2013, \aap, 559, A80

\bibitem[{Lega {et~al.}(2014)Lega, Crida, Bitsch, \&
  Morbidelli}]{lega_migration_2014}
Lega, E., Crida, A., Bitsch, B., \& Morbidelli, A. 2014, \mnras, 440, 683

\bibitem[{Levermore(1984)}]{levermoreRelatingEddingtonFactors1984}
Levermore, C.~D. 1984, \jqsrt, 31, 149

\bibitem[{Levermore \&
  Pomraning(1981)}]{levermoreFluxlimitedDiffusionTheory1981}
Levermore, C.~D. \& Pomraning, G.~C. 1981, \apj, 248, 321

\bibitem[{Lin \& Papaloizou(1985)}]{linDynamicalOriginSolar1985}
Lin, D. N.~C. \& Papaloizou, J. 1985, On the dynamical origin of the solar
  system. (University of Arizona Press)

\bibitem[{LLNL(2022)}]{LLNLunits}
LLNL. 2022, Units C++ runtime library, \url{https://github.com/LLNL}

\bibitem[{{Lodato}(2008)}]{lodato2008}
{Lodato}, G. 2008, \nar, 52, 21

\bibitem[{Lust(1952)}]{Lust1952entwicklung}
Lust, R. 1952, Zeitschrift für Naturforschung A, 7, 87

\bibitem[{Lynden-Bell \& Pringle(1974)}]{lynden-bellEvolutionViscousDiscs1974}
Lynden-Bell, D. \& Pringle, J.~E. 1974, \mnras, 168, 603

\bibitem[{Masset(2000)}]{masset_fargo_2000}
Masset, F. 2000, \aaps, 141, 165

\bibitem[{Masset(2002)}]{massetCoorbitalCorotationTorque2002}
Masset, F. 2002, \aap, 387, 605

\bibitem[{Masset(2015)}]{massetGfargoFargoGpu2015}
Masset, F. 2015, Astrophysics Source Code Library, ascl:1509.008

\bibitem[{Masset(2017)}]{masset_coorbital_2017}
Masset, F. 2017, \mnras, 472, 4204

\bibitem[{{Menou} \& {Goodman}(2004)}]{menou2004low_mass}
{Menou}, K. \& {Goodman}, J. 2004, \apj, 606, 520

\bibitem[{{Meyer} \& {Meyer-Hofmeister}(1983)}]{meyer1983}
{Meyer}, F. \& {Meyer-Hofmeister}, E. 1983, \aap, 121, 29

\bibitem[{{Mignone} {et~al.}(2007){Mignone}, {Bodo}, {Massaglia}, {Matsakos},
  {Tesileanu}, {Zanni}, \& {Ferrari}}]{mignone2007pluto}
{Mignone}, A., {Bodo}, G., {Massaglia}, S., {et~al.} 2007, \apjs, 170, 228

\bibitem[{Mignone {et~al.}(2019)Mignone, Flock, \&
  Vaidya}]{mignone_particle_2019}
Mignone, A., Flock, M., \& Vaidya, B. 2019, \apjs, 244, 38

\bibitem[{Mihalas \&
  Mihalas(1984)}]{mihalasFoundationsRadiationHydrodynamics1984}
Mihalas, D. \& Mihalas, B.~W. 1984, Foundations of radiation hydrodynamics
  (Oxford University Press)

\bibitem[{Mineshige \&
  Osaki(1983)}]{mineshigeDiskinstabilityModelOutbursts1983}
Mineshige, S. \& Osaki, Y. 1983, Publications of the Astronomical Society of
  Japan, 35, 377

\bibitem[{Morohoshi \& Tanaka(2003)}]{morohoshi_gravitational_2003}
Morohoshi, K. \& Tanaka, H. 2003, \mnras, 346, 915

\bibitem[{{Mu{\~n}oz} {et~al.}(2019){Mu{\~n}oz}, {Miranda}, \&
  {Lai}}]{munoz2019quadropole}
{Mu{\~n}oz}, D.~J., {Miranda}, R., \& {Lai}, D. 2019, \apj, 871, 84

\bibitem[{Müller(2013)}]{mullerPlanetsBinaryStar2013}
Müller, T. W.~A. 2013, Dissertation, Universität Tübingen

\bibitem[{Müller \& Kley(2012)}]{mullerCircumstellarDisksBinary2012}
Müller, T. W.~A. \& Kley, W. 2012, \aap, 539, A18

\bibitem[{Müller \& Kley(2013)}]{mullerModellingAccretionTransitional2013}
Müller, T. W.~A. \& Kley, W. 2013, \aap, 560, A40

\bibitem[{Müller {et~al.}(2012)Müller, Kley, \&
  Meru}]{mullerTreatingGravityThindisk2012}
Müller, T. W.~A., Kley, W., \& Meru, F. 2012, \aap, 541, A123

\bibitem[{Nakagawa {et~al.}(1986)Nakagawa, Sekiya, \&
  Hayashi}]{nakagawa_settling_1986}
Nakagawa, Y., Sekiya, M., \& Hayashi, C. 1986, Icarus, 67, 375

\bibitem[{O'Neill(2018)}]{neill2018jsf}
O'Neill, M. 2018, Bob Jenkins's Small PRNG Passes PractRand (And More!),
  \url{https://www.pcg-random.org/posts/bob-jenkins-small-prng-passes-practrand.html}

\bibitem[{Paardekooper {et~al.}(2023)Paardekooper, Dong, Duffell, Fung, Masset,
  Ogilvie, \& Tanaka}]{paardekooper_planet-disk_2023}
Paardekooper, S., Dong, R., Duffell, P., {et~al.} 2023, Astronomical Society of
  the Pacific, 534, 685

\bibitem[{Paardekooper {et~al.}(2011)Paardekooper, Baruteau, \&
  Kley}]{paardekooper_torque_2011}
Paardekooper, S.-J., Baruteau, C., \& Kley, W. 2011, \mnras, 410, 293

\bibitem[{Paardekooper \& Mellema(2006)}]{paardekooper_halting_2006}
Paardekooper, S.~J. \& Mellema, G. 2006, \aap, 459, L17

\bibitem[{Paardekooper \& Papaloizou(2008)}]{paardekooper_disc_2008}
Paardekooper, S.~J. \& Papaloizou, J. C.~B. 2008, \aap, 485, 877

\bibitem[{Picogna \& Kley(2015)}]{picogna_how_2015}
Picogna, G. \& Kley, W. 2015, \aap, 584, A110

\bibitem[{Picogna {et~al.}(2018)Picogna, Stoll, \&
  Kley}]{picognaParticleAccretionPlanets2018b}
Picogna, G., Stoll, M. H.~R., \& Kley, W. 2018, \aap, 616, A116

\bibitem[{Pierens \& Huré(2005)}]{pierens_how_2005}
Pierens, A. \& Huré, J.~M. 2005, \aap, 433, L37

\bibitem[{{Pierens} \& {Nelson}(2008)}]{pierens2008constraints}
{Pierens}, A. \& {Nelson}, R.~P. 2008, \aap, 482, 333

\bibitem[{Price {et~al.}(2018)Price, Wurster, Tricco, Nixon, Toupin, Pettitt,
  Chan, Mentiplay, Laibe, Glover, Dobbs, Nealon, Liptai, Worpel, Bonnerot,
  Dipierro, Ballabio, Ragusa, Federrath, Iaconi, Reichardt, Forgan, Hutchison,
  Constantino, Ayliffe, Hirsh, \& Lodato}]{price_phantom_2018}
Price, D.~J., Wurster, J., Tricco, T.~S., {et~al.} 2018, Publications of the
  Astronomical Society of Australia, 35, e031

\bibitem[{{Pringle}(1981)}]{pringle1981}
{Pringle}, J.~E. 1981, \araa, 19, 137

\bibitem[{Regály {et~al.}(2012)Regály, Juhász, Sándor, \&
  Dullemond}]{regalyPossiblePlanetformingRegions2012a}
Regály, Z., Juhász, A., Sándor, Z., \& Dullemond, C.~P. 2012, \mnras, 419,
  1701

\bibitem[{Rein \& Liu(2012)}]{rein_rebound_2012}
Rein, H. \& Liu, S.-F. 2012, \aap, 537, A128

\bibitem[{Rein \& Spiegel(2015)}]{rein_ias15_2015}
Rein, H. \& Spiegel, D.~S. 2015, \mnras, 446, 1424

\bibitem[{Rendon~Restrepo \&
  Barge(2022)}]{rendonrestrepoMorphologyDynamicalStability2022}
Rendon~Restrepo, S. \& Barge, P. 2022, \aap, 666, A92

\bibitem[{Rendon~Restrepo \&
  Barge(2023)}]{rendonrestrepoSelfgravityThindiscSimulations2023}
Rendon~Restrepo, S. \& Barge, P. 2023, \aap, 675, A96

\bibitem[{{Rendon Restrepo} {et~al.}(2022){Rendon Restrepo}, {Barge}, \&
  {Vavrik}}]{2023_rendon_restrepoBarge_rossbi3d}
{Rendon Restrepo}, S., {Barge}, P., \& {Vavrik}, R. 2022, arXiv e-prints,
  arXiv:2207.04252

\bibitem[{{Rendon Restrepo} \&
  {Gressel}({2023})}]{2023_rendon_gressel_PPVII_poster}
{Rendon Restrepo}, S. \& {Gressel}, O. {2023}, {2D simulations of dust trapping
  by self-gravitating vortices}, {Protostars and Planets VII, poster PF-07-003}

\bibitem[{Rometsch {et~al.}(2020)Rometsch, Rodenkirch, Kley, \&
  Dullemond}]{rometsch_migration_2020}
Rometsch, T., Rodenkirch, P.~J., Kley, W., \& Dullemond, C.~P. 2020, \aap, 643,
  A87

\bibitem[{Rometsch {et~al.}(2021)Rometsch, Ziampras, Kley, \&
  Béthune}]{rometschSurvivalPlanetinducedVortices2021b}
Rometsch, T., Ziampras, A., Kley, W., \& Béthune, W. 2021, \aap, 656, A130

\bibitem[{Sauer(2012)}]{sauer2012numerical}
Sauer, T. 2012, Numerical Analysis (Pearson Education)

\bibitem[{{Shakura} \& {Sunyaev}(1973)}]{shakura1973alpha}
{Shakura}, N.~I. \& {Sunyaev}, R.~A. 1973, \aap, 500, 33

\bibitem[{Shu(1992)}]{shu_physics_1992}
Shu, F.~H. 1992, The physics of astrophysics. {Volume} {II}: {Gas} dynamics.
  (University Science Books)

\bibitem[{{Sod}(1978)}]{sod1978}
{Sod}, G.~A. 1978, Journal of Computational Physics, 27, 1

\bibitem[{{Speith} \& {Kley}(2003)}]{speith2003}
{Speith}, R. \& {Kley}, W. 2003, \aap, 399, 395

\bibitem[{{Stone} \& {Norman}(1992)}]{stone1992zeus1}
{Stone}, J.~M. \& {Norman}, M.~L. 1992, \apjs, 80, 753

\bibitem[{Stone {et~al.}(2020)Stone, Tomida, White, \&
  Felker}]{stone_athena_2020}
Stone, J.~M., Tomida, K., White, C.~J., \& Felker, K.~G. 2020, \apjs, 249, 4

\bibitem[{{Tscharnuter} \&
  {Winkler}(1979)}]{tscharnuter1979artificial_viscosity}
{Tscharnuter}, W.~M. \& {Winkler}, K. H.~A. 1979, Computer Physics
  Communications, 18, 171

\bibitem[{{Vaidya} {et~al.}(2015){Vaidya}, {Mignone}, {Bodo}, \&
  {Massaglia}}]{vaidya2015pvte}
{Vaidya}, B., {Mignone}, A., {Bodo}, G., \& {Massaglia}, S. 2015, \aap, 580,
  A110

\bibitem[{{van Leer}(1977)}]{van1977towards}
{van Leer}, B. 1977, Journal of Computational Physics, 23, 276

\bibitem[{{Von Neumann} \& {Richtmyer}(1950)}]{neumann1950shocks}
{Von Neumann}, J. \& {Richtmyer}, R.~D. 1950, Journal of Applied Physics, 21,
  232

\bibitem[{{Warner}(2003)}]{BrianWarner2003}
{Warner}, B. 2003, {Cataclysmic Variable Stars} (Cambridge University Press)

\bibitem[{Wilson {et~al.}(2012)Wilson, Aruliah, Titus~Brown, Chue~Hong, Davis,
  Guy, Haddock, Huff, Mitchell, Plumbley, Waugh, White, \&
  Wilson}]{wilson_best_2012}
Wilson, G., Aruliah, D.~A., Titus~Brown, C., {et~al.} 2012, arXiv e-prints,
  arXiv:1210.0530

\bibitem[{Woitke \& Helling(2003)}]{woitkeDustBrownDwarfs2003a}
Woitke, P. \& Helling, C. 2003, \aap, 399, 297

\bibitem[{{Woodward} \& {Colella}(1984)}]{woodward1984art_vis_and_godunov}
{Woodward}, P. \& {Colella}, P. 1984, Journal of Computational Physics, 54, 115

\bibitem[{Youdin \& Lithwick(2007)}]{youdinParticleStirringTurbulent2007}
Youdin, A.~N. \& Lithwick, Y. 2007, Icarus, 192, 588

\bibitem[{Zhu {et~al.}(2014)Zhu, Stone, Rafikov, \& Bai}]{zhu_particle_2014}
Zhu, Z., Stone, J.~M., Rafikov, R.~R., \& Bai, X.-n. 2014, \apj, 785, 122

\bibitem[{Ziampras {et~al.}(2023{\natexlab{a}})Ziampras, Nelson, \&
  Rafikov}]{ziamprasModellingPlanetinducedGaps2023}
Ziampras, A., Nelson, R.~P., \& Rafikov, R.~R. 2023{\natexlab{a}}, \mnras, 524,
  3930

\bibitem[{Ziampras {et~al.}(2023{\natexlab{b}})Ziampras, Paardekooper, \&
  Nelson}]{ziampras_buoyancy_2023}
Ziampras, A., Paardekooper, S.-J., \& Nelson, R.~P. 2023{\natexlab{b}}, \mnras,
  525, 5893

\end{thebibliography}
%%%%%%%%%%%%%%%%%%%%%%%%%%%%%%%%%%%%%%%%

%%%%%%%%%%%%%%%%%%%%%%%%%%%%%%%%%%%%%%%%
%%%%%%%%%%%%%%%%%%%%%%%%%%%%%%%%%%%%%%%%
%%%%%%%%%%%%%%%%%%%%%%%%%%%%%%%%%%%%%%%%
\appendix
%%%%%%%%%%%%%%%%%%%%%%%%%%%%%%%%%%%%%%%%
%%%%%%%%%%%%%%%%%%%%%%%%%%%%%%%%%%%%%%%%
%%%%%%%%%%%%%%%%%%%%%%%%%%%%%%%%%%%%%%%%

%%%%%%%%%%%%%%%%%%%%%%%%%%%%%%%%%%%%%%%%
\section{Parallel scaling}
%%%%%%%%%%%%%%%%%%%%%%%%%%%%%%%%%%%%%%%%
\label{sec:app_scaling_test}

This section presents the scaling of the code with the number of cores used in the simulation.
The scaling was measured on the Tübingen compute cluster BINAC on 1 to 16 nodes with 28 cores each of an Intel Xeon E5-2630v4 CPU connected via an InfiniBand network.

\begin{figure}
	\begin{center}
	\includegraphics[width=\linewidth]{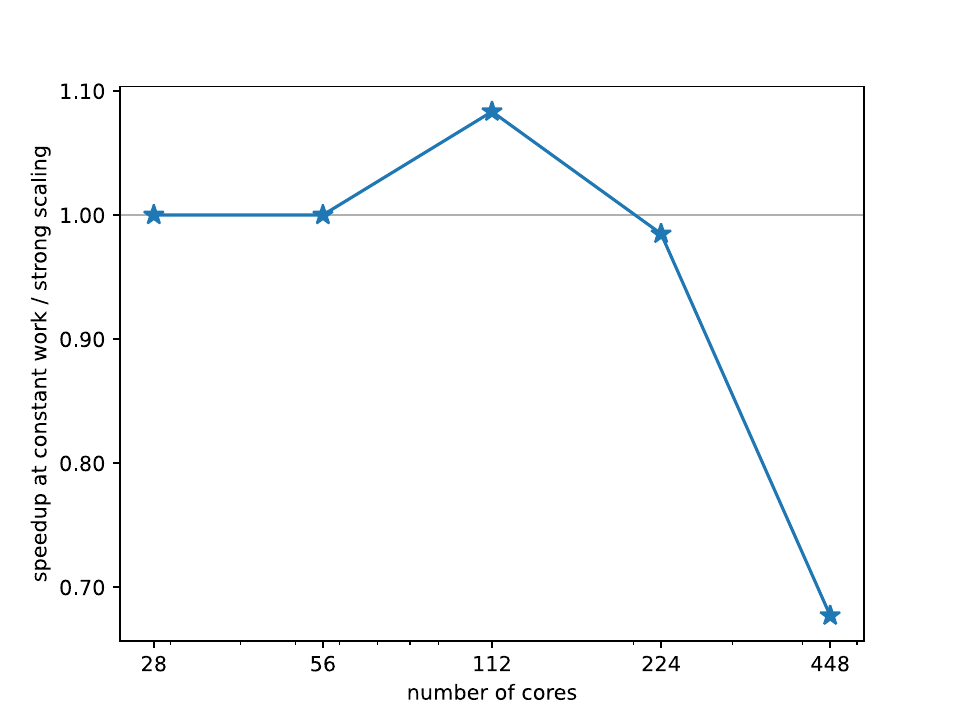}
	\caption{Speedup factor with number of cores on the Tübingen compute cluster BINAC.\label{fig:scaling}}
	\end{center}
\end{figure}

The test was performed using a locally isothermal disk with an embedded Saturn-mass planet and a grid size of $N_r \times N_\phi = 1024 \times 2048$.
The strong scaling speed-up, i.e. the time required for a constant workload (same grid size) divided by the number of cores, is nearly perfect up to 224 cores, see Fig.~\ref{fig:scaling}.

For 112 cores, there appears to be a super-linear speed-up. 
The code is parallelized by dividing the disk up into radial subdomains, each assigned to an MPI process (one for each NUMA node used). 
Each subdomain consists of several consecutive rings. 
These domains are then processed by several OpenMP threads (7 in the case of this test).
The case of 112 cores might be a sweet spot where the CPU cache is utilized more efficiently for this specific size of the subdomain, resulting in the super-linear speed up.
For higher core counts, the speed-up declines, which might be due to the increased communication overhead or because the grid size is not large enough to keep the cores busy.

%%%%%%%%%%%%%%%%%%%%%%%%%%%%%%%%%%%%%%%%
\section{Simulations of equilibrium disks}
%%%%%%%%%%%%%%%%%%%%%%%%%%%%%%%%%%%%%%%%
\label{sec:app_equilibrium_disk}

To simulate a steady-state accretion disk, we prescribe the densities and velocities of a locally isothermal model at the outer boundary with an additional wave-killing zone where we damp to only the velocities of the isothermal mode.
To achieve a steady state, we need to consider all forces in the centrifugal balance.

The model disk is described by the following equations.
The scale height is given by
\begin{equation}
        \label{eq:scaleheigt}
        H = h\,r = h_0 \,r \,\left( \frac{r}{R_0}\right)^F \,,
\end{equation}
with the flaring index $F$ and $R_0$ the radius where $h = h_0$.
The surface density is
\begin{equation}
        \label{eq:surfacedensity}
        \Sigma = \Sigma_0 \left( \frac{r}{R_0}\right)^{-S} \,,
\end{equation}
with the free parameter $S$.
The sound speed is
\begin{equation*}
\label{eq:soundspeed}
 c_{s,\,\mathrm{iso}} = h v_\text{K}\,,
\end{equation*}
with $v_\text{K}$ and $\Omega_\text{K}$ denote the Keplerian orbital velocity and frequency.
Assuming an ideal equation of state, the pressure is
\begin{equation*}
         \label{eq:pressure}
          P = c_{s,\,\mathrm{iso}}^2 \Sigma
\end{equation*}
and the energy density is
\begin{equation}
         \label{eq:energy}
          e = \frac{1}{\gamma - 1} \Sigma h^2 v_\text{K}^2
\end{equation}
where $r$ is the distance to the center of mass.
For angular momentum transport we use the $\alpha$-viscosity \citep{shakura1973alpha} prescription:
\begin{equation*}
         \label{eq:viscosity}
          \nu = \alpha H c_{s,\,\mathrm{iso}} \sqrt{\gamma}\,.
        \end{equation*}
%
% Where $\gamma$ is the adiabatic index and
Here, $\alpha$ is the viscosity parameter.

The gravitational interaction between the central star and the disk is computed using a smoothed gravitational potential of a point mass
\begin{equation}
\label{eq:potential}
 \Phi = -\frac{\mathrm{G}M}{\sqrt{r^2 + \epsilon^2}} = -\frac{\mathrm{G}M}{r} \frac{1}{\sqrt{1 + (\frac{\epsilon}{r})^2}}\,,
\end{equation}
with a smoothing length $\epsilon = \alpha_\text{sm} H$.
Note that this smoothing length also has a radial dependence. It's radial 
derivative is given by
\begin{equation*}
\frac{\partial \epsilon}{\partial r} = (F+1) \frac{\epsilon}{r}
\end{equation*}
With that in mind, we can compute the radial derivative of the smoothed	potential
\begin{equation}\label{eq:smoothed_potential_derivative}
\frac{\partial \Phi}{\partial r} = -\frac{\mathrm{G}M}{r^2}\frac{1 + (F+1) (\frac{\epsilon}{r})^2}{\left.\sqrt{1 + (\frac{\epsilon}{r})^2}\right.^3}\,,
\end{equation}
where we identify the second term on the r.h.s as the effect of the gravitational
force due to smoothing.

Following is a derivation of the equilibrium azimuthal velocity of the disk.
Assume, that the disk is axially symmetric and in a steady state, thus, in the radial momentum equation Eq.~\eqref{eq:momentum_r_eq_polar} we have $\partial/\partial t = 0$ and $\partial/\partial \phi = 0$. Additionally, assume, that $u_r$ is much smaller than $u_\phi$ and that radial changes of $u_r$ are small, so $u_r \partial u_r/\partial r$ can be neglected.
Furthermore, assume that viscous effects can be neglected, $f_r = 0$.
Eq.~\eqref{eq:momentum_r_eq_polar} then reduces to
\begin{align}\label{eq:centrifugal_balance}
	-\frac{u_\phi^2}{r} 
	&=-\frac{1}{\Sigma} \frac{\partial P}{\partial r}+ k_r+\frac{f_r}{\Sigma}\,,
\end{align}
Now assume that the gravitational forces are the only external force $k_r = - \partial \Phi / \partial r + g_r$ with the gravitational potential due to point masses $\Phi$ and the radial acceleration due to the gravity of the disk $g_r$. Then, multiply by $-r$ to arrive at
\begin{align*} \label{eq:eq_omega_condition}
	u_\phi^2 &= \frac{r}{\Sigma} \frac{\partial P}{\partial r} + r \frac{\partial \Phi}{\partial r} - r g_r
	= \frac{r\,c_s^2}{P} \frac{\partial P}{\partial r} + r \frac{\mathrm{G}M_*}{r^2} \, f_g - r g_r \\
	&= c_s^2 \frac{\partial \log P}{\partial \log r} + v_\text{K} \, f_g - r g_r\,.
\end{align*}
In the second step, we used $P = c_s^2 \Sigma$ and factored out the squared Keplerian velocity $v_K^2 = \mathrm{G}M_*/r$ from the gravitational potential term.
The new factor $f_g$ contains the information about the spatial dependence of the smoothing length and of higher multipole moments of the gravitational potential (for the quadrupole term see Eq.~\eqref{eq:quadrupole_term}).
For a non-smoothed gravitational potential of a point mass or one with a smoothing length without spacial dependence $f_g = 1$.
Typically, $f_g$ is close to unity.
Now we divide by $r^2$, take the square root, and use $c_s = h\,v_\text{K}$ to obtain the angular velocity
\begin{align*}
 \Omega = \sqrt{\Omega_\text{K} \left[ h^2 \frac{\partial \log P}{\partial \log r} + f_g \right] - r\,g_r}\,.
\end{align*}

Finally, we insert the expressions for $h$ (Eq.~\eqref{eq:scaleheigt}) and $\Sigma$ (Eq.~\eqref{eq:surfacedensity}), and differentiate the smoothed gravitational potential (see Eq.~\eqref{eq:smoothed_potential_derivative}) to obtain
\begin{equation} \label{eq:eq_omega} \Omega = 
\sqrt{\Omega_\text{K}^2 \left[ \left(2F - S - 1\right) h^2 + \frac{1+(F+1)h^2\alpha_\text{sm}^2}{(1+ h^2 \alpha_\text{sm}^2)^{(3/2)}} + \mathcal{Q} \right] - \frac{g_r}{r}}\,.
\end{equation}
The second term in square brackets would be equal to 1 if the effect of the gravitational smoothing is neglected and the formula would then be the common solution for a pressure-supported disk.

The quadrupole term $\mathcal{Q}$ (see Eq.~\eqref{eq:quadrupole_term}) is only needed for a disk around binary stars.
In this case, the approximation used for computing the quadrupole term is linearly independent of the other contribution to the centrifugal balance Eq.~\eqref{eq:centrifugal_balance}.
Therefore the contribution of the quadrupole moment can simply be added to the term inside the square brackets.

The radial velocity in a steady-state viscous accretion disk is given by \citep[e.g.,][]{lodato2008}:
\begin{equation}
	\label{eq:eq_vr_condition}
	\Sigma u_r \frac{\partial (r^2 \omega)}{\partial r} = \frac{1}{r} \frac{\partial}{\partial r} \left(\nu \Sigma r^3
\frac{d \omega}{d r}\right).
\end{equation}
While analytical solutions exist to Eq.~\eqref{eq:eq_vr_condition}, we found them to be impractical and chose to solve the equation numerically for $u_r$, using a five-point stencil derivative \citep[][p. 250]{sauer2012numerical} to generate a lookup table that is evaluated with linear interpolation.
To ensure that the in-falling mass rate has the exact prescribed value the density must only be created at the outer boundary.
Beware of using the wave-damping zones at the same time, because they can also create mass.

%%%%%%%%%%%%%%%%%%%%%%%%%%%%%%%%%%%%%%%%
\section{Leap-Frog like scheme}
%%%%%%%%%%%%%%%%%%%%%%%%%%%%%%%%%%%%%%%%
\label{appendix:leap_frog_scheme}
We implemented a Leap-Frog-like scheme for the time-stepping, trying to solve a numeric instability that arose in the simulation of circum-binary disk simulations.
It increases the accuracy of the source terms step by splitting it into two halves, one before and one after the transport step.
This corresponds to a kick-drift-kick Leap-Frog scheme known from N-body integrators.

During testing, we found that the biggest source of errors is in the transport step and improving the source terms step brings little improvement in most cases.
The gas feedback on the N-body system is already accurate with a single kick because the hydro time step is significantly smaller than the N-body time step would allow.
The scheme only noticeably improves the accuracy of the code for simulations where the transport step is not important, such as the heating and cooling test presented in Appendix.\,\ref{sec:heating_cooling_test}.
There, the simulation benefits from the higher temporal resolution of the source terms step.
However, this scenario would be identical to using a smaller time step for the whole simulation.

The rough outline of this scheme is that a kick-drift-kick scheme is used for the gas (when identifying the source terms as kick and transport as drift) and a drift-kick-drift scheme for the N-body system.

First, we advance the N-body system and the dust particles by half a time step from $t_0$ to $t_0 + 1/2 dt$ and use them at this position to update their velocities and compute their interaction with the gas. 
We then apply the source terms with half a time step and perform the transport step with a full time step. 
After transport, we compute the second half of the source terms and N-body-gas interactions and with the fully updated velocities evolve the N-body system and gas particles to the full time step.

The steps in the scheme are as follows:
\begin{itemize}
\item advance N-body by $\Delta t /2$ to $t + \Delta t /2$
\item update N-body velocities by $\Delta t /2$ from disk feedback
\item advance particles to $t + \Delta t / 2$ from interactions with N-body and gas
\item gas source terms with $\Delta t /2$
\item gas transport gas by $\Delta t$
\item update N-body velocities by $\Delta t /2$ from disk feedback
\item gas source terms with $\Delta t /2$
\item advance particles to $t + \Delta t$ from interactions with N-body and gas
\item advance N-body with previously calculated accelerations by $\Delta t /2$ to $t + \Delta t$
\end{itemize}

The N-body advance steps include accretion onto the N-body objects.

Ultimately, we found that the benefits of using this scheme are only minor.
The transport step is the largest source of error except in artificial test cases where the source terms are the only relevant part of the simulation.

The only noteworthy difference we could find is that the scheme needs fewer hydro steps compared to the default scheme because the CFL conditions for the source terms can be relaxed.
This can lead to less numerical diffusion but at slightly longer runtimes.
For example, for a typical simulation of a circumstellar disk in a close binary, the Leap-Frog scheme needed 25\% fewer iterations compared to the default scheme at 5\% increased runtime.

%%%%%%%%%%%%%%%%%%%%%%%%%%%%%%%%%%%%%%%%
%%%%%%%%%%%%%%%%%%%%%%%%%%%%%%%%%%%%%%%%
%%%%%%%%%%%%%%%%%%%%%%%%%%%%%%%%%%%%%%%%
\section{Test suite} \label{app:testsuite}
%%%%%%%%%%%%%%%%%%%%%%%%%%%%%%%%%%%%%%%%
%%%%%%%%%%%%%%%%%%%%%%%%%%%%%%%%%%%%%%%%
%%%%%%%%%%%%%%%%%%%%%%%%%%%%%%%%%%%%%%%%

This section describes some of the tests that are included in the test suite of \fargocpt.
The test suite can be run by executing the \texttt{run\_tests.sh} script within the \texttt{tests} directory.
For a list of the tests, please refer to Sect.~\ref{sec:testsuite}.

\subsection{Steady-state accretion test}
%%%%%%%%%%%%%%%%%%%%%%%%%%%%%%%%%%%%%%%%
\label{app:steady_state_accretion_test}

\begin{figure}
	\begin{center}
	\includegraphics[width=\linewidth]{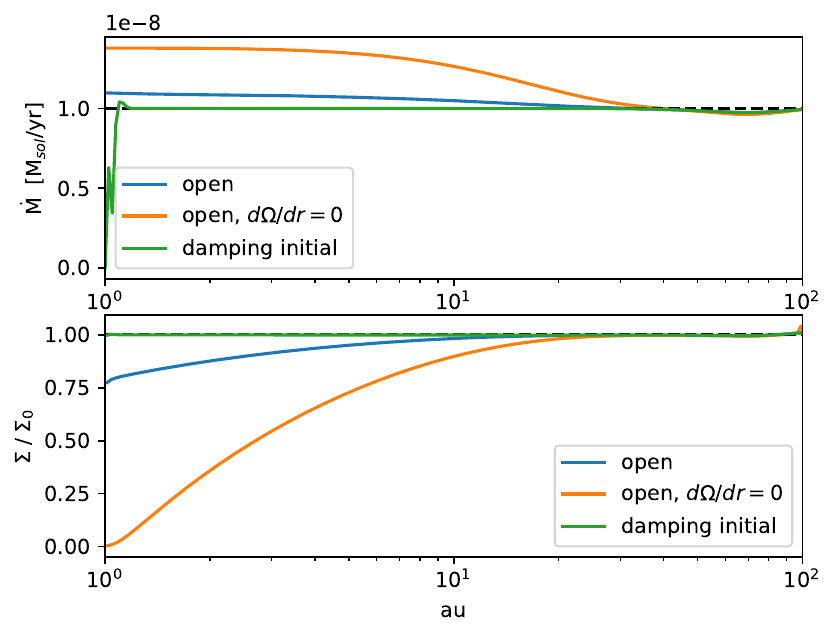}
	\caption{The outflow boundary removes more mass than supplied by the disk. The figure shows the accretion rate through the disk and surface density for different choices of boundary conditions.
	The top and bottom panels show the mass per unit of time flowing through a ring of a given radius and the ratio of surface density to its initial value, respectively.\label{fig:massflow_open}}
	\end{center}
\end{figure}

To test our boundary conditions (BCs), we initialize an infinite disk with a constant mass flow rate throughout the whole domain.
We use a simple locally isothermal model without any gravitational potential smoothing ($\Phi = \sqrt{\mathrm{G}M / r}$)
to simplify the equations in Sect.\,\ref{sec:app_equilibrium_disk}. Using $\Sigma(r) = 600.55 (\frac{r}{1\,\mathrm{au}})^{-1/2} \mathrm{g\,cm^{-2}}$, $h = 0.05$, $\alpha = 10^{-3}$, $F = 0$
around a $1\,\mathrm{M}_\odot$ star should result in $\dot{M} = 10^{-8} \mathrm{M}_\odot / \mathrm{yr}$. At the outer edge, we set the surface density
and velocities of the model in the ghost cells and damp to the velocities of the model near the boundaries (from 100 to 64\,au, with a damping time factor of $\beta=3$, see Eq.~\eqref{eq:damping_timescale}).
By only setting the density in the ghost cells and not damping to it, we keep precise control over the amount of mass flowing into the domain. In Fig.\,\ref{fig:massflow_open} 
we showcase the model on a domain from 1 to 100\,au 
and $\mathrm{N_r} \times \mathrm{N_\phi} = 192 \times 270$ resolution (square cells), the plot is taken after a time of $5\cdot 10^5$ orbits at $r = 1$.
At the outer boundary, our inflow condition deviates from the analytical model by 3\% for the mass flow rate and 0.4\% for the surface density.
At the inner boundary, a simple open boundary (blue line in Fig.\,\ref{fig:massflow_open}) leads to a larger radial velocity (and thereby mass flow rate) than the viscous speeds.
This causes the disk to be drained of mass inwards to outwards. Setting a no torque condition for the azimuthal velocity ($\mathrm{d}\Omega / \mathrm{d}r = 0$, orange line)
increases the radial velocity and mass drain at the inner boundary. We do not recommend this option, as it has also caused problems in other simulations.
When we damp the radial velocity and surface density to the initial values from 1 to 2\,au (green line), the mass rate is constant throughout the whole domain, except the few cells at the inner boundary; but those are well within the damping zone and do not affect the inner domain.

\begin{figure}
	\begin{center}
	\includegraphics[width=\linewidth]{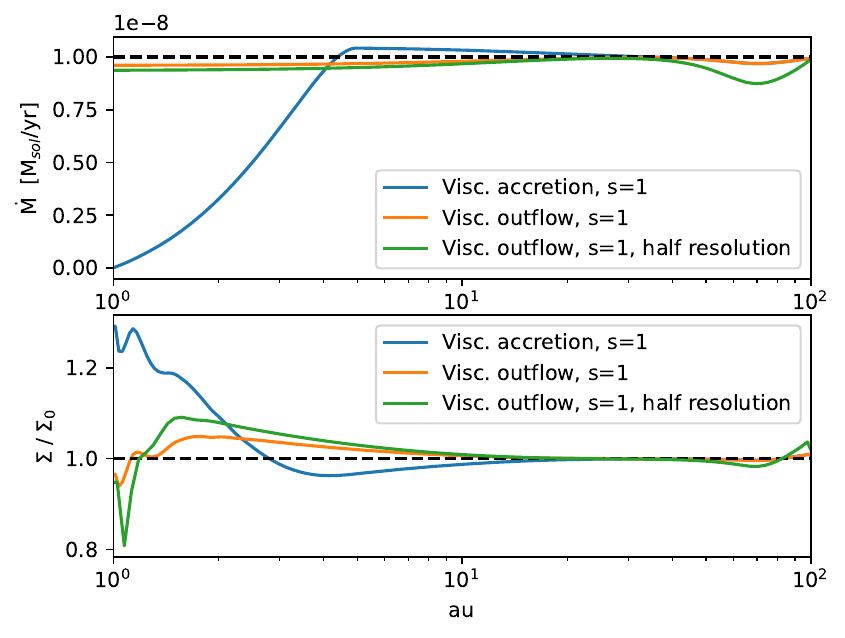}
	\caption{Viscous accretion through and at the inner boundary for different resolutions and methods.
	The panels are analogous to Fig.~\ref{fig:massflow_open}.
	The dip in mass flow rate for the blue line in the inner region is due to the method removing mass from the domain and thus disturbing the equilibrium state.
	\label{fig:massflow_viscous}}
	\end{center}
\end{figure}

In Fig.\,\ref{fig:massflow_viscous} we test our viscous outflow boundary (see Sect.~\ref{sec:boundary_conditions_vr}) and our viscous accretion function (see Sect.~\ref{sec:planet_accretion}), each with a
viscous enhancement factor of $s=1$. The viscous accretion function removes mass from 1 to 5\,au, which is not
measured in the mass flow rate inside the domain, the mass piles up near the inner boundary due to the reflective BCs used. 
Note that the accretion function is not intended as a BC but for accreting objects inside the simulation domain.
Both functions keep the surface density and mass flow rate relatively close to
the analytical model inside the domain. 
These functions utilize the viscosity of the gas to remove mass from the domain. 
Other mechanisms can drive accretion but are not measured by our $\alpha$ viscosity, such as angular momentum transfer by spirals due to a massive planet.
These have to be accounted for in the accretion enhancement factor $s$. This factor is an approximation for the increased accretion due to 
shearing at the boundary layer between the disk and the star (compare eq.\,46 in \citet{lodato2008}):
\begin{equation}
	\dot{M} = \frac{3 \pi \nu \Sigma}{1 - \sqrt{R_\mathrm{in} / R}} \approx 3 \pi \nu \Sigma s
\end{equation}
Using the viscous outflow condition allows to change the influence on the disk from a reflective-like boundary $s \lessapprox 3$
to an open like boundary $s \gtrapprox 10$. A value of $s=5$ was found to be suitable for accretion on a star with Jupiter-type companion
\cite{pierens2008constraints}.

Generally, all our tests show deviations from the analytical solution, which we suspect is due to the 
numerical errors when the equations are solved on the grid. Even small discrepancies in the velocity on the grid and
analytical solution can then lead to larger density pile-ups that we find in our tests. To test this, we 
repeated our viscous inflow simulation with half the resolution ($\mathrm{N_r} \times \mathrm{N_\phi} = 96 \times 135$)
which is the green line in Fig.\,\ref{fig:massflow_viscous}. This run shows slightly larger errors in the mass flow rate and 
significantly larger errors in the density profile close to the inner boundary.
%
%%%%%%%%%%%%%%%%%%%%%%%%%%%%%%%%%%%%%%%%
\subsection{Shock tube}
%%%%%%%%%%%%%%%%%%%%%%%%%%%%%%%%%%%%%%%%
\label{sec:shock_tube}

\begin{figure}
	\begin{center}
	\includegraphics[width=\linewidth]{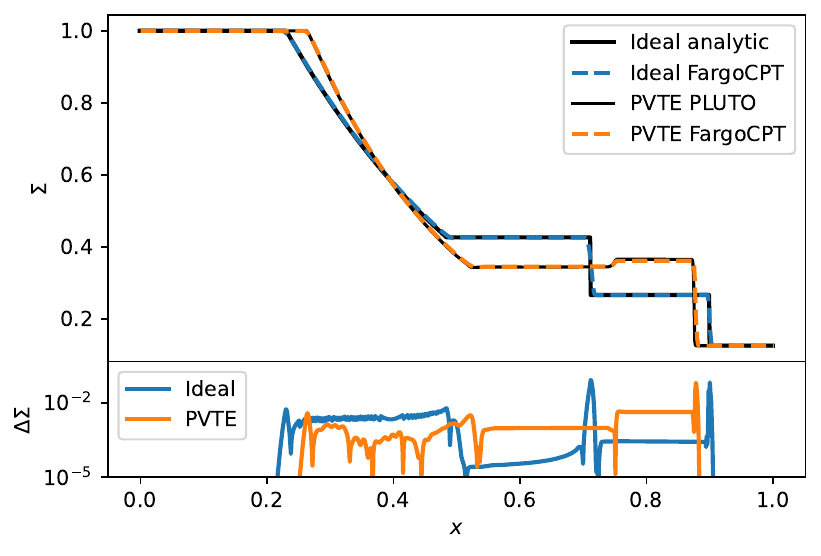}
	\caption{\label{fig:shocktube} Gas surface densities of the shock tube test with 100 cells for different combinations of artificial viscosity and integration schemes as well as for the caloric equation of state (PVTE) by \citet{vaidya2015pvte} with 1000 cells. The simulation time is $t = 0.228$ in both cases. The top and bottom panels show the surface density and the deviations from the reference case respectively, respectively.\label{fig:sod}
	}
	\end{center}
\end{figure}

We added the Sod shock tube test by \cite{sod1978} to our code, see Fig.\,\ref{fig:sod}, which is a classic test for the transport step, updated due to pressure forces and artificial viscosity.
The test is split into two parts, the first part is the classical shock tube with a perfect gas, meaning that the adiabatic index
for the sound speed and pressure are equal and constant and the second part is for an ideal gas with the caloric equation of state (PVTE) by \cite{vaidya2015pvte}.
For these tests, we approximate a Cartesian grid by spacing the radial cells arithmetically between $R_\mathrm{min} = 1000$ and $R_\mathrm{max} = 1001$.
As the setup is axisymmetric and the viscosity is zero, there is no interaction due to the azimuthal dimension.
All units are set to 1 for this test.
For the classical shock tube, we follow \cite{stone1992zeus1} and initialize the right half of the domain with $\Sigma_0 = 1$ and $e_0 = 2.5$ and the left half with $\Sigma_0 = 0.125$ and $e = 0.25$.
How to analytically solve the setup at $t > 0$ is described in \cite{hawley1984sod}.
The analytical solution as well as the results from our code with 100 cells for different combinations of integrator schemes and artificial viscosity
is shown in Fig\,\ref{fig:sod}. For space reasons, we only show the results for the densities. The full results can be viewed by executing the Python script of the test case in our code.
All four combinations of integration schemes and artificial viscosity reproduce the analytical solution and converge to it for higher resolutions without meaningful differences in quality and performance.

We used the same initial conditions for the caloric equation of state shock tube test. But in this case,
the units used are important and there is no known analytical solution. We copied the units from the shock tube test in \cite{vaidya2015pvte}
and used it as a reference to our results.
Both, the setup and the units are supplied with the official \textsc{Pluto} code inside the HD test problems and 1000 radial cells are used. The results are shown as the second set of lines in Fig.\,\ref{fig:sod}
where our \fargo run used the TW artificial viscosity and the standard integration scheme and agrees with the \textsc{Pluto} results.
%

%%%%%%%%%%%%%%%%%%%%%%%%%%%%%%%%%%%%%%%%
\subsection{Heating and cooling test}
\begin{figure}
	\begin{center}
	\includegraphics[width=\linewidth]{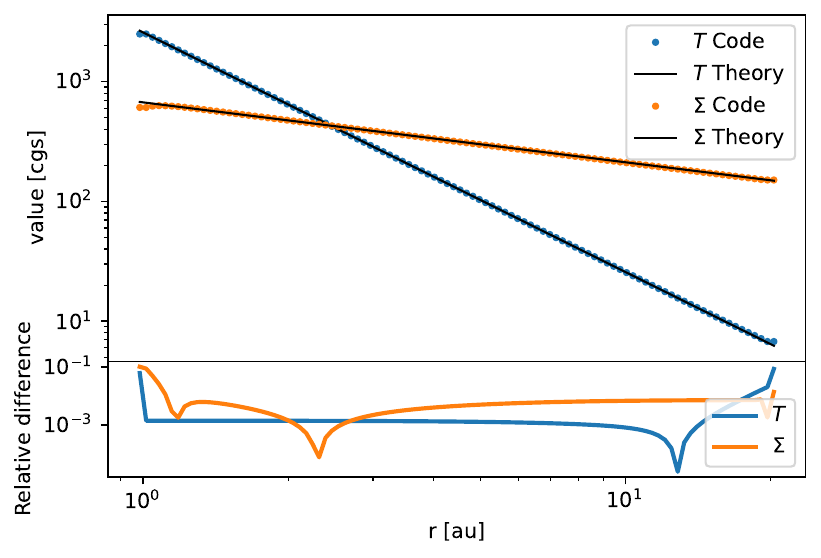}
	\caption{Gas surface density and temperature profile in the hydrostatic equilibrium 
	for the model from \cite{dangelo_thermohydrodynamics_2003} together with the expected theoretical profiles.\label{fig:TemperatureTest}}
	\end{center}
\end{figure}

%%%%%%%%%%%%%%%%%%%%%%%%%%%%%%%%%%%%%%%%
\label{sec:heating_cooling_test}
To evaluate our viscous heating and radiative cooling modules,
we used the setup and model presented in \cite{dangelo_thermohydrodynamics_2003}, see their Sect.~3.1.
The model simplifies the radiative cooling module such that an analytical formula
for the density and temperature profile for a disk in hydrostatic equilibrium
can be derived.
The formula for the effective opacity in \eqref{eq:effective_opacity} is changed to
\begin{align}
	\tau_\text{eff} = \frac{3}{8}\tau
\end{align}
with $\tau = 1/2 \kappa \Sigma$.
The opacity of the material in the disk is computed as
\begin{align}
	\kappa = 2\times 10^{-6}\, T^2 \mathrm{\;cm^2\,g^{-1}\,K^{-2}}.
\end{align}
The kinematic viscosity is set to a constant $\nu = 5 \times 10^{16} \mathrm{\;cm^2\,s^{-1}}$.
The setup consists of a central star $M = 1 \mathrm{M_\odot}$ and a domain ranging from 1 to 20\,au 
with reflective boundaries. To reach the equilibrium state faster, the radial velocities at the boundaries
are damped to zero. The initial surface density and temperature are set as constant
$\Sigma = 197\;\mathrm{g\,cm^{-2}}$ and $T = 352\,\mathrm{K}$. The simulation is then evolved to $10^4$ orbital periods at $r = 1\,\mathrm{au}$.
The expected equilibrium profiles for this specific setup are given by \citep{dangelo_thermohydrodynamics_2003}:
\begin{align}
	\Sigma(r) &= 300 \sqrt{\frac{5\,\mathrm{au}}r}\;\mathrm{g\,cm^{-2}}\\
	T(r) &= 104 \left(\frac{5\,\mathrm{au}}{r}\right)^{2}\;\mathrm{K}\\
\end{align}
and are plotted alongside the results from our code in Fig.\,\ref{fig:TemperatureTest}. Apart from deviations at the boundaries,
the theoretical profiles are reproduced by the code well.
%
%%%%%%%%%%%%%%%%%%%%%%%%%%%%%%%%%%%%%%%%
\subsection{Viscous spreading ring test}
\begin{figure}
	\begin{center}
	\includegraphics[width=\linewidth]{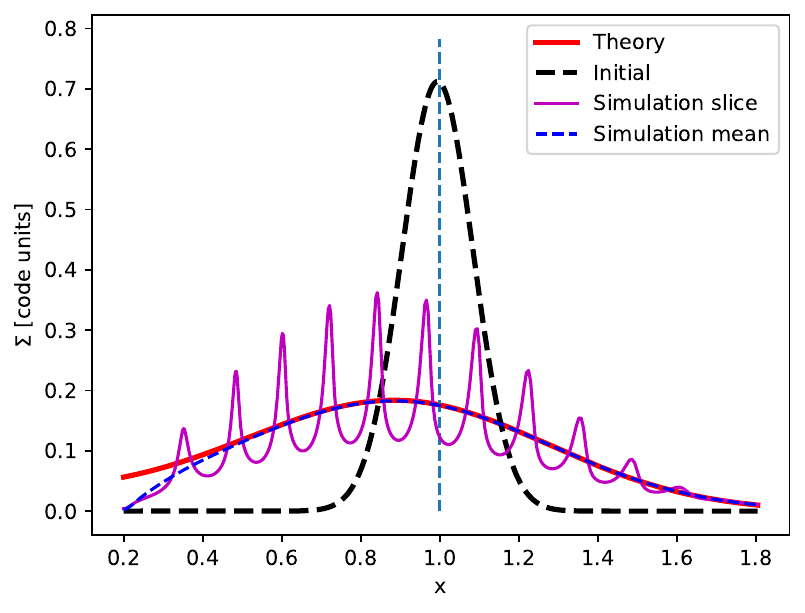}
	\caption{Viscous spreading ring test from \cite{speith2003} on a logarithmic grid $N_r \times N_\phi = 512\times256$.
	The ring is initialized with $\Sigma(x, \tau = 0.016)$ (black line) and then evolved to $\tau = 439.82$. The vertical dashed cyan line
	marks the position of the initial ring at time $\tau = 0$.
	The blue line is the azimuthal averaged surface density, which matches the analytical formula (red line) well.
	The viscous instability can be seen as waves in a density slice along the $x$ axis (magenta line).
	\label{fig:spreading_ring}}
	\end{center}
\end{figure}

%%%%%%%%%%%%%%%%%%%%%%%%%%%%%%%%%%%%%%%%
\label{sec:spreading_ring}
The viscous spreading ring is suited to test the ability of the code to transport angular momentum
due to the radial shearing inside disks. A pressure-less ring is initialized and then evolved in time.
Such a setup has been studied and solved analytically by \cite{Lust1952entwicklung,pringle1981} among others.
A gas ring of mass $m$ at radius $R_0$ from the central star evolves in terms of dimensionless radius $x = R/R_0$ and time $\tau = 12 \nu t R_0^{-2}$ as:
\begin{align}
\label{eq:ring}
\Sigma(x,\tau) = \frac{m}{\pi R_0^2} \tau^{-1} x^{-1/4} \;\mathrm{exp}\left[-(1+x^2)/ \tau \right] I_{1/4}(2x / \tau),
\end{align}
where $I_{1/4}$ is the modified Bessel function of the first kind. The ring is also subject to a viscous instability developing on top the the ring \citep{speith2003}. We initialized a spreading ring according to \eqref{eq:ring} at $\tau_0 = 0.16$ in a setup where $\mathrm{G}M_c = R_0 = 1$
on a logarithmic grid with $N_r \times N_\phi = 512\times256$ in a domain ranging from $0.2$ to $1.8$
with a constant kinematic viscosity of $\nu = 4.77\cdot 10^{-5}$. Fig\,\ref{fig:spreading_ring} shows the azimuthally averaged 
surface density at $\tau = 439.82$ and it agrees well with the analytical model from \eqref{eq:ring}. There are clear deviations from the analytical model
close to the inner boundary, which is due to the strict outflow boundaries used in the simulation. The viscous instability becomes
visible as density waves when taking a slice along the $x$ axis (magenta line). This instability was studied numerically in detail by \citet{joseph_measuring_2023},
where it was found that it should always develop as a one-armed trailing spiral spanning the whole domain.
%
%%%%%%%%%%%%%%%%%%%%%%%%%%%%%%%%%%%%%%%%

%%%%%%%%%%%%%%%%%%%%%%%%%%%%%%%%%%%%%%%%
\subsection{Dust diffusion test}
%%%%%%%%%%%%%%%%%%%%%%%%%%%%%%%%%%%%%%%%
\label{app:dust_diffusion_test}

\begin{figure}
	\begin{center}
	\includegraphics[width=\linewidth]{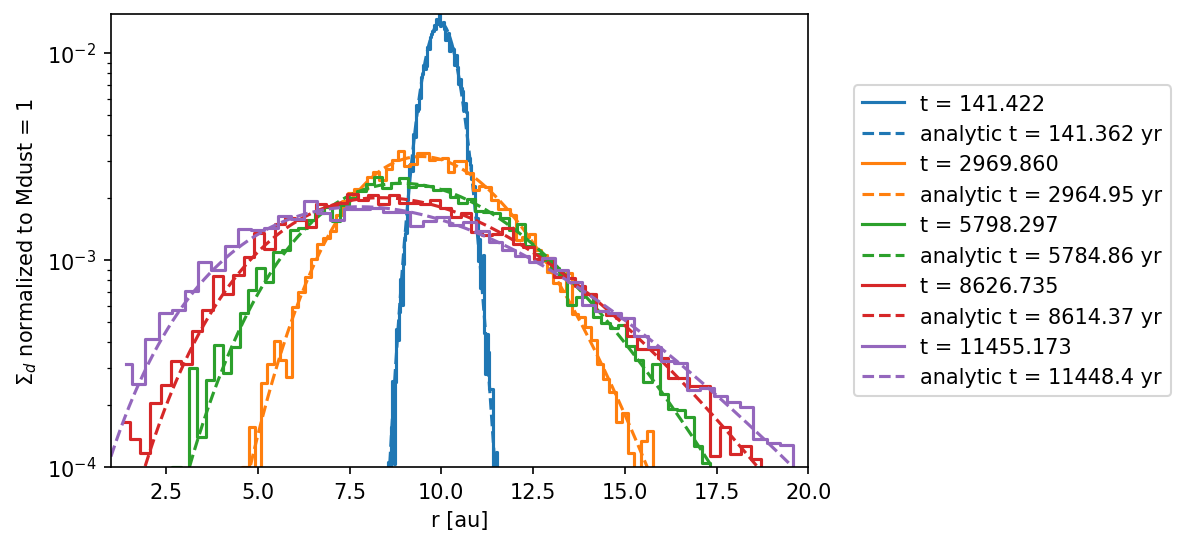}
	\caption{Dust diffusion test. Comparison of dust diffusion with Lagrangian super-particles in \fargocpt with numerical integration of the 1D advection-diffusion equation using an implicit method.\label{fig:dust_diffusion_test}}
	\end{center}
\end{figure}

This section presents the test case for the dust diffusion module.
We compare our stochastic implementation against a numerical solution of the 1D advection-diffusion equation.
The setup is inspired by the test case in \citet{charnoz_three-dimensional_2011} (see their Eq.~(28) for the 1D advection-diffusion equation).
We tried to exactly replicate their setup as described in their Sect.~3.3 with the goal of replicating their Fig. 5.
However, using the parameters available in their text and using educated guesses for the remaining model parameters we could not match the curves in their Figure.
Thus, we changed our reference to a 1D simulation with the DISKLAB code developed by Cornelis Dullemond and Til Birnstiel which can be used to solve the 1D advection-diffusion equation using an implicit method.
The comparison of the dust surface density calculated from the dust particle location histogram with the surface density from the DISKLAB simulation is shown in Fig.~\ref{fig:dust_diffusion_test}.
The dust surface density is normalized such that the total dust mass equals 1.
There is an excellent agreement between the results of the two approaches.

Notably, the 2D correction from Eq.~\eqref{eqn:dust_diffusion_2d_correction} to account for kicks in the azimuthal direction is needed to match the results from the advection-diffusion equation.

The parameters for the setup are as follows.
The simulation tracks 10.000 particles with a physical particle size of $10^{-5}$\,cm and a material density of $2.65$\,g/cm$^2$ on their orbit around a 0.5 solar-mass star.
Turbulence is parameterized by a viscous $\alpha = 0.01$.
The surface density of the gas follows $\Sigma(r) = 20\,\mathrm{g/cm}^2\, \left(\mathrm{r/1au}\right)^{-1}$.
At $r=10$\,au, where the particles are initially launched on circular orbits, the Stokes number is $\mathrm{St} = 2.08\times 10^{-5}$.
The grid spans from 1\,au to 40\,au with 1235 radial cells spaces logarithmically and 726 equally spaced azimuthal cells.
The time step is fixed at 0.1 in code units which corresponds to 5.81 days.

%%%%%%%%%%%%%%%%%%%%%%%%%%%%%%%%%%%%%%%%
\subsection{Dust drift test}
%%%%%%%%%%%%%%%%%%%%%%%%%%%%%%%%%%%%%%%%
\label{app:dust_drift_test}

This test repeats the test from \citet{picogna_how_2015} Appendix C.1, first suggested by \citet{zhu_particle_2014}, by comparing the dust drift velocity of individual particles with different stokes numbers to an analytical prediction.

The equilibrium radial dust drift velocity is expected to be \citep[][Eq.~(C.1)]{picogna_how_2015}
\begin{align}\label{eq:dust_drift_velocity}
	v_\text{drift} = \frac{\mathrm{St} v_{r,\mathrm{gas}} - \eta v_K}{\mathrm{St} + \mathrm{St}^{-1}}\,,
	\quad \eta = - h^2 \left(\frac{\mathrm{dlog}\Sigma}{\mathrm{dlog}r} + \frac{\mathrm{dlog}T}{\mathrm{dlog}r}\right)\,.
\end{align}
In this case, the gas is not evolved and its radial velocity $u_{r,\mathrm{gas}}$ is kept zero in the whole domain, such that only the second term contributes to the drift velocity.
Note that this definition of $\eta$ follows from \citet{nakagawa_settling_1986} Eq.~(1) for a locally isothermal disk by correcting for a missing '/' in their formula, such that $r \Omega_K^2$ is in the denominator.

\begin{figure}
	\begin{center}
	\includegraphics[width=\linewidth]{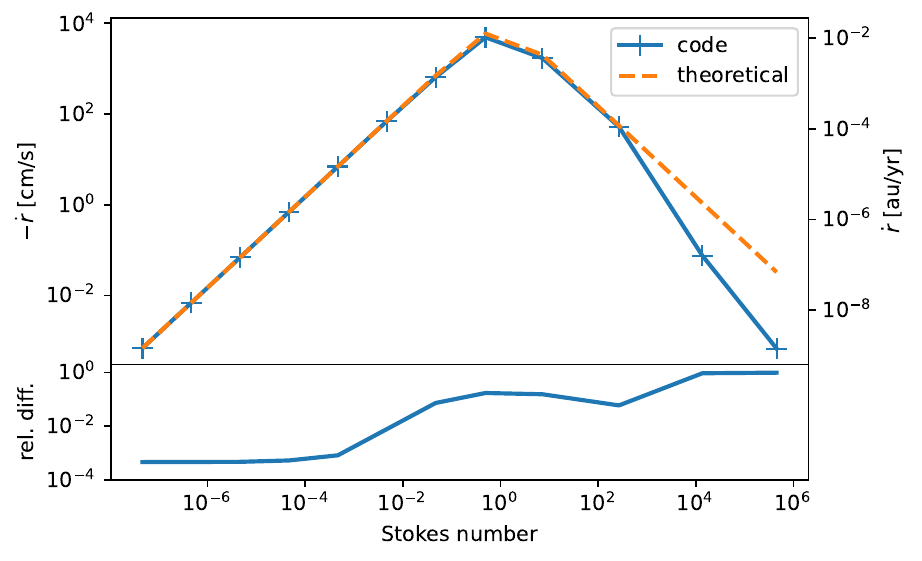}
	\caption{Results of the dust drift test. The top and bottom panel shows the value and deviation from the theoretical prediction of the drift speed as a function of the Stokes number, respectively. The large deviations for Stokes numbers around and larger than unity are due to oscillations in the integration. See Fig.~\ref{fig:dust_trajectories}. \label{fig:dust_drift}}
	\end{center}
\end{figure}

\begin{figure}
	\begin{center}
	\includegraphics[width=\linewidth]{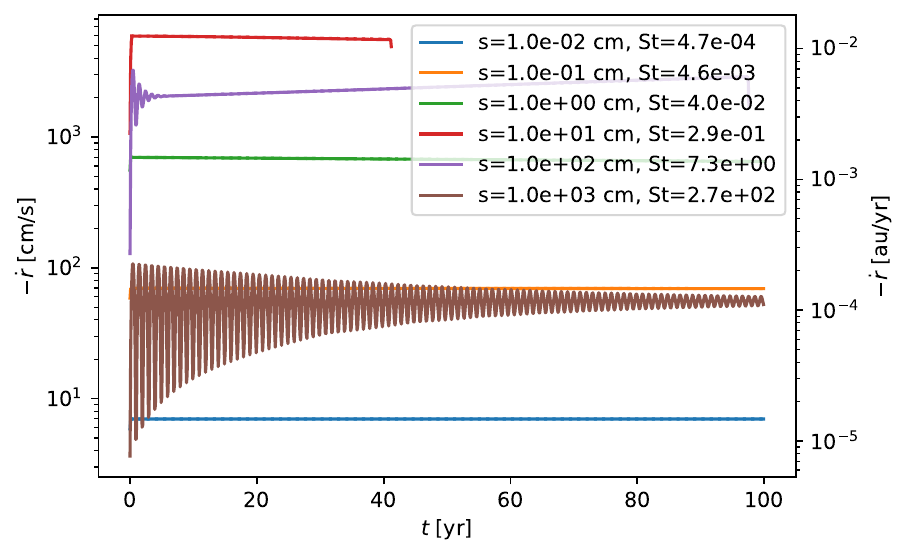}
	\caption{Selected dust particle trajectories from the dust drift test. The panel shows the drift velocity as a function of time with the particle size and Stokes number encoded by color. The black lines show the expected value from Eq.~\eqref{eq:dust_drift_velocity}. For larger dust particles with Stokes equal or greater than unity, oscillations occur due to the integration method. 
	The particles represented by the red and purple lines leave the domain at the inner boundary during the simulation.
	\label{fig:dust_trajectories}}
	\end{center}
\end{figure}

The resulting comparison of particle drift velocities is compared to the analytical predictions for the exponential midpoint integrator in Fig.~\ref{fig:dust_drift}.
The time evolution of these velocities is shown in Fig.~\ref{fig:dust_trajectories} where the oscillations of the velocity around the mean value can be observed for particles with $\mathrm{St} > 1$, in line with the findings of \citet{picogna_how_2015} Fig.~C.2 and \citet{zhu_particle_2014} Fig.~23.

The simulation models a protoplanetary disk around a \(1\,\mathrm{M}_\odot\) star, with an initial surface density of \(\Sigma(r) = 88.872\, \mathrm{g/cm}^2 \left(\mathrm{r/1au}\right)^{-1}\) at \(r=1\,\mathrm{au}\). The disk, with zero explicit viscosity and a constant aspect ratio of \(h=0.05\), extends from \(0.5\,\mathrm{au}\) to \(3\,\mathrm{au}\), and is governed by an isothermal equation of state.
Dust particles with material density $2.65$\,g/cm$^3$ and sizes from $10^{-8}$ to $10^{2}$\,m are initialized on circular orbits at $r=1\,\mathrm{au}$.

%%%%%%%%%%%%%%%%%%%%%%%%%%%%%%%%%%%%%%%%
\subsection{Planet torque}
%%%%%%%%%%%%%%%%%%%%%%%%%%%%%%%%%%%%%%%%
\label{app:planet_torque_test}

\begin{figure}
	\begin{center}
	\includegraphics[width=\linewidth]{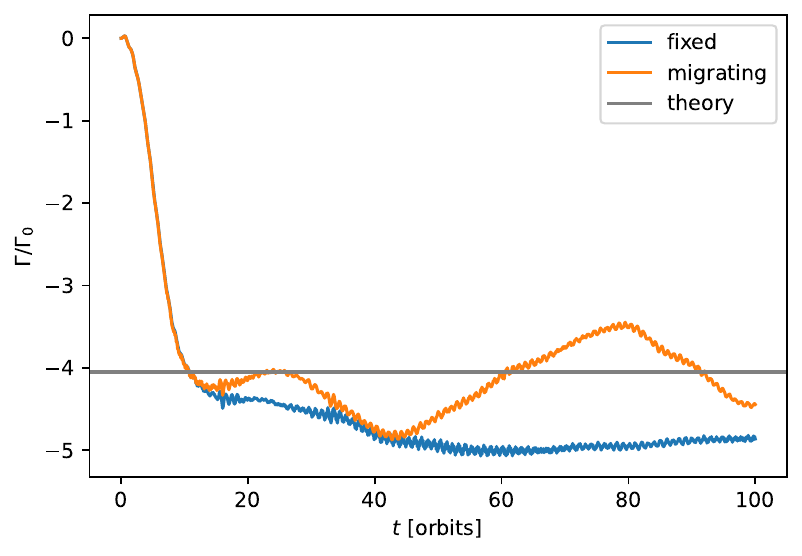}
	\caption{Torque of a small mass planet compared to theoretical prediction. The torque needs some time to build while the initially smooth disk adjusts to the presence of the planet.\label{fig:planet_torque}}
	\end{center}
\end{figure}

Linear Lindblad torque from \citep{paardekooper_torque_2011} (their Eq.~(14))
\begin{align}
\frac{\Gamma_\mathrm{L}}{\Gamma_0} = - \frac{1}{\gamma} \left( 2.5 + 1.7 \beta - 0.1 \alpha \right) \left( \frac{0.4}{b/h} \right)\,,
\end{align}
with the adiabatic index $\gamma$, $\beta = \frac{\mathrm{dlog}T}{\mathrm{dlog}r}$, $\alpha = \frac{\mathrm{dlog}\Sigma}{\mathrm{dlog}r}$, aspect ratio $h$, and the smoothing length factor $b = r_\mathrm{sm}/r$.
The torque normalization is $\Gamma_0 = (q/h)^2 \Sigma_p r_p^4 \Omega_p^2$ with the planet-to-star mass ratio $q$, the surface density at the planet location $\Sigma_p$, the planetary orbital radius $r_p$ and the planetary orbital angular velocity $\Omega_p$.

Figure~\ref{fig:planet_torque} shows a comparison of the torque as measured in two simulations and the theoretical prediction.
One simulation calculates the torques for a planet that is on a fixed orbit, while in the other simulation, the planet is allowed to move.
For the fixed planet, the resulting torque is overestimated while the torque for the moving planet oscillates around the expected value.

It should be noted that the boundary conditions can have a substantial impact on the torque in this test.
Here, reflective boundary conditions in combination with wave-damping zones towards the inner and outer boundaries had to be used.

The simulation features a $2\cdot10^{-5}\,$M$_\odot$ planet at $r=1\,$au around a $1\,$M$_\odot$ star.
The locally isothermal disk is initialized with $\Sigma(r) = 0.000376 \left(\mathrm{r/1au}\right)^{-1.5} = 3340.84\, \mathrm{g/cm}^2 \, \left( \mathrm{r/1au}\right)^{-1.5}$ with zero explicit viscosity and a constant aspect ratio of $h=0.05$.
The resolution is such that the scale height is resolved by 6 cells at the location of the planet which corresponds to 219 x 753 cells.

%%%%%%%%%%%%%%%%%%%%%%%%%%%%%%%%%%%%%%%%
\subsection{Flux-Limited-Diffusion test}
%%%%%%%%%%%%%%%%%%%%%%%%%%%%%%%%%%%%%%%%
\label{appendix:FLD_test}

This section describes the test of the flux-limited diffusion (FLD) module.
This test is a simple 1D diffusion test with constant opacity and two different temperatures at the inner and outer boundaries.

\begin{figure}
	\begin{center}
	\includegraphics[width=\linewidth]{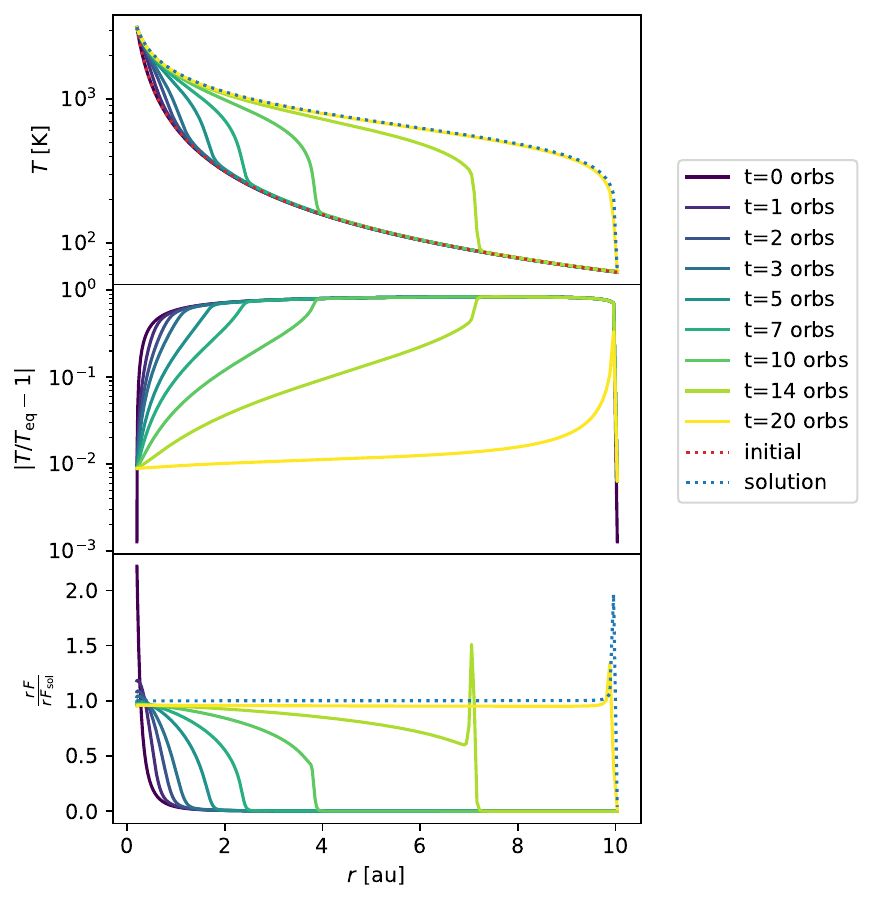}
	\caption{Equilibration of a radial temperature profile due to radiation transport in the midplane with a constant opacity. 
	This is used to test the FLD module.
	The panels show, from top to bottom, the radial temperature profile, the deviation from the equilibrium solution, and the azimuthally integrated flux through a ring at the respective radius, normalized by the equilibrium flux. Time is indicated by the color of the lines.
	 \label{fig:FLD_1D_test}}
	\end{center}
\end{figure}

Figure~\ref{fig:FLD_1D_test} shows the results of this test.
The panels show, from top to bottom, the radial temperature profile, the deviation from the equilibrium solution, and the azimuthally integrated flux through a ring at the respective radius, normalized by the equilibrium flux. Time is indicated by the color of the lines.
The azimuthally integrated flux is expected to be constant for an equilibrium disk.

The numerical criterion for passing this test is that the maximum deviation from the equilibrium solution is smaller than 0.1 inside of $r < 9.5\,\text{au}$.
The numerical solution shows boundary effects, because of which the pass criterion is relatively loose.
Please see the center panel of Fig.~\ref{fig:FLD_1D_test} for the radial profile and time evolution of the deviation.

%%%%%%%%%%%%%%%%%%%%%%%%%%%%%%%%%%%%%%%%
\subsection{Diffusion equation solver test}
%%%%%%%%%%%%%%%%%%%%%%%%%%%%%%%%%%%%%%%%
\label{appendix:sor_solver_test}

This test is aimed at the 2D diffusion part of the FLD solver and tests the solution of the diffusion equation with a constant diffusion coefficient.
The diffusion equation with constant coefficient
\begin{align}
	\frac{\partial x}{\partial t} = K \Delta x
\end{align}
has an analytical solution in 2D in the form of a Gaussian profile with a prefactor containing the time.
We use this analytical solution both as an initial condition and as the solution to compare.
In this test, we treat the temperature as an arbitrary variable and manually set the diffusion coefficient to a constant value.
The usual schedule for the time stepping is ignored and the initial condition is loaded from a file at the start of the diffusion test and the result is written out directly after the test.
We perform a specified number of iterations with a fixed time step.

With an infinite domain and a $\delta$ distribution as an initial condition, the analytical solution in two dimensions is
\begin{align}
	x(\vec{r}, t) = \frac{x_0}{4\pi t K} \exp\left( - \frac{|\vec{r} - \vec{r_0}|^2}{4Kt} \right) + c\,,
\end{align}
where $c$ is a constant offset.

We use a domain size of $r\in\{0.01,2\}\,$cm with 1000 uniformly spaced radial cells and 1500 azimuthal cells, a constant diffusion coefficient of $K = 1\,\text{cm}^2/\text{s}$, the center of the Gaussian profile $\vec{r}_0 = (1\,\text{cm}, 0)$, an offset $c=0.1$, and an initial time of $t_0 = 10^{-3}\,\text{s}$.
We then evolve the diffusion equation until $t = 2\times 10^{-3}\,\text{s}$ with 10 steps of $\Delta t = 10^{-4}\,\text{s}$.

\begin{figure}
	\begin{center}
	\includegraphics[width=\linewidth]{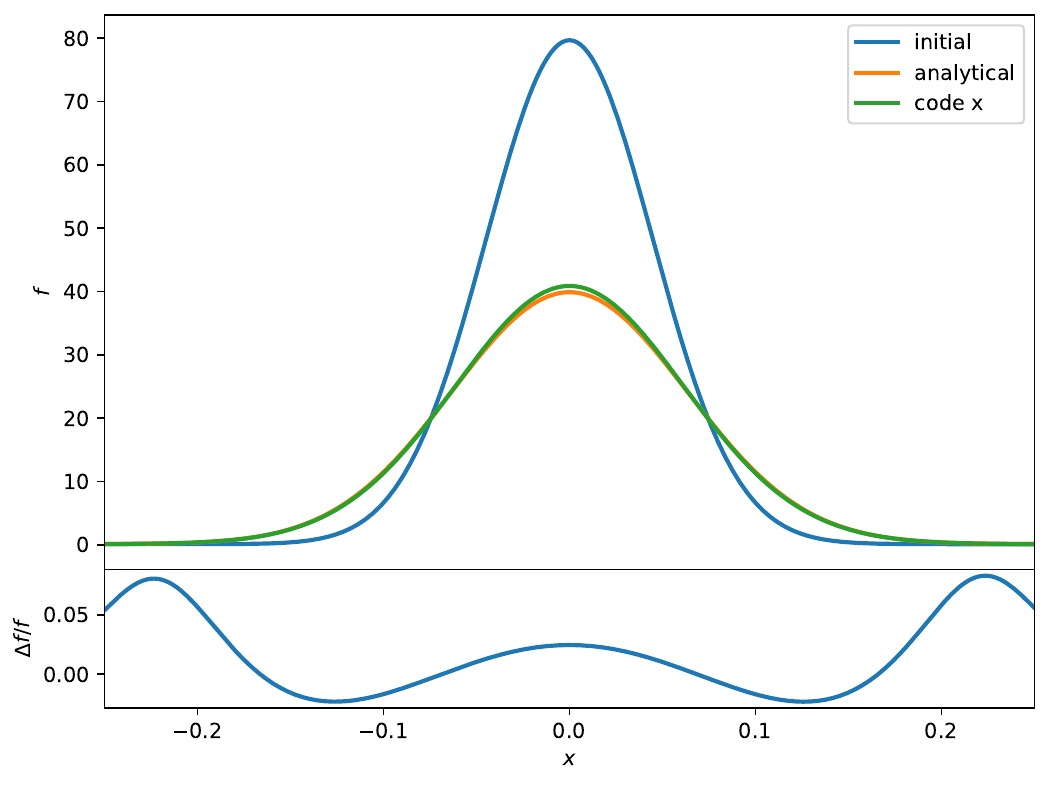}
	\caption{2D diffusion of the analytical solution to the diffusion equation with constant diffusion coefficient. The top panel shows radial cuts through the center of the 2D distributions: the initial condition and the analytical and numerical solutions. The bottom panel shows the relative deviation between the numerical and the analytical solution.\label{FLD_diffusion_test}}
	\end{center}
\end{figure}

A radial cut through the center of the Gaussian profiles and the deviation from the analytical solution is shown in Fig.~\ref{FLD_diffusion_test}.
The top panel shows the radial cut minus the offset and the bottom panel shows the relative deviation from the analytical solution.

The criterion for passing the test is that the integrated absolute deviation $\Delta = \sum_i^{N_\text{cells}} A_i \left| f_i^\text{code} - f_i^\text{analytical} \right|$ with the cell area $A_i$ is smaller than the threshold of $4\times10^{-2}$ at the given resolution.

There are deviations from the analytical solution at the center of the numerical solution.
It tends to be slightly higher than the analytical solution.
We suspect that this is due to boundary effects.
Increasing the resolution helps to reduce the deviation at the center, but it stays up to the resolution of 1000 radial and 1500 azimuthal cells, for which $\delta = 1.5\times10^{-2}$.
For runtime reasons, the resolution for the test suite is chosen lower at 100 times 150 cells, for which $\delta = 3.73\times10^{-2}$ and the threshold is chosen just above this value at $4\times10^{-2}$.

%%%%%%%%%%%%%%%%%%%%%%%%%%%%%%%%%%%%%%%%
\subsection{Self-gravity solver test}
%%%%%%%%%%%%%%%%%%%%%%%%%%%%%%%%%%%%%%%%
\label{appendix:sg_solver_test}

This test is aimed at verifying the self-gravity solver based on the Fourier method, as described in Sect.~\ref{sec:self-gravity}.
It is separated into two parts: a test of the solver in the radial direction and one in the azimuthal direction.
We test the implementation with the symmetric smoothing length given in Eq.~\eqref{eq:sg_smoothing_length_result}.

As a comparison, we recompute the gravitational acceleration from the surface density by direct summation according to
\begin{align}
\vec{a}(\vec{r}) = - \sum_{n=1}^{N_\text{rad}} \sum_{k=1}^{N_\text{az}} \frac{\text{G}\, A_{nk}\, \Sigma_{nk}}{\left(d^2 + \epsilon(r, r_{nk}, h)^2\right)^{3/2}} \vec{d}
\end{align}
where $\vec{d} = \vec{r}_{nk} - \vec{r}$, $d = |\vec{d}|$, and $A_\text{nk}$ is the cell area.
The quantities with subscript are the cell center values loaded from the simulation output.
These recomputed values are then compared to the output from the Fourier-method-based SG solver in the code.

In both cases, the test is performed on a 2D grid with $N_\text{rad} \times N_\text{az} = 128 \times 256$ cells, with a logarithmic radial grid spanning from 1 to 12.5\,au.

\begin{figure}
	\begin{center}
	\includegraphics[width=\linewidth]{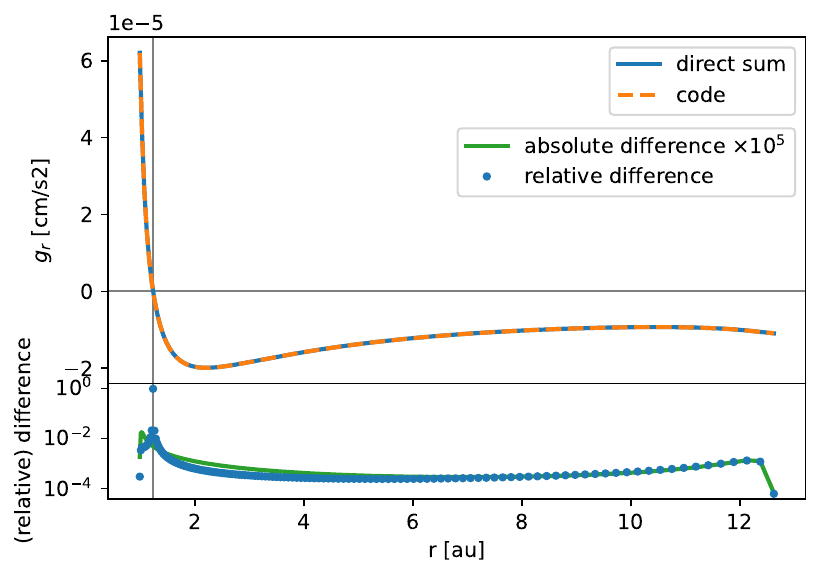}
	\caption{Comparison of the radial SG acceleration against results from direct summation. The top and bottom panels show the radial SG acceleration as a function of radius obtained with the Fourier method (code) and direct summation, and the relative and absolute differences between both curves, respectively. The horizontal gray line marks the zero value of acceleration and the vertical gray line indicates the crossing of zero. \label{fig:sg_solver_test}}
	\end{center}
\end{figure}

For the radial test, the surface density is axisymmetric and given by $\Sigma(r) = 200\,\mathrm{g/cm}^2 \left(\mathrm{r/1au}\right)^{-1}$ and the aspect ratio is constant throughout the disk with $h=0.05$.
Fig.~\ref{fig:sg_solver_test} shows the radial acceleration of the disk due to the self-gravity of the disk.
The top panel shows the radial SG acceleration as a function of radius for the direct summation $g_r^\text{ds}$ and the Fourier method $g_r^{F}$.
The bottom panel shows their relative difference $| g_r^\text{F}/g_r^\text{ds} - 1 |$ and absolute difference $| g_r^\text{F} - g_r^\text{ds}|$.
The two methods agree well which illustrates that the Fourier method works as intended.
The relative difference is below $0.0014$ for $r>2$\,au.
This value is used as a threshold in the pass-fail test.
We exclude the zone inwards of 2\,au because the acceleration has a crossing of zero there which enlarges the relative difference at this location.

For the azimuthal case, we initialize the surface density with two Gaussian peaks at $r_0 = 4$\,au at two different azimuths
\begin{align}
\Sigma(r, \phi) = \Sigma_0 \sum_{i\in\{1,2\}} \exp\left( - \frac{(r - r_0)^2}{2\sigma_r^2} - \frac{(\phi - \phi_i)^2}{2\sigma_\phi^2} \right)\,,
\end{align}
with $\sigma_r = 1$\,au, $\sigma_\phi = 0.3$\,rad, $\phi_1 = \pi$, $\phi_2 = \pi/2$ and $\Sigma_0 = 50\,\mathrm{g/cm}^2$.

\begin{figure}
	\begin{center}
	\includegraphics[width=\linewidth]{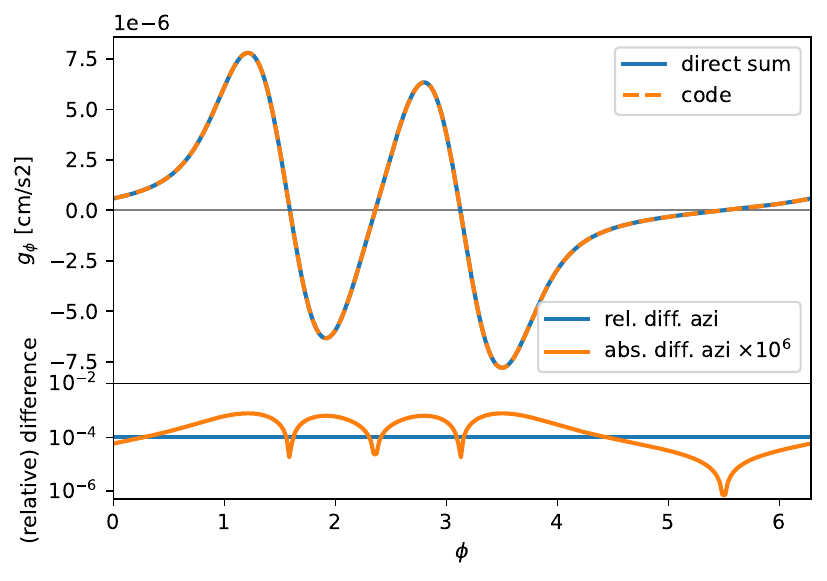}
	\caption{Like Fig.~\ref{fig:sg_solver_test} but showing the azimuthal SG acceleration as a function of azimuth at $r=4\,\text{au}$.\label{fig:sg_solver_test_azi}}
	\end{center}
\end{figure}

Fig.~\ref{fig:sg_solver_test_azi} shows the azimuthal acceleration due to the self-gravity of the disk as a function of azimuth at $r=4$\,au.
In this case, the match is even better than in the radial case and relative deviations are constant and smaller.

The difference between the radial and azimuthal direction might stem from a subtle difference in the implementation of the Fourier-based SG solver.
A property of this solver is that it implicitly treats all directions as periodic.
In the azimuthal direction, our simulation grid is periodic which makes the Fourier method directly applicable.
In the radial direction, however, a trick has to be used and the grid needs to be enlarged at the outer radial boundary to twice the size with cells containing zero density.
We suspect that this trick causes the radial acceleration to be less accurate than the azimuthal acceleration.

\end{document}